\documentclass{article}
\usepackage{authblk}
\usepackage[utf8]{inputenc}
\usepackage[english]{babel}
\usepackage[shortlabels]{enumitem}
\usepackage[margin=0.9in]{geometry}

\usepackage{pdfpages}
\usepackage{booktabs}

\usepackage{mathtools}
\usepackage{dsfont}
\usepackage{amsmath}
\usepackage{physics}
\usepackage{hyperref}
\usepackage{graphicx, caption}
\usepackage{tcolorbox}

\usepackage{complexity}
\usepackage{float}

\usepackage{cite}

\usepackage{nicematrix}

\renewcommand\vec{\mathbf}

\usepackage{wrapfig}
\usepackage{lscape}
\usepackage{rotating}
\usepackage{epstopdf}

\usepackage{amsthm}
\usepackage{amssymb}

\usepackage{tikz}
\usetikzlibrary{quantikz2}

\theoremstyle{definition}
\newtheorem{definition}{Definition}[section]
\newtheorem{exmp}[definition]{Example}
\newtheorem{theorem}{Theorem}
\newtheorem{corollary}[definition]{Corollary}
\newtheorem{lemma}[definition]{Lemma}
\newtheorem{remark}[definition]{Remark}
\newenvironment{claim}[1]{\par\noindent\underline{Claim:}\space#1}{}
\newenvironment{claimproof}[1]{\par\noindent\underline{Proof:}\space#1}{\hfill $\blacksquare$}

\newcommand{\ketf}[1]{|{#1})}

\begin{document}
	\title{Ternary tree transformations are equivalent to linear encodings of the Fock basis}
	
	\author[1]{Mitchell Chiew}
    \author[2]{Brent Harrison}
	\author[3]{Sergii Strelchuk}
	\affil[1]{\small DAMTP, Centre for Mathematical Sciences, University of Cambridge, Cambridge CB3 0WA, UK}
    \affil[2]{\small Department of Physics and Astronomy, Dartmouth College, Hanover, New Hampshire 03755, USA}
	\affil[3]{\small Department of Computer Science, University of Oxford, OX1 3QD, UK}

	\maketitle
	\begin{abstract}
		We consider two approaches to designing fermion--qubit mappings: (1) ternary tree transformations, which use Pauli representations of the Majorana operators that correspond to root-to-leaf paths of a tree graph and  (2) linear encodings of the Fock basis, such as the Jordan--Wigner and Bravyi--Kitaev transformations, which store linear binary transformations of the fermionic occupation number vectors in the computational basis of qubits.
		These approaches have emerged as distinct concepts, with little notational consistency between them.
		In this paper, we propose a universal description of fermion-qubit mappings, which reveals the relationship between ternary tree transformations and linear encodings. 
		Using our notation, we show that every ternary tree transformation is equivalent to a linear encoding of the Fock basis.


	\end{abstract}

\section{Introduction}

A fermion--qubit mapping is the core ingredient of any attempt to study the behaviour of fermionic particles with quantum computers using methods of second quantisation. Throughout the short history of the sub-field, many different encodings of the fermionic algebra have emerged, each with their own strengths and weaknesses \cite{Jordan:1928wi,bravyi_fermionic_2002,verstraete2005mapping,whitfield2016local,jiang_optimal_2020, Vlasov_2022,jiang2019majorana,derby2021compact,chien2022optimizing}. The central goal is to estimate the properties of a fermionic system, such as the ground state energy or simulation of the particles' dynamics over time, from the Hamiltonian of the system. In second quantisation, the Hamiltonians that describe the interactions between fermions take the general form
\begin{align}
	H_{\text{fermion}} = \sum_{i,j=0}^{n-1} (c_{ij}) \hat{a}_i ^\dagger \hat{a}_j + \frac{1}{2} \sum_{i,j,k,l=0}^{n-1} (c^{kl}_{ij}) \hat{a}^\dagger _i \hat{a}^\dagger _j \hat{a}_k \hat{a}_l\, , \label{eqn:ham}
\end{align} 
where the creation and annihilation operators $\{\hat{a}_i^{(\dagger)}\}_{i=0}^{n-1}$ act on a $(2^n)$--dimensional fermionic Hilbert space $\mathcal{H}_\text{fermion}$ and satisfy the canonical anticommutation relations
\begin{equation} \label{eqn:cars1}
	\{\hat{a}_i^\dagger , \hat{a}_j \} = \delta_{ij}, \, \quad \{\hat{a}_i , \hat{a}_j\} = 0\, , \quad \text{for all} \quad \, i,j \in [n]\, .
\end{equation}
Existing classical simulation algorithms scale exponentially with the system size for many problems of interest, proportional to the dimension of the Hilbert space, and in fact simulation tasks generally correspond to \QMA--complete computational problems \cite{kempe_complexity_2006, mcclean2014exploiting, ogorman_electronic_2021}.

Quantum computers offer ways to meet the computational challenges of fermionic simulation, with $n$ qubits encapsulating
a $(2^n)$--dimensional Hilbert space.
There are many strategies to study properties of Hamiltonians on quantum computers, such as phase estimation \cite{kitaev1995quantum,aspuru2005simulated,wang2008quantum,abrams1999quantum} and variational quantum eigensolvers \cite{mcclean2014exploiting, peruzzo2014variational, mcclean2016theory}.
Studying quantum algorithms for simulation of fermionic systems in the qubit (spin--$\frac{1}{2}$) space always requires, in
some form or another, a translation between the fermionic and qubit Hilbert spaces.

\begin{figure}[btp]
	\centering
	\includegraphics[width=0.9\linewidth]{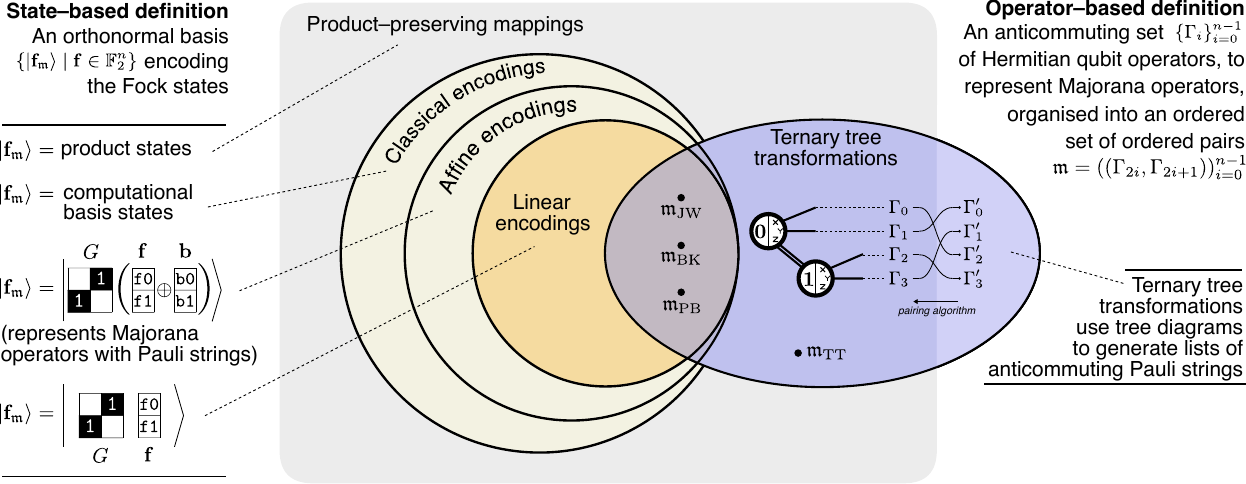}
	\caption{The state--based and operator--based approaches have led to seemingly distinct categories of fermion--qubit mappings. The mapping with the minimum average Pauli weight for Majorana operators is the ternary tree transformation $\mathfrak{m}_{\text{TT}}$; however it is unclear how ternary tree transformations relate to linear encodings of the Fock basis, another well--defined classes of fermion--qubit mappings.}
	\label{fig:litreview}
\end{figure}

Fermion--qubit mappings refer to the unitary maps between fermionic and qubit states and operators. With matching dimensions, an $n$--qubit quantum register can fully encode the dynamics of an $n$--mode fermionic system, preserving the properties of a fermionic Hamiltonian. The Jordan--Wigner transformation \cite{Jordan:1928wi} is the canonical example of an $n$--qubit mapping, and encodes fermionic occupation numbers  into the computational basis states. The representations of creation and annihilation operators under the Jordan--Wigner transformation are $\mathcal{O}(n)$--weight $n$--qubit Pauli strings, which offer an exponential improvement over the spatial requirements of classical methods, but still fall short of the limited scope of fault--tolerant operations on the quantum technology of today. New fermion--qubit mappings have emerged to reduce the weight of the qubit operators.
Allowing for additional ancilla qubits, locality--preserving fermion--qubit mappings encode terms of the fermionic Hamiltonian into operators acting on at most $\mathcal{O}(1)$ qubits, at the expense of using $\mathcal{O}(n)$ additional ancilla qubits \cite{verstraete2005mapping, derby2021compact, whitfield2016local, bravyi_fermionic_2002, chen2018exact, chen2023equivalence}.

This work focuses on the application of fermion--qubit mappings to encode fermionic Hamiltonians using the same number of qubits as there are modes in the system, which we term ``ancilla--free'' fermion--qubit mappings. The Bravyi--Kitaev transform \cite{bravyi_fermionic_2002} marked a potential improvement upon the Jordan--Wigner transformation, producing representations of the annihilation operators as sums of Pauli strings with maximum weight $(\lceil \log_2n \rceil + 1)$. 
The Bravyi--Kitaev and Jordan--Wigner transformations share the property of linearly encoding the Fock vectors into classical bit-strings: each Fock state $\ketf{\vec{f}} \in \mathcal{H}_\text{fermion}$ with mode occupation vector $\vec{f} \in \mathbb{F}_2^{\otimes n}$ corresponds to a computational basis state $\ket{\vec{q}}$ for $\vec{q} \in \mathbb{F}_2^{\otimes n}$ through an invertible binary matrix multiplication $\vec{q}  = G \vec{f}$ for $G \in \text{GL}_n(\mathbb{F}_2)$.
Mappings with this property, which we call ``linear encodings of the Fock basis", correspond to classical data structures \cite{harrison_sierpinski_2024} and  invertible binary matrices \cite{seeley_bravyi-kitaev_2012, wang_resource-optimized_2021, wang2023evermore, steudtner2018fermion, steudtner2019quantum}. In linear encodings, fermionic operations manifest through update, parity check, and flip rules on superpositions of the classical $n$--bit vectors representing the fermionic system.
Searching over the space of invertible binary matrices \cite{wang_resource-optimized_2021, wang2023evermore} led to the discovery of linear encodings that produce Hamiltonians for molecular compounds with operators weights lower still than the Bravyi--Kitaev transformation. For example, the pruned Sierpinski tree transform \cite{harrison_sierpinski_2024} is a another linear encoding that implements the update rules of fermionic interactions in only ${\sim} \lceil\log_3 (2n+1)\rceil$ fundamental quantum gates.

Ternary tree transformations \cite{Vlasov_2022, miller2023bonsai, miller2024treespilation} marked a paradigm shift towards thinking primarily of the encoded operators of a fermion--qubit mapping, rather than the encoded form of the Fock states.
The operator--first perspective enabled the discovery of a mapping with the provably minimal average Pauli weight of ${\sim}\lceil \log_3(2n+1)\rceil$ for Majorana operators \cite{jiang_optimal_2020, Vlasov_2022}, and seemingly expanded fermion--qubit mappings into new territory. The defining feature of ternary--tree transformations is the branching structure of the encoded Pauli operators, which the Bonsai algorithm \cite{miller2023bonsai} exploits to reduce SWAP gate requirements of fermionic simulation on real-world quantum hardware. Treespilation \cite{miller2024treespilation} performs a computational search over a well-behaved subset of ternary tree transformations to minimise the CNOT cost for preparing qubit ansatzes for molecular simulation.

Despite recent progress, a comprehensive framework for fermion--qubit mappings remains elusive, and existing analytic results did not focus on the equivalence of the state-- and operator--based definitions or make clear that there is a significant relation between linear encodings of the Fock basis and ternary tree transformations. The numerous descriptions of fermion--qubit mappings have led to a large number of seemly distinct approaches, as Figure \ref{fig:litreview} illustrates, but differing terminologies can also conceal equivalences which could simplify the decision--making process. For one, the identical operator weighting of the pruned Sierpinski tree transformation's update rules \cite{harrison_sierpinski_2024} and the Pauli weight of the operators in the complete ternary tree transformation \cite{jiang_optimal_2020, Vlasov_2022} suggests that, perhaps, the two are one and the same.

\begin{figure}[btp]
	\centering
    \includegraphics[width=0.9\linewidth]{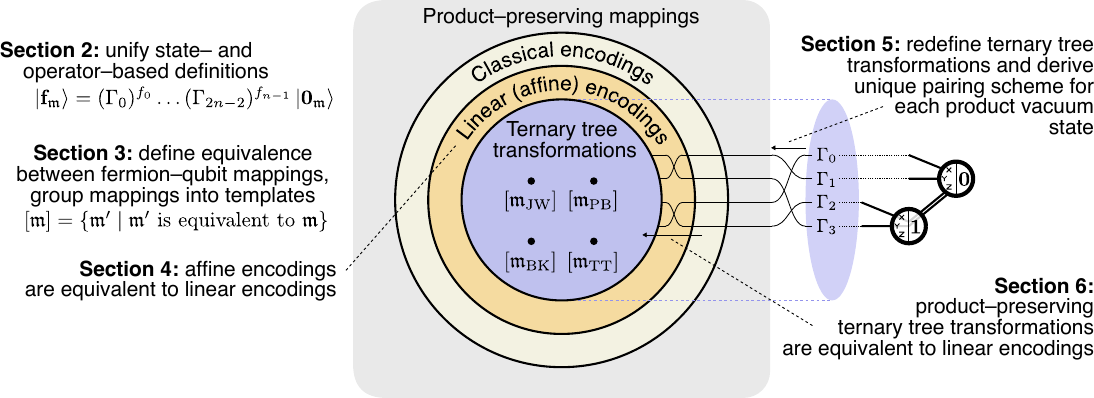}
	\caption{This paper provides a unified description of ancilla--free fermion--qubit mappings and reveals that all product--preserving ternary tree transformations are linear encodings of the Fock basis, up to labelling and sign choices for Pauli strings.}
	\label{fig:thispaper}
\end{figure}

In this paper, we give a comprehensive definition of fermion--qubit mappings which unifies previous approaches. This allows us to prove that ternary tree transformations with product--state encodings fall within a subset of linear encodings of the Fock basis up to a trivial equivalence. Figure \ref{fig:thispaper} demonstrates the simplification of the concepts in Figure \ref{fig:litreview} which our approach makes possible. Our findings are summarised as follows.
    
	\begin{itemize}
		\item \textbf{Section \ref{sec:prelim}: A unified definition for fermion--qubit mappings.} We give two equivalent definitions for a fermion--qubit mapping $\mathfrak{m}$, first as an ordered sequence of Hermitian, unitary operator pairs and then as an orthonormal basis of qubit states.
        A unitary operator, unique to each fermion--qubit mapping, links the two definitions.

		\item \textbf{Section \ref{sec:equiv}: Equivalence relations for fermion--qubit mappings.}
		We establish a practical notion of an equivalence relation on the set of fermion--qubit mappings that represent Majorana operators with Pauli strings.
		
		\item \textbf{Section \ref{app:ibm}: Classical, affine and linear encodings of the Fock basis.} 
		We prove that a classical encoding of the Fock basis with Pauli operators is an affine transformation 
        of the Fock basis. We show that the affine encodings of the Fock basis are equivalent, in the sense of Section \ref{sec:equiv}, to linear encodings 
        up the signs of the Pauli operators.
		
		\item \textbf{Section \ref{sec:treebased}: Ternary tree transformations.}
		We define ternary tree transformations to a rigorous standard using our notation, revealing that the choice of a single Fock state completely determines the operator pairing scheme of a ternary tree transformation.
		
		\item \textbf{Section \ref{sec:cbptree}: Product--preserving ternary tree transformations are equivalent to linear encodings of the Fock basis.}
		Our main result shows that each product--preserving ternary tree transformation is equivalent to a linear encoding of the Fock basis.
		We provide a formula to determine the specific linear encoding of the Fock basis that corresponds to each ternary tree graph.

        \item \textbf{Appendix \ref{sec:appendix}} provides a glossary of notation and \textbf{Appendix \ref{app:affine}} reviews stabiliser tableaux of affine encodings.
		
	\end{itemize}

	\section{A unified definition for fermion--qubit mappings}\label{sec:prelim}

	The purpose of this section is to unify two different approaches to defining fermion--qubit mappings. The first method, to encode the Fock basis in qubit states, led to the Jordan--Wigner \cite{Jordan:1928wi}, Bravyi--Kitaev \cite{bravyi_fermionic_2002} and parity basis \cite{seeley_bravyi-kitaev_2012} transformations along with other linear encodings \cite{steudtner2018fermion, wang2023evermore}. The second method, to represent the Majorana operators as Pauli operators, led to the ternary tree transformation \cite{jiang_optimal_2020} and its generalisations \cite{miller2023bonsai, miller2024treespilation}.
	

    The mappings we consider here are \textit{ancilla--free}, in that they map $n$--mode fermionic systems to $n$--qubit systems. Mappings can use ancilla qubits to unitarily embed fermionic Hamiltonians into $m$--qubit systems for $m>n$, and such constructions are the focus of active investigation \cite{verstraete2005mapping, derby2021compact, whitfield2016local, bravyi_fermionic_2002, chen2018exact, chen2023equivalence}. However, any ancilla--qubit mapping is the restriction of some {$m$--mode ancilla--\textit{free}} fermion--qubit mapping (see e.g.\ Section 5 of \cite{bravyi_fermionic_2002}), meaning there is no fundamental mathematical distinction between ancilla--free and mappings that use ancillas. We include the phrase ``ancilla--free" in parentheses throughout the paper to stress that our work follows conceptually from previous results that apply to $n$--qubit mappings designed for $n$--mode systems, such as linear encodings of the Fock basis and ternary tree transformations.

    Throughout this work we label quantum states of fermionic systems with curved ket notation $\ketf{\psi}$ as in \cite{obrien_local_2024,bultinck2017fermionic, harrison_sierpinski_2024}. Qubit systems have quantum states with the conventional notation $\ket{\psi}$.
		
	\begin{definition}\textit{(Fock basis of fermionic state space.)}\label{defn:fock} Given an $n$--mode fermionic system with state space $\mathcal{H}_\text{fermion}$ and annihilation operators $\{\hat{a}_i\}_{i=0}^{n-1}$, choose $\ketf{\Omega_\text{vac}}\in \mathcal{H}_\text{fermion}$ as a simultaneous 0--eigenstate of the number operators $\{\hat{a}_i^\dagger \hat{a}_i\}_{i=0}^{n-1}$. Any two simultaneous 0--eigenstates of the number operators are proportional, differing by only a phase $e^{i\phi}$ for some $\phi \in [0,2\pi)$. We define the \textit{Fock basis of $\mathcal{H}_\emph{fermion}$} to be the set $\{\ketf{\vec{f}}\, | \, \vec{f}\in \mathbb{F}_2^n\} \subset \mathcal{H}_\text{fermion}$, where
		\begin{align}
			\ketf{\vec{f}} = \ketf{f_{0}, f_{1} , \dots , f_{n-2} , f_{n-1}} \coloneqq (\hat{a}_0^\dagger)^{f_0} (\hat{a}_1^\dagger)^{f_1} \dots (\hat{a}_{n-1}^\dagger)^{f_{n-1}} \ketf{\Omega_{\text{vac}}}\, . \label{eqn:fockbasis}
		\end{align}
		The physical significance of the Fock basis is that each state $\ketf{\vec{f}} \in \mathcal{H}_\text{fermion}$ has well-defined occupancy: if the system is in state $\ketf{\vec{f}}$, then mode $i$ is occupied by a fermion if and only if $f_i=1$.
	\end{definition}
	
	Note that altering the ordering of annihilation operators in Equation \ref{eqn:fockbasis} would result in an alternative Fock basis $\{\ketf{\vec{f}}' \, | \, \vec{f}\in \mathbb{F}_2^n\}$ with $\ketf{\vec{f}}' = -\ketf{\vec{f}}$ for a subset of the strings $\vec{f} \in \mathbb{F}_2^n$. While this is unimportant to the physics on a fermionic level, it is important to specify an ordering here to properly define fermion--qubit mappings. As an example,  Remark \ref{rem:ordering} demonstrates that the fermionic mode ordering affects the relative sign of the Fock states of the Jordan--Wigner transformation.
	
	\begin{definition}\textit{(Majorana operators.)} \label{defn:maj}
			Let $\{\hat{a}_i\}_{i=0}^{n-1}$ be the annihilation operators of an $n$--mode fermionic system. Then, the \textit{Majorana operators} $\{\hat{\gamma}_i\}_{i=0}^{2n-1}$ satisfy $\hat{a}_i = \frac{1}{2}(\hat{\gamma}_{2i} + i \hat{\gamma}_{2i+1})$, and are an alternative set of generators for the fermionic algebra. The canonical anticommutation relations of Equation \ref{eqn:cars1} translate to
			\begin{align}\label{eqn:majcars}
				\{\hat{\gamma}_i, \hat{\gamma}_j\} = 2\delta_{ij} \, , \quad \hat{\gamma}_i^\dagger = \hat{\gamma}_i \, .
			\end{align}
	\end{definition}

	\begin{lemma} \label{lem:occnobasis}
		\emph{(Defining the Fock basis in terms of Majorana operators.)} Let $\mathcal{H}_\text{fermion}$ be the state space of an $n$--mode fermionic system with  annihilation operators $\{\hat{a}_i\}_{i=0}^{n-1}$ and vacuum state $\ketf{\Omega_\text{vac}}$. Then, the Fock basis of $\mathcal{H}_\text{fermion}$ has the following equivalent definitions:
		\begin{align}
			\ketf{\vec{f}} &= (\hat{a}_0^\dagger)^{f_0} (\hat{a}_1^\dagger)^{f_1} \dots (\hat{a}_{n-1}^\dagger)^{f_{n-1}} \ketf{\Omega_{\text{vac}}}
			\\
			&= (\hat{\gamma}_0)^{f_0} (\hat{\gamma}_2)^{f_1} \dots (\hat{\gamma}_{2n-2})^{f_{n-1}} \ketf{\Omega_\text{vac}} \label{eqn:onbasis1}\\
			&= (-i\hat{\gamma}_1)^{f_0} (-i\hat{\gamma}_3)^{f_1} \dots (-i\hat{\gamma}_{2n-1})^{f_{n-1}} \ketf{\Omega_\text{vac}} \quad \text{for all} \quad \vec{f} \in \mathbb{F}_2^n\, . \label{eqn:onbasis2}
		\end{align}
		\begin{proof}
			Note that the vacuum state $\ketf{\Omega_\text{vac}}$ is a simultaneous 0--eigenstate of $\hat{a}_i^\dagger \hat{a}_i$, and that $\hat{a}_i^\dagger \hat{a}_i = \frac{1}{2} (\mathds{1} + i \hat{\gamma}_{2i} \hat{\gamma}_{2i+1})$. Therefore $\ketf{\Omega_\text{vac}}$ is a simultaneous $(+1)$--eigenstate of $-i\hat{\gamma}_{2i}\hat{\gamma}_{2i+1}$ for all $i=0,1,\dots,n-1$. Combine this with the fact that $\left(\hat{\gamma}_{2i}\right)^2 = \left(\hat{\gamma}_{2i+1}\right)^2 = \mathds{1}$ to obtain
			\begin{align}
				\hat{a}_{i}^\dagger \ketf{\Omega_\text{vac}} =  \frac{1}{2} (\hat{\gamma}_{2i} - i \hat{\gamma}_{2i+1}) \ketf{\Omega_\text{vac}} &= \frac{1}{2} \hat{\gamma}_{2i} (\mathds{1} - i \hat{\gamma}_{2i} \hat{\gamma}_{2i+1}) \ketf{\Omega_\text{vac}} =  \hat{\gamma}_{2i} \ketf{\Omega_\text{vac}} \label{eqn:onbstrick} \\
				&= \frac{1}{2} (-i\hat{\gamma}_{2i+1}) (-i \hat{\gamma}_{2i} \hat{\gamma}_{2i+1} + \mathds{1}) \ketf{\Omega_\text{vac}} = -i\hat{\gamma}_{2i+1} \ketf{\Omega_\text{vac}}\, , \label{eqn:onbstrick2}
			\end{align}
			for all $i \in \{0,1,\dots, n-1\}$.  
			
			Let $\vec{f} \in \mathbb{F}_2^n$ have $k$ nonzero bits $f_{i_1}, f_{i_2}, \dots, f_{i_k} = 1$ where $0 \leq i_1 < \dots <i_{k} \leq n-1$ for some $k\in \{1,\dots,n\}$, with all other bits zero. The Fock basis state $\ketf{\vec{f}}$ has the form
			\begin{align}
				\ketf{\vec{f}} &= 
				\hat{a}_{i_1}^\dagger \hat{a}_{i_2}^\dagger  \dots \hat{a}_{i_{k-1}}^\dagger
				\hat{a}_{i_{k}}^\dagger \ketf{\Omega_\text{vac} } = 
				\left(\hat{a}_{i_1}^\dagger \hat{a}_{i_2}^\dagger \dots \hat{a}_{i_{k-1}}^\dagger \right) \hat{\gamma}_{2i_k} \ketf{\Omega_\text{vac}}\, , \label{eqn:l=k-1}
			\end{align}
			using Equation \ref{eqn:onbstrick}. Suppose that there is some $l<k$ for which
			\begin{align}
				\ketf{\vec{f}} &= \left(\hat{a}_{i_1}^\dagger \hat{a}_{i_2}^\dagger \dots \hat{a}_{i_l}^\dagger\right) \hat{\gamma}_{2i_{l+1}} \hat{\gamma}_{2i_{l+2}} \dots \hat{\gamma}_{2i_{k-1}} \hat{\gamma}_{2i_k} \ketf{\Omega_\text{vac}}\, .
			\end{align}
			Equation \ref{eqn:l=k-1} shows that such an $l$ exists, with value $l=k-1$. Since $\hat{a}_{i_l}^\dagger$ is a linear combination of the Majorana operators $\hat{\gamma}_{2i_l}$ and $\hat{\gamma}_{2i_l+1}$, it anticommutes with any $\hat{\gamma}_{2j}$ where $j \neq i_l$, and so
			\begin{align}
				\ketf{\vec{f}} &= (-1)^{k-(l+1)} \left(\hat{a}_{i_1}^\dagger \hat{a}_{i_2}^\dagger \dots \hat{a}_{i_{l-1}}^\dagger \right) \hat{\gamma}_{2i_{l+1}} \hat{\gamma}_{2i_{l+2}} \dots \hat{\gamma}_{2i_{k-1}} \hat{\gamma}_{2i_k} \left( \hat{a}_{i_l}^\dagger \right)\ketf{\Omega_\text{vac}}\label{eqn:l1}\\
				&= (-1)^{k-(l+1)} \left( \hat{a}_{i_1}^\dagger \hat{a}_{i_2}^\dagger \dots \hat{a}_{i_{l-1}}^\dagger \right) \hat{\gamma}_{2i_{l+1}} \hat{\gamma}_{2i_{l+2}} \dots \hat{\gamma}_{2i_{k-1}} \hat{\gamma}_{2i_k} \hat{\gamma}_{2i_l} \ketf{\Omega_\text{vac}}\\
				&= 
				\left(\hat{a}_{i_1}^\dagger \hat{a}_{i_2}^\dagger \dots \hat{a}_{i_{l-1}}^\dagger \right) \hat{\gamma}_{2i_l} \hat{\gamma}_{2i_{l+1}} \hat{\gamma}_{2i_{l+2}} \dots \hat{\gamma}_{2i_{k-1}} \hat{\gamma}_{2i_k} \ketf{\Omega_\text{vac}}\, , \label{eqn:l2}
			\end{align}
			again using Equation \ref{eqn:onbstrick} and anticommuting $\hat{\gamma}_{2i_l}$ back to the position of $\hat{a}_{i_{l}}^\dagger$, which cancels out the phase of $(-1)^{k-(l+1)}$. Relabel $l \leftarrow (l{-}1)$ and repeat steps \ref{eqn:l1}--\ref{eqn:l2}, proceeding via induction until $l=0$. At the end, we obtain the result that
			\begin{align}
				\ketf{\vec{f}
                } = \hat{\gamma}_{2i_1} \hat{\gamma}_{2i_2} \dots \hat{\gamma}_{2i_k} \ketf{\Omega_\text{vac}} \, ,
			\end{align}
			which proves the result for Equation \ref{eqn:onbasis1}.
			If we had used Equation \ref{eqn:onbstrick2} rather than \ref{eqn:onbstrick}, we would have obtained
			\begin{align}
				\ketf{\vec{f}} = \hat{a}_{i_1}^\dagger \hat{a}_{i_2}^\dagger \dots \hat{a}_{i_{k-1}}^\dagger \hat{a}_{i_k}^\dagger \ketf{\Omega_\text{vac}} = \left(
				\hat{a}_{i_1}^\dagger \hat{a}_{i_2}^\dagger \dots \hat{a}_{i_{k-1}}^\dagger
				\right)(-i\hat{\gamma}_{2i_k+1}) \ketf{\Omega_\text{vac}}\, ,
			\end{align}
			instead of Equation \ref{eqn:l=k-1}. Following identical steps gives
			\begin{align}
				\ketf{\vec{f}} = (-i\hat{\gamma}_{2i_1+1})(-i\hat{\gamma}_{2i_2+1}) \dots (-i\hat{\gamma}_{2i_k+1}) \ketf{\Omega_\text{vac}}\, ,
			\end{align}
			which proves the result of Equation \ref{eqn:onbasis2}.
		\end{proof}
	\end{lemma}
	
    Physical operators on $n$ qubits are the $2^n {\times} 2^n$ unitary matrices $\mathcal{U}_n$. A subgroup of the unitary operators $\mathcal{U}_n$, the Pauli group $\mathcal{P}_n = \langle X, Y, Z \rangle^{\otimes n}$ consists of \textit{Pauli strings}, which are $n$--fold tensor products of the single--qubit Pauli operators.
	
	\begin{definition}\label{defn:jw}
		\textit{(Operator--based definition of the Jordan--Wigner transformation.)} The Jordan--Wigner transformation serves as the canonical fermion--qubit mapping. One can specify the Jordan--Wigner transformation as the following identification of each of the $2n$ Majorana operators with Hermitian Pauli strings:
		\begin{alignat}{4}
			\gamma_0 &= X_0\, , \quad &\gamma_2 = Z_0 X_1 \, ,\quad  &  \dots \, , \quad  & \gamma_{2n-2} &= \left( \bigotimes_{k=0}^{n-2} Z_k \right) X_{n-1}\, ,\\
			\gamma_1 &=  Y_0\, , \quad &\gamma_3 = Z_0 Y_1 \, , \quad & \dots \, , \quad & \gamma_{2n-1} &= \left( \bigotimes_{k=0}^{n-2} Z_k \right) Y_{n-1}\, .
		\end{alignat}
	\end{definition}

	\begin{definition}\label{defn:mapping}
		\textit{(Operator--based definition of fermion--qubit mappings.)} An  \textit{$n$--mode (ancilla--free) fermion--qubit mapping} $\mathfrak{m}$ is an ordered list of ordered pairs of anticommuting, unitary and Hermitian qubit operators
		\begin{align}
			\mathfrak{m} = \big( (\Gamma_0, \Gamma_1), (\Gamma_2, \Gamma_3) , \dots , (\Gamma_{2n-2}, \Gamma_{2n-1}) \big) \label{eqn:pairing}
		\end{align}
		where the representation of the $i$th Majorana operator $\hat{\gamma}_i$ under $\mathfrak{m}$ is the unitary Hermitian matrix $\Gamma_i \in \mathcal{U}_n$, which is the image of the $i$th Jordan--Wigner Pauli string $\gamma_i$ under conjugation by some unitary $U_\mathfrak{m} \in \mathcal{U}_n$:
		\begin{align} \label{eqn:conjg}
			U_\mathfrak{m} : {\gamma}_i \longmapsto \Gamma_i = U_\mathfrak{m} \gamma_i U_\mathfrak{m}^\dagger \quad \text{for all} \quad i \in \{0,1,\dots,2n-1\}\, .
		\end{align}
		The conjugation relations in Equation \ref{eqn:conjg} define $U_\mathfrak{m}$ up to a global phase and ensure that the operators $\{\Gamma_i\}_{i=0}^{2n-1}$ satisfy the same anticommutations as the Majorana operators do in Equation \ref{eqn:majcars}. It is true that $U_\mathfrak{m}e^{i\theta}$ gives the same operators $(U_\mathfrak{m} e^{i\theta})({\gamma}_i)(U_\mathfrak{m}e^{i \theta})^\dagger$ as $U_\mathfrak{m}$ for any $\theta \in [0,2\pi)$. However, Definition \ref{defn:occno} of the vacuum state and Fock basis of a fermion--qubit mapping $\mathfrak{m} = ((\Gamma_{2i}, \Gamma_{2i+1}))_{i=0}^{n-1}$ fixes a unique $U_\mathfrak{m}$. 
	\end{definition}

	\begin{figure}[btp]
		\centering
		\includegraphics[width=\linewidth]{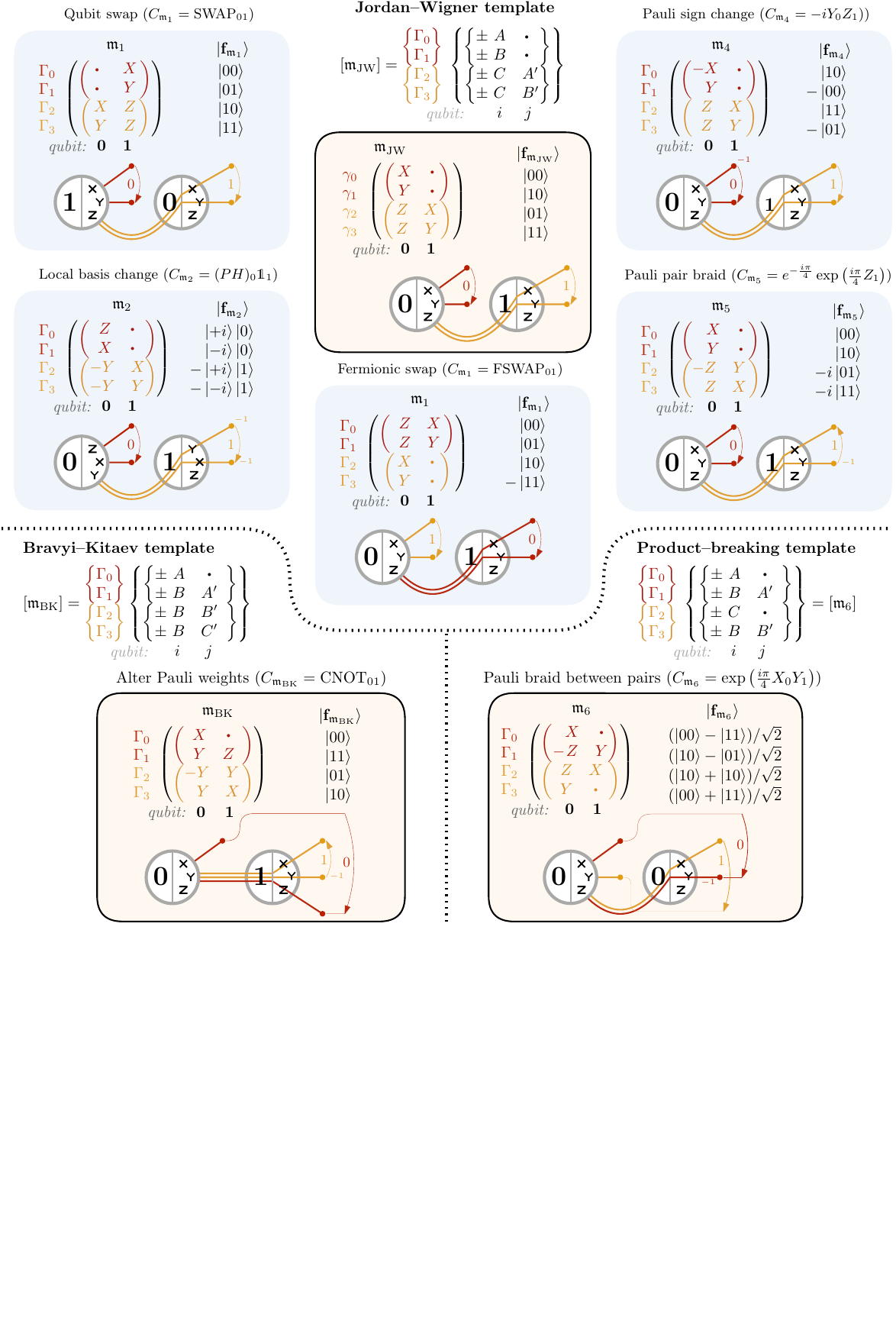}
		\caption{The operator-- and state--based descriptions as well as the diagrams and templates for several two--mode ancilla--free fermion--qubit mappings. On display are the two--mode Jordan--Wigner transformation $\mathfrak{m}_\text{JW}$, five equivalent mappings $\mathfrak{m}_1$---$\mathfrak{m}_5$, the two--mode Bravyi--Kitaev transformation $\mathfrak{m}_\text{BK}$ and the two--mode product--breaking transformations $\mathfrak{m}_6$. The similar diagram structure of $\mathfrak{m}_\text{JW}$ and $\mathfrak{m}_1$---$\mathfrak{m}_5$ reveals their equivalence to each other and distinction from the Bravyi--Kitaev and product--breaking templates. The Clifford operator $C_\mathfrak{m}$ is the unique matrix to satisfy $C_\mathfrak{m} \ket{\vec{f}} = \ket{\vec{f}_\mathfrak{m}}$ for each of the mappings $\mathfrak{m}$; as a consequence, it also satisfies $C_\mathfrak{m} \gamma_i C_\mathfrak{m}^\dagger = \Gamma_i$.}
		\label{fig:somemappings}
	\end{figure}

	The purpose of specifying a fermion--qubit mapping $\mathfrak{m}$ as an ordered set of ordered pairs of Pauli operators is to encode the $n$ annihilation operators via the relations
	\begin{align}
		\hat{a}_i &\longmapsto  A_i \coloneqq \frac{1}{2} \left(\Gamma_{2i} + i \Gamma_{2i+1}\right)  \quad \text{for} \quad i \in \{0,1,\dots,n-1\}\, . \label{eqn:annihilation}
	\end{align}

	\begin{definition}\textit{(The vacuum state, vacuum stabilisers, and Fock basis of a fermion--qubit mapping.)} \label{defn:occno} \\
		Let $\mathfrak{m} = ((\Gamma_{2i},\Gamma_{2i+1}))_{i=0}^{n-1}$ be a fermion--qubit mapping. Define the \textit{vacuum state of} $\mathfrak{m}$, which we denote $\ket{\vec{0}_\mathfrak{m}} \in \mathcal{H}_2^{\otimes n}$, to be any one of the simultaneous 0--eigenstates of the number operators $\{A_i^\dagger A_i\}_{i=0}^{n-1}$ of $\mathfrak{m}$.	It is equivalent to define $\ket{\vec{0}_\mathfrak{m}}$ to be any one of the simultaneous $(+1)$--eigenstates of the operators $\{-i\Gamma_{2i}\Gamma_{2i+1}\}_{i=0}^{n-1}$, which we call the \textit{vacuum stabilisers of} $\mathfrak{m}$; this follows from the relation $A_i^\dagger A_i = \frac{1}{2}(\mathds{1}^{\otimes n} + i \Gamma_{2i} \Gamma_{2i+1})$. The qubit state $\ket{\vec{0}_{\mathfrak{m}}}$ is the representation under $\mathfrak{m}$ of the fermionic vacuum state $\ketf{\Omega_\text{vac}}$. As in the fermionic case, any two simultaneous 0--eigenstates differ only by some phase $e^{i\phi}$ for $\phi \in [0,2\pi)$.
		
		The \textit{Fock basis of $\mathfrak{m}$} is the orthonormal set $\{\ket{\vec{f}_\mathfrak{m}} \, | \, \vec{f}\in \mathbb{F}_2^n\} \subset \mathcal{H}_2^{\otimes n}$, where the \textit{Fock state $\ket{\vec{f}_\mathfrak{m}}$ of $\mathfrak{m}$} has definition
		\begin{align}
			\ket{\vec{f}_\mathfrak{m}} \coloneqq \left(A_0^\dagger\right)^{f_0} \left(A_1^\dagger\right)^{f_1} \dots \left(A_{n-1}^\dagger\right)^{f_{n-1}} \ket{\vec{0}_\mathfrak{m}}\, .
			\label{eqn:occnobasis}
		\end{align}
		By Lemma \ref{lem:occnobasis}, equivalent definitions for the Fock states of $\mathfrak{m}$ are
		\begin{align}
			\ket{\vec{f}_\mathfrak{m}} &= \left(\Gamma_{0}\right)^{f_0} \left(\Gamma_2\right)^{f_1} \dots \left(\Gamma_{2n-2}\right)^{f_{n-1}} \ket{\vec{0}_\mathfrak{m}} \label{eqn:occno} = \left(-i\Gamma_{1}\right)^{f_0}  \left(-i\Gamma_3\right)^{f_1} \dots \left( -i\Gamma_{2n-1} \right)^{f_{n-1}} \ket{\vec{0}_\mathfrak{m}}\, .
		\end{align}
		We tend to use the formula $\ket{\vec{f}_\mathfrak{m}}=(\Gamma_0)^{f_0} (\Gamma_2)^{f_1} \dots (\Gamma_{2n-2})^{f_{n-1}}\ket{\vec{0}_{\mathfrak{m}}}$ for its simplicity.
		The Fock state $\ket{\vec{f}_\mathfrak{m}}$ is an $(f_i)$--eigenstate of the number operator $A_i^\dagger A_i$ of $\mathfrak{m}$; equivalently, it is a $((-1)^{f_i})$--eigenstate of the vacuum stabiliser $-i\Gamma_{2i}\Gamma_{2i+1}$ of $\mathfrak{m}$.
	\end{definition}

	\begin{exmp} \textit{(Fock basis of the Jordan--Wigner transformation.)}
		The Jordan--Wigner transformation is the list of pairs of anticommuting, Hermitian Pauli operators from Definition \ref{defn:jw} of the form $\mathfrak{m}_\text{JW}		=	((\gamma_{2i}, \gamma_{2i+1}))_{i=0}^{n-1}$,
		and thus has vacuum stabilisers of the form
		\begin{align}
			-i\gamma_{2i} \gamma_{2i+1} = -i\left( \left(\bigotimes_{k=0}^{i-1} Z_k \right) X_i \right) \left( \left( \bigotimes_{k=0}^{i-1} Z_k \right) Y_i \right) = Z_i\, . 
		\end{align}
		The simultaneous 0--eigenstates of the vacuum stabilisers $\{Z_i\}_{i=0}^{n-1}$ of $\mathfrak{m}_{\text{JW}}$  are the states $ \{ e^{i \phi} \ket{0}^{\otimes n}\, | \, \phi \in  [0,2\pi)\}$. By convention, we choose $\ket{0}^{\otimes n}$ to be $\ket{\vec{0}_{\mathfrak{m}_\text{JW}}}$ without loss of generality. Using Equation \ref{eqn:occno}, the Fock states of the Jordan--Wigner transformation are identical to the corresponding computational basis states:
		\begin{align}\label{eqn:jwstates}
			\ket{\vec{f}_{\mathfrak{m}_\text{JW}}} = (\gamma_0)^{f_0} (\gamma_2)^{f_1} \dots (\gamma_{2n-2})^{f_{n-1}} \ket{0}^{\otimes n} = (X_0)^{f_0} (Z_0 X_1)^{f_1} \dots (Z_0 Z_1 ... Z_{n-2} X_{n-1} )^{f_{n-1}} \ket{0}^{\otimes n}  = \ket{\vec{f}}\, ,
		\end{align}
		where $\ket{\vec{f}} =  \ket{f_{0}} \ket{f_{1}} \dots \ket{f_{n-1}}$ is the computational basis state corresponding to the bit-string $\vec{f}$.
	\end{exmp}

	\begin{remark}\textit{(Ordering of the creation operators in the Fock basis of a mapping.)} \label{rem:ordering}
		Recall that the ordering of the creation operators in the definition of the Fock basis in Equation \ref{eqn:fockbasis} is arbitrary on the fermionic level. However, the only ordering of the product of $(\gamma_{2i})^{f_i}$ operators in Equation \ref{eqn:jwstates} that would have yielded $\ket{\vec{f}_{\mathfrak{m}_\text{JW}}} = \ket{\vec{f}} $ for all $\vec{f}\in \mathbb{F}_2^n$ is the ordering $(\gamma_0)^{f_0} \dots (\gamma_{2n-2})^{f_{n-1}}$. If it were any other order, there would be bit-strings $\vec{f} \in \mathbb{F}_2^n$ and indices $i \in [n]$ for which the $Z$--chain $(\bigotimes_{k=0}^{i-1} Z_k)$ in Equation \ref{eqn:jwstates} would act on an odd number of qubits in the state $\ket{1}$; for these bit--strings, this would mean that $\ket{\vec{f}_{\mathfrak{m}_{\text{JW}}}} = -\ket{\vec{f}}$.
	\end{remark}

	For each bit-string $\vec{f}\in \mathbb{F}_2^n$, the fermion--qubit mapping $\mathfrak{m}$ employs a qubit state $\ket{\vec{f}_\mathfrak{m}} \in \mathcal{H}_2^{\otimes n}$ as the encoding of the fermionic state $\ketf{\vec{f}} \in \mathcal{H}_\text{fermion}$. 	In fact, the Fock basis $\{\ket{\vec{f}_\mathfrak{m}} \mid \vec{f} \in \mathbb{F}_2^n\}$ of a fermion--qubit mapping $\mathfrak{m}$ fully determines the representations of the Majorana operators $\{\Gamma_i\}_{i=0}^{2n-1}$ of $\mathfrak{m}$. An equivalent definition for a fermion--qubit mapping is via its Fock basis:
	
	\begin{definition}\label{defn:statesfirst} \textit{(State--based definition of fermion--qubit mappings.)}
	An \textit{$n$--mode (ancilla--free) fermion--qubit mapping} $\mathfrak{m}$ is an encoding of the Fock basis as an orthonormal basis $\{ \ket{\vec{f}_{\mathfrak{m}}} \mid \vec{f} \in \mathbb{F}_2^n\}$, where the qubit state $\ket{\vec{f}_\mathfrak{m}} \in \mathcal{H}_2^{\otimes n}$ is the representation of the fermionic state  $\ketf{\vec{f}} \in \mathcal{H}_\text{fermion}$.
	\end{definition}
	
	Using the Jordan--Wigner transformation as a reference point,
	Lemma \ref{lem:Cfix} demonstrates that Definition  \ref{defn:mapping} for fermion--qubit mappings is equivalent to Definition \ref{defn:statesfirst}. The link is via a unique unique unitary operator $U_\mathfrak{m}\in \mathcal{U}_n$ satisfying
	\begin{align}
		\mathcal{H}_\text{fermion}  \ni   \ketf{\vec{f}}  \quad \xmapsto{\mathfrak{m}_{\text{JW}}} \quad  \ket{\vec{f}}
		\quad \xmapsto{U_\mathfrak{m}} \quad U_\mathfrak{m}\ket{\vec{f}}  =  \ket{\vec{f}_\mathfrak{m}} \in  \mathcal{H}_2^{\otimes n} \quad \text{for all } \vec{f}   \in   \mathbb{F}_2^n\, . \label{eqn:statebased}
	\end{align}
	
	\begin{lemma}
		\textit{(Equivalence of the operator--based description $\mathfrak{m}=((\Gamma_{2i},\Gamma_{2i+1}))_{i=0}^{2n-1}$ and the state--based description $\{\ket{\vec{f}_\mathfrak{m}} \mid \vec{f}\in \mathbb{F}_2^n\}$ of fermion--qubit mappings.)}
		\label{lem:Cfix}
		Let $\mathfrak{m}$ be a fermion--qubit mapping as per Definition \ref{defn:statesfirst}, denoting the encoding $\{\ket{\vec{f}_\mathfrak{m}} \mid \vec{f}\in \mathbb{F}_2^n\}$ of the Fock basis. The
        unitary operator $U_\mathfrak{m} \in \mathcal{U}_n$ that performs the map
		\begin{align}
			{U}_\mathfrak{m}: \ket{\vec{f}} \longmapsto \ket{\vec{f}_\mathfrak{m}} \quad \text{for all} \quad  \vec{f} \in \mathbb{F}_2^n\, \label{eqn:statesfirst}
		\end{align}
        defines operators $\Gamma_i \coloneqq U_{\mathfrak{m}} \gamma_i U_\mathfrak{m}^\dagger$ such that $\mathfrak{m} = ((\Gamma_{2i}, \Gamma_{2i+1}))_{i=0}^{n-1}$ as per Definition \ref{defn:mapping}.		
		
		\begin{proof}
			

            Suppose that $\mathfrak{m}$ is a fermion--qubit mapping with respect to Definition \ref{defn:statesfirst} with encoding $\{\ket{\vec{f}_\mathfrak{m}} \mid \vec{f} \in \mathbb{F}_2^n\}$ of the Fock basis. Let $U_\mathfrak{m} \in \mathcal{U}_n$ be the unitary operator in Equation \ref{eqn:statesfirst}, and define $\Gamma_i \coloneqq U_\mathfrak{m} \gamma_i U_\mathfrak{m}^\dagger$ for all $i \in [2n]$. Note that
            \begin{align}
                \ket{\vec{f}_\mathfrak{m}} \coloneqq U_\mathfrak{m} \ket{\vec{f}} = U_\mathfrak{m} \left( X_0^{f_0} X_1^{f_1} \dots X_{n-1}^{f_{n-1}}\ket{\vec{0}} \right) &= \left(U_\mathfrak{m}(\gamma_0)^{f_0} U_\mathfrak{m}^\dagger \right) 
                \dots \left(U_\mathfrak{m} (\gamma_{2n-2})^{f_{n-1}} U_\mathfrak{m}^\dagger \right) U_\mathfrak{m} \ket{\vec{0}}
                \\&= \Gamma_0^{f_0} \Gamma_1^{f_1} \dots \Gamma_{2n-2}^{f_{n-1}} \ket{\vec{0}_\mathfrak{m}}\, ,
            \end{align}
            which is the definition of $\ket{\vec{f}_\mathfrak{m}}$ according to Definition \ref{defn:mapping}.
	\end{proof}\end{lemma}

The $n$--qubit Clifford group $\mathcal{C}_n$ is the subgroup of $\mathcal{U}_n$ that conjugates the Pauli group to itself, i.e.\ $\mathcal{C}_n = \{C \in \mathcal{U}_n \mid C \mathcal{P}_n C^\dagger = \mathcal{P}_n\}$. Fermion--qubit mappings $\mathfrak{m} = ((\Gamma_{2i}, \Gamma_{2i+1}))_{i=0}^{2n-1}$ for which the defining unitary matrix $U_\mathfrak{m}$ is a Clifford operator have Majorana representations $\Gamma_i \in \mathcal{P}_n$ that are Pauli strings. We emphasise our focus on this class of mappings by denoting the unitary by $U_\mathfrak{m} = C_\mathfrak{m}$ to denote its membership of the Clifford group.

Pauli operator mappings make up nearly all of the existing results: not only the ancilla--free linear encodings and ternary tree transformations that concern this work, but also those that use ancilla qubits such as locality--preserving mappings. Using Lemma \ref{lem:Cfix}, each mapping we consider will be of the form $\mathfrak{m} = ((\Gamma_{2i}, \Gamma_{2i+1}))_{i=0}^{n-1}$ for $\Gamma_i \in \mathcal{P}_n$, and will encode the Fock basis into an orthonormal set of stabiliser states $\{\ket{\vec{f}_\mathfrak{m}} \mid \vec{f}\in \mathbb{F}_2^n\}$.
 
	\begin{exmp}
		Figure \ref{fig:somemappings} depicts the two--mode Jordan--Wigner transformation both as a set of Pauli operator pairs $\mathfrak{m}_\text{JW}$ and as the encoded states  \{$\ket{\vec{f}_{\mathfrak{m}_\text{JW}}} \mid \vec{f} \in (00,10,01,11)\}$. In accompaniment are five similar mappings $\mathfrak{m}_1$---$\mathfrak{m}_5$ in operator-- and state--based forms, along with the corresponding Clifford operators $C_{\mathfrak{m}_1}$---$C_{\mathfrak{m}_5}$. Also making an appearance is the two--mode Bravyi--Kitaev transformation $\mathfrak{m}_\text{BK}$ as well as the two--mode product--breaking mapping $\mathfrak{m}_6$, so--called because its Fock states are entangled.
		
		Note that we write the order of the Fock basis states from top-to-bottom as $\ket{00_\mathfrak{m}}, \ket{10_\mathfrak{m}}, \ket{01_\mathfrak{m}}, \ket{11_\mathfrak{m}}$ in Figure \ref{fig:somemappings}, which is a slight change from the conventional order of $\ket{00_\mathfrak{m}}, \ket{01_\mathfrak{m}}, \ket{10_\mathfrak{m}}, \ket{11_\mathfrak{m}}$.
	\end{exmp}

\section{Equivalence relations for fermion--qubit mappings} \label{sec:equiv}
	
Focussing on fermion--qubit mappings with Pauli representations of the Majorana operators, in Section \ref{sec:equivalence}, we introduce a diagrammatic notation with a practical application. The equivalence relation factors out the Clifford operations of qubit and fermionic mode relabelling, braids within Pauli operator pairs, sign change of Pauli operators and local Pauli basis orientation, which all represent arbitrary labelling choices by the users of quantum computing hardware. Our notion of equivalence categorises two--mode fermion--qubit mappings into Jordan--Wigner and Bravyi--Kitaev templates, and a third template consisting of transformations with entangled Fock bases.

\subsection{Labelling symmetries}\label{sec:equivalence}

This section introduces diagrammatic notation for (ancilla--free) fermion--qubit mappings with Pauli representations of the Majorana operators, which extends the scope of the ternary tree diagrams in \cite{miller2023bonsai, miller2024treespilation}. Taking the symmetry group of the diagrammatic notation establishes a robust definition of equivalence between these fermion--qubit mappings. Our definition identifies redundancies in the sign of Pauli operators and their ordering within pairs, the labels of the single--qubit Pauli operators themselves, and the enumeration of fermionic modes and qubits. The resulting equivalence classes that strip these redundancies away become `templates' for different fermion--qubit mapping types.

\begin{definition}\textit{(Diagrammatic notation for ancilla--free fermion--qubit mappings.)} \label{defn:diagram}
	Let $\mathfrak{m} = ((\Gamma_{2i},\Gamma_{2i+1}))_{i=0}^{n-1}$ be an (ancilla--free) $n$--mode fermion--qubit mapping where the Majorana representations are Pauli operators $\Gamma_i \in \mathcal{P}_n$. The \textit{diagram} of $\mathfrak{m}$ is a visual tool to represent the pairs $(\Gamma_{2i}, \Gamma_{2i+1})$ of Pauli operators for $i \in [n]$, and results from following the instructions (see Figure \ref{fig:somemappings} for examples):
	\begin{enumerate}
		\item \textit{Qubit labels:} draw one circle for each qubit, and label them with the values $i \in [n]$,
		\item \textit{Local basis labels:} inscribe the Pauli labels $X,Y,Z$ from top to bottom in each circle,
		\item[\textbullet\phantom{.}] \textbf{Representation of Pauli operator pairs as strings:} draw each Pauli operator $\Gamma_i$ as a string intersecting the circles with labels corresponding to the qubits in its support, such that the string departs circle $j$ at the level of the label $X$, $Y$ or $Z$ according to the local action of $\Gamma_i$ on qubit $j$. For each pair $(\Gamma_{2i}, \Gamma_{2i+1})$, draw a curve joining the end of the string representing $\Gamma_{2i}$ to the end of the string representing $\Gamma_{2i+1}$,
		\item \textit{Pauli ordering within pairs:} place a directed arrow on the curve connecting the ends of the strings of each pair $(\Gamma_{2i}, \Gamma_{2i+1})$ so that the curve points away from the string for $\Gamma_{2i}$ and towards the string for $\Gamma_{2i+1}$,
		\item \textit{Pauli signs:}  if the Pauli operator $\Gamma_i$ is such that $-\Gamma_i \in \mathcal{P}_n/K$, indicate this by writing $(-1)$ at the leftmost end of the string representing $\Gamma_i$,
		\item \textit{Fermionic labels:} mark the arrow joining the ends of the lines representing $\Gamma_{2i}$ and $\Gamma_{2i+1}$ with the label $i$.
	\end{enumerate}
The five numbered steps of this process correspond to labelling tasks, which are necessary to identify $\mathfrak{m}$ uniquely from its diagram.
\end{definition}

The diagrammatic notation of Definition \ref{defn:diagram} is a generalisation of the ternary tree diagram notation from \cite{jiang_optimal_2020, miller2023bonsai, miller2024treespilation}, which is limited in the sense that it can only describe mappings with diagrams that have an underlying tree--like structure. Our diagrammatic notation can describe any fermion--qubit mapping with Pauli representations of the Majorana operators, although ternary tree transformations remain our main focus in this paper.

\begin{exmp}\textit{(Examples of fermion--qubit mapping diagrams.)} \label{exmp:similar}
Each of the mappings $\mathfrak{m}_\text{JW}, \mathfrak{m}_\text{BK}, \mathfrak{m}_1$---$\mathfrak{m}_6$ in Figure  \ref{fig:somemappings} have an accompanying diagram. Note that the diagrams of the mappings $\mathfrak{m}_1$---$\mathfrak{m}_5$ all have a similar structure to the diagram for $\mathfrak{m}_\text{JW}$, where the mapping $\mathfrak{m}_i$ differs from $\mathfrak{m}_\text{JW}$ only in the $i$th labelling step of Definition \ref{defn:diagram}.
\end{exmp}

Example \ref{exmp:similar} points out that two distinct fermion--qubit mappings with Pauli representations of the Majorana operators may have identical diagrams if one ignores the labelling steps in Definition \ref{defn:diagram}, which we will build into a notion of equivalence. We argue that the relative multiplication relations between Pauli strings and their underlying pairs is the only factor relevant to quantum computation, because the labelling steps correspond to choices that a user could make essentially mentally at the start of a simulation algorithm. However, any change that is more substantial than this alters the tensor product structure of the Pauli operators, which would affect the time and depth of subsequent quantum computation. We illustrate the set of mappings equivalent to the two--mode Jordan--Wigner transformation in Definition \ref{defn:temp1}.

\begin{definition}\textit{(Template of the two--mode Jordan--Wigner transformation.)}\label{defn:temp1}
	The set of ancilla--free fermion--qubit mappings with diagrams identical to the diagram of the two--qubit Jordan--Wigner transformation $\mathfrak{m}_\text{JW} = ((X_0, Y_0), (Z_0 X_1, Z_0Y_1))$ up to labelling is the \textit{template}
	\begin{align}
		[\mathfrak{m}_\text{JW}] &=
		\bigg\{ \quad
		\underbrace{\big\{
		\overbrace{\{A_i, B_i\}}, 
		 \overbrace{\{C_i A'_j, C_iB'_j\}}^{\mathclap{\textit{Pauli ordering within pairs} \qquad \qquad }}
		 \big\}}_{\mathclap{\textit{Fermionic labels}}}
		 \quad \big\vert \quad 
		 \underbrace{A \neq B \neq C\, , \,  A' \neq B' \in \pm\{X,Y,Z\}}_{\mathclap{\textit{Local Pauli basis labels, Pauli signs}}}
		 \, , \quad 
		 \underbrace{i \neq j \in\{0,1\}}_{\mathclap{\textit{Qubit labels}}} \quad \bigg\}\, \label{eqn:templatefull}
	\end{align}
	where the annotations in Equation \ref{eqn:templatefull} indicate how the mappings in the template may differ from $\mathfrak{m}_\text{JW}$ by the five labelling steps in Definition \ref{defn:templates} symbolically and with the use of unordered pairs and lists. In Figure \ref{fig:somemappings}, mappings $\mathfrak{m}_1$---$\mathfrak{m}_5$ are all elements of $[\mathfrak{m}_\text{JW}]$.
\end{definition}

%

Definition \ref{defn:templates} formalises the notion of label--invariant templates to the set of all $n$--mode ancilla--free fermion--qubit mappings, which corresponds to an  equivalence relation.

	\begin{definition}\textit{(Templates and equivalence of fermion--qubit mappings.)}\label{defn:templates}
		Let $\mathfrak{m} = ((\Gamma_{2i},\Gamma_{2i+1}))_{i=0}^{n-1}$ be an $n$--mode fermion--qubit mapping with Pauli representations of the Majorana operators. The \textit{template of} $\mathfrak{m}$ is the set of fermion--qubit mappings $[\mathfrak{m}]$, where $\mathfrak{m}' = ((\Gamma_{2i}',\Gamma_{2i+1}'))_{i=0}^{n-1} \in [\mathfrak{m}]$ if and only if there exists a Clifford operator $C=\prod_k C_k \in \mathcal{C}_n$ comprising a sequence of Clifford operations, where each Clifford $C_k$ implements one of the following:
		\begin{enumerate}
			\item \textit{Qubit swap:} a permutation $\sigma \in S_n$ of the qubit labels (i.e.\ $C_k :A_i\mapsto  A_{\sigma(i)}$ for $A \in \{X,Y,Z\}$ and $i \in [n]$),
			\item \textit{Local basis change:} a re-orientation of the local Pauli basis of a qubit (i.e.\ $C_k: (X_i,Y_i,Z_i) \mapsto (Y_i, Z_i, X_i)$, or $C_k: (X_i,Y_i,Z_i) \mapsto (Z_i, X_i, Y_i)$, or  $C_k:(X_i,Y_i,Z_i) \mapsto (Y_i, X_i, -Z_i)$ and so on, for some $i \in [n]$),
			\item \textit{Pauli pair braid:} a braid of a Majorana pair (i.e.\ $C_k:(\Gamma_{2i}, \Gamma_{2i+1}) \mapsto (\pm \Gamma_{2i+1},  \mp \Gamma_{2i})$ for some $i \in [n]$),
			\item \textit{Pauli sign change:} the change of sign of a Pauli operator (i.e.\ $C_k:\Gamma_i \mapsto (-1)^{\gamma_i} \Gamma_i$ for some $\gamma_i \in \{0,1\}$ for each $i \in [n])$, 
			\item \textit{Fermionic swap:} a permutation $\rho \in S_n$ of the fermionic mode labels (i.e.\ $C_k:(\Gamma_{2i}, \Gamma_{2i+1}) \mapsto (\Gamma_{2\rho(i)}, \Gamma_{2\rho(i)+1})$ for all $i \in [n]$),
	\end{enumerate}
			such that $\Gamma_i' = C\Gamma_i C^\dagger$ for all $i \in [2n]$. We say that two fermion--qubit mappings with Pauli representations of the Majorana operators $\mathfrak{m}_1$ and $\mathfrak{m}_2$ are \textit{equivalent} if and only if $[\mathfrak{m}_1] = [\mathfrak{m}_2]$.
	\end{definition}
	
With the exception of the local basis change and Pauli pair braid operations, the five types of Clifford operation in Definition \ref{defn:templates} correspond to the five labelling steps in Definition \ref{defn:temp1}. We choose to use the physically--motivated local basis change and Pauli pair braid symmetries to generate the equivalence classes of fermion--qubit mappings rather than the more abstract relabelling of single--qubit Pauli matrices and reordering operators within pairs. The end result is the same because of the freedom to change the signs of Pauli operators.

For example, the sequence of the Pauli braid $\exp(\frac{\pi}{4} \Gamma_{2i} \Gamma_{2i+1})$ implementing $(\Gamma_{2i}, \Gamma_{2i+1}) \mapsto (-\Gamma_{2i+1}, \Gamma_{2i})$ \cite{McLauchlan2024newtwistmajorana}, followed by the Pauli sign change $\Gamma_{2i+1} \mapsto -\Gamma_{2i+1}$ reproduces the pair reordering symmetry $(\Gamma_{2i}, \Gamma_{2i+1}) \mapsto (\Gamma_{2i+1}, \Gamma_{2i})$ of fermion--qubit mapping diagrams:
\begin{align}
	(\Gamma_{2i}, \Gamma_{2i+1}) \xmapsto{\text{Pauli pair braid}} (-\Gamma_{2i+1}, \Gamma_{2i}) \xmapsto{\text{Pauli sign change}} (\Gamma_{2i+1}, \Gamma_{2i})\, .
\end{align}
This decision is due to notational convenience in Sections \ref{sec:treebased} and \ref{sec:cbptree}. Similarly, combining the local basis change $(X_i,Y_i,Z_i) \mapsto (Y_i,X_i,-Z_i)$ with sign changes of any Pauli operators involving $Z_i$ will have the effect of relabelling the Pauli operators on qubit $i$ according to $(X_i,Y_i,Z_i) \mapsto (Y_i,X_i,Z_i)$.

\begin{exmp}\textit{(Two--mode ancilla--free fermion--qubit mappings.)}
In addition to the Jordan--Wigner template $[\mathfrak{m}_\text{JW}]$ from Definition \ref{defn:temp1}, there are two other templates for two--mode mappings. One, the Bravyi--Kitaev template, is of the form
\begin{align}
	[\mathfrak{m}_\text{BK}] &=
\bigg\{ \quad
\big\{
	\{A_i, B_i A_i'\}, 
	\{B_i B_j', B_i C_j'\}
	\big\}
\,\,  \big\vert \, \,  
A \neq B \, , \,  A' \neq B' \neq C' \in \pm \{X,Y,Z\}
\, , \quad 
{i \neq j \in\{0,1\}} \quad \bigg\}\,,
\end{align}
where $\mathfrak{m}_\text{BK} = ((X_0, Y_0Z_1), (-Y_0Y_1, Y_0X_1))$ is the two--qubit Bravyi--Kitaev transformation as Figure \ref{fig:somemappings} displays. The other template is the equivalence class of the mapping $\mathfrak{m}_6 = ((X_0, -Z_0Y_1), (Z_0 X_1, Y_0))$,
\begin{align}
	[\mathfrak{m}_6] &=
	\bigg\{ \quad
	\big\{
	\{A_i, B_i A_j'\}, 
	\{C_i, B_i B_j'\}
	\big\}
	\quad \big\vert \quad 
	A \neq B \neq C \, , \,   A' \neq B'  \in  \pm \{X,Y,Z\}
	\, , \quad 
	{i \neq j \in\{0,1\}} \quad \bigg\}\,,
\end{align}
which we call the `product--breaking template' because the vacuum state of any mapping in $[\mathfrak{m}_6]$ is entangled.
\end{exmp}

\section{Classical, affine and linear encodings of the Fock basis} \label{app:ibm}
We now turn to defining classical, affine and linear encodings of the Fock basis, a prominent hierarchy of fermion--qubit mappings which store fermionic occupation numbers in the bit-strings of the phaseless computational basis. Section \ref{sec:affine} proves in Theorem \ref{thm:affine1} that all classical encodings with Pauli representations of the Majorana operators are affine encodings and vice versa, and in Corollary \ref{cor:affinelin} we prove that affine encodings are equivalent to linear encodings with respect to the equivalence relation from Section \ref{sec:equiv}.

Linear encodings have appeared previously in the literature \cite{bravyi2017taperingqubitssimulatefermionic, steudtner2018fermion, wang2023evermore}, and this section expands the topic by introducing the explicit formula for the Majorana representatives of any affine encoding.
Additionally, this section  lays the essential groundwork for unveiling the state--based definition of ternary tree transformations in Section \ref{sec:cbptree}.

	\begin{definition} \textit{(Notation involving the computational basis.)} \label{defn:compbasis}
		Let $\mathfrak{C}_n$ denote the computational basis $\{\ket{\vec{f}} \, | \, \vec{f} \in \mathbb{F}_2^n\} = \{\ket{0}^{\otimes n}, \ket{0}^{\otimes (n-1)}\ket{1}, \dots , \ket{1}^{\otimes n}\}$. For a quantum state $\ket{\psi} \in  \mathcal{H}_2^{\otimes n}$, we say that $\ket{\psi}$ is \textit{in the computational basis} if and only if $\ket{\psi} \in \mathfrak{C}_n$. For any $\theta \in [0, 2\pi)$, define 
		\begin{align} e^{i \theta} \mathfrak{C}_n \coloneqq \{e^{i \theta} \ket{0}^{\otimes n}, e^{i \theta} \ket{0}^{\otimes (n-1)}\ket{1}, \dots , e^{i \theta} \ket{1}^{\otimes n}\}\, .\end{align}
		We use $-\mathfrak{C}_n$ as shorthand to denote $(-1) \mathfrak{C}_n$; similarly $\pm \mathfrak{C}_n$ denotes $\mathfrak{C}_n \cup (-\mathfrak{C}_n)$ and $\pm i \mathfrak{C}_n$ denotes $i \mathfrak{C}_n \cup (-i) \mathfrak{C}_n$.
	\end{definition}
	
	\begin{definition}\textit{(Classical, affine and linear encodings of the Fock basis.)}\label{defn:classlin}
		Let $\mathfrak{m}$ be an $n$--mode ancilla--free fermion--qubit mapping with Fock basis $\{\ket{\vec{f}_\mathfrak{m}} \mid \vec{f} \in \mathbb{F}_2^n\}$.
		\begin{enumerate}[label=\alph*)]
			\item We say that $\mathfrak{m}$ is a \textit{classical encoding of the Fock basis}, or that $\mathfrak{m}$ \textit{classically encodes the Fock basis}, if $\ket{\vec{f}_\mathfrak{m}} \in \mathfrak{C}_n$ for all $\vec{f} \in \mathbb{F}_2^n$.
            \item We say that $\mathfrak{m}$ is an \textit{affine encoding of the Fock basis}, or that $\mathfrak{m}$ \textit{affinely encodes the Fock basis}, if $\ket{\vec{f}_\mathfrak{m}} = \ket{G(\vec{f} \oplus \vec{b})}$ for all $\vec{f} \in \mathbb{F}_2^n$, where $G \in \text{GL}_n(\mathbb{F}_2)$ is some invertible binary matrix and $\vec{b} \in \mathbb{F}_2^n$.
			\item We say that $\mathfrak{m}$ is a \textit{linear encoding of the Fock basis}, or that $\mathfrak{m}$ \textit{linearly encodes the Fock basis}, if $\ket{\vec{f}_\mathfrak{m}} = \ket{G\vec{f}}$ for all $\vec{f} \in \mathbb{F}_2^n$, where $G \in \text{GL}_n(\mathbb{F}_2)$ is some invertible binary matrix.
		\end{enumerate}

	\end{definition}

	
	It is a persistent result in the literature \cite{dehaene_clifford_2003, Picozzi_2023} that unitary transformations of the form $\ket{\vec{f}} \mapsto \ket{\vec{G(\vec{f} \oplus\vec{b})}}$ are in fact Clifford transformations, and so all affine encodings represent Majorana operators with Pauli strings $\Gamma_i \in \mathcal{P}_n$. While Example \ref{exmp:linear} assumes this knowledge, the purpose of Section \ref{sec:affine} is to provide an explicit proof of the statement.
 
\begin{figure}[btp]
		\centering
		\includegraphics[width=0.9\linewidth]{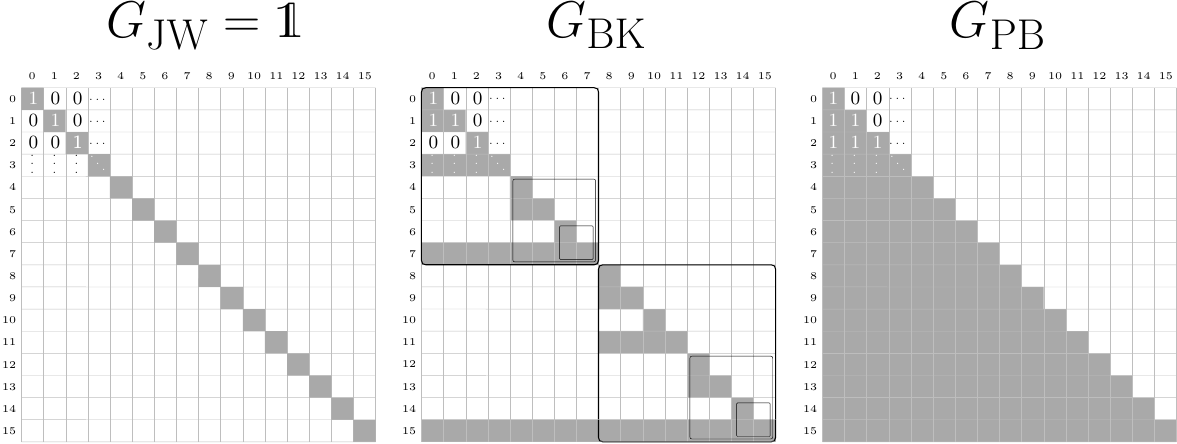}
		\caption{The invertible binary matrices $G_\text{JW}=\mathds{1}$, $G_\text{BK}$ and $G_\text{PB}$ that define the Fock bases of the Jordan--Wigner, Bravyi--Kitaev and parity basis transformations via $\ket{\vec{f}_\mathfrak{m}} = \ket{G \vec{f}}$, respectively, for $n=16$. Shaded squares indicate entries that are equal to 1.}
		\label{fig:bkmat}
	\end{figure}

	
	\begin{exmp} \textit{(Linear encodings and Clifford operators.)} \label{exmp:linear}
		If $\mathfrak{m}$ is a linear encoding of the Fock basis with $\ket{\vec{f}_\mathfrak{m}} = \ket{G\vec{f}}$, then the Clifford operator $C_\mathfrak{m}$ has the property $C_\mathfrak{m} \ket{\vec{f}} = \ket{G\vec{f}}$.
		\begin{enumerate}[label=\alph*)]
			\item The Clifford transformation
			\begin{align}
				\mathds{1}^{\otimes n}: \ket{\vec{f}} \longmapsto \ket{\vec{f}}
			\end{align}
			corresponds to the Jordan--Wigner transformation $\mathfrak{m}_\text{JW} =((\gamma_{2i}, \gamma_{2i+1}))_{i=0}^{n-1}$ with $\ket{\vec{f}_\mathfrak{m}} = \ket{\vec{f}}$.
			\item If $n$ is a power of 2, there is a unique Clifford transformation
			\begin{align}
				C_{\mathfrak{m}_\text{BK}} : \ket{\vec{f}} \longmapsto \ket{G_\text{BK} \vec{f}}\, , \label{eqn:bkcliff}
			\end{align}
			where $G_\text{BK}$ is a recursively-defined invertible binary matrix; see Figure \ref{fig:bkmat} for an example when $n=16$. The fermion--qubit mapping $\mathfrak{m}_\text{BK} = ((C_{\mathfrak{m}_\text{BK}} \gamma_{2i} C_{\mathfrak{m}_\text{BK}}^\dagger , C_{\mathfrak{m}_\text{BK}} \gamma_{2i+1} C_{\mathfrak{m}_\text{BK}}))_{i=0}^{n-1}$ is exactly the Bravyi--Kitaev transformation \cite{bravyi_fermionic_2002}  with $\ket{\vec{f}_{\mathfrak{m}_\text{BK}}} = \ket{G_{\text{BK}} \vec{f}}$ as it appears in \cite{seeley_bravyi-kitaev_2012}.
			\item There is a Clifford transformation
			\begin{align}
				C_{\mathfrak{m}_\text{PB}} : \ket{\vec{f}} \longmapsto \ket{G_\text{PB} \vec{f}}\, , \label{eqn:pbcliff}
			\end{align}
			where $G_\text{PB}$ is the matrix with lower--triangular entries equal to 1, and all other entries zero. The fermion--qubit mapping that corresponds to $C_{\mathfrak{m}_\text{PB}}$ is exactly the parity basis transformation \cite{seeley_bravyi-kitaev_2012}.
			
		\end{enumerate}
	\end{exmp}
	
	\begin{exmp}\textit{(An affine encoding of the Fock basis that is not linear.)} \label{exmp:affine}
		The two--mode fermion--qubit mapping
		\begin{align}
			\mathfrak{m} = ((\Gamma_0, \Gamma_1), (\Gamma_2, \Gamma_3)) = ((X_0, -Y_0), (-Z_0X_1, -Z_0Y_1))
		\end{align}
		has vacuum stabilisers $-i\Gamma_0 \Gamma_1 = -Z_0$ and $-i\Gamma_2\Gamma_3= Z_1$, with simultaneous $(+1)$--eigenstates $\{ e^{i \phi}\ket{10} \mid \phi \in [0,2\pi)\}$. By Definition \ref{defn:templates}, this mapping is equivalent to the two--mode Jordan--Wigner transformation: that is, $\mathfrak{m} \in [\mathfrak{m}_\text{JW}]$. Choosing $\ket{\vec{0}_\mathfrak{m}}=\ket{10}$, the Fock basis of $\mathfrak{m}$ is
		\begin{align}
			\ket{00_\mathfrak{m}} &= \ket{10} \, , \quad \ket{10_\mathfrak{m}} = \ket{00}\, , \quad \ket{01_\mathfrak{m}} = \ket{11}\, , \quad \ket{11_\mathfrak{m}} = \ket{01}\, . \label{eqn:affinebasis}
		\end{align}
		Note that the relation $\ket{\vec{f}_\mathfrak{m}} = \ket*{\vec{f} \oplus 10}$ summarises Equation \ref{eqn:affinebasis}. Therefore $\mathfrak{m}$ is an affine encoding of the Fock basis, but it is not a linear encoding. 
	\end{exmp}

	The affine encoding of the Fock basis in Example \ref{exmp:affine} demonstrates a fermion--qubit mapping in  which the bit-string $10$ separates the encoded Fock state $\ket{\vec{f} \oplus 10}$ from the occupation vector of the fermionic Fock state $\ketf{\vec{f}}$ it represents.
	In Theorem \ref{thm:affine1}, we prove that the set of all classical encodings of the Fock basis that use Pauli strings to represent the Majorana operators is equal to the set of all affine encodings. 
	It is helpful to define the update, flip, parity and remainder sets of invertible binary matrices:
	
	\begin{definition} \label{defn:upf}\textit{(Generalised update, parity, flip and remainder sets from \cite{steudtner2018fermion}.)} Let $G\in \text{GL}_n(\mathbb{F}_2)$.
		For each $i \in \{0,1,\dots,n-1\}$, define the \textit{generalised} \textit{update}, \textit{flip}, \textit{parity} and \textit{remainder} sets of $i$ as follows:
		\begin{enumerate}
			\item The \textit{update set of $i$} is $U(i) = \{j \in \{0,1,\dots,n-1\} \, | \, G_{ji}= 1 \}$; i.e.\ the set $U(i)$ contains the indices of the rows of $G$ that have non-zero elements in column $i$.
			\item The \textit{flip set of $i$} is $F(i) = \{j \in \{0,1,\dots,n-1\}\, | \, (G^{-1})_{ik} = 1\}$; i.e.\ the set $F(i)$ contains the indices of the columns of $G^{-1}$ that have non-zero elements in row $i$.
			\item The \textit{parity set of $i$} is $P(i) = \{ j \in \{0,1,\dots,n-1\}\, | \, (\Pi G^{-1})_{ik} = 1 \}$, where $\Pi$ is the lower-triangular matrix of all 1s:
			\begin{align}
				\Pi = \begin{pmatrix}
					0 & 0 & 0 & \dots & 0 & 0 \\
					1 & 0 & 0 & \dots & 0 & 0 \\
					\vdots & \vdots & \vdots & \ddots & \vdots & \vdots \\
					1 & 1& 1 & \dots & 0 & 0 \\
					1 & 1& 1 & \dots & 1 & 0 
				\end{pmatrix} \in \text{M}_n(\mathbb{F}_2)\, ;
			\end{align}
			i.e.\ the set $P(i)$ contains the index of each column of $G^{-1}$ for which the sum of the first $(i{-}1)$ elements of that column is nonzero modulo 2. It is equivalent to define $P(i) = F(0) \, \triangle \, F(1) \, \triangle \, \dots \, \triangle \, F(i-1)$, where $\triangle$ denotes the symmetric difference.
			\item The \textit{remainder set of $i$} is $R(i) = F(i) \, \triangle \, P(i)$.
		\end{enumerate}
	\end{definition}

	\begin{lemma}\label{lem:upf2} \textit{(Original appearance in \cite{harrison2024mapping}.)}
		For any invertible binary matrix $G \in \text{GL}_n(\mathbb{F}_2)$, the update, parity, flip and remainder sets of Definition \ref{defn:upf} satisfy
		\begin{enumerate}[label=\alph*)]
			\item $|U(i) \cap F(i) | $ is odd for all $i \in \{0,1,\dots, n-1\}$,  \label{lem:upfa}
			\item $|U(i) \cap P(i) |$ is even for all $i \in \{0,1,\dots,n-1\}$, and \label{lem:upfb}
			\item $|U(i) \cap R(i) |$ is odd for all $i \in \{0,1,\dots,n-1\}$. \label{lem:upfc}
		\end{enumerate}
		\begin{proof}
			The relevant fact is that the rows of $G^{-1}$ and the columns of $G$ are orthonormal.  Because $G^{-1} G=\mathds{1}$, the matrix-vector product of the $i$th row of $G^{-1}$ and the $i$th column of $G$ must be 1 (mod 2). Thus, there must be an odd number of shared elements of $F(i)$ and $U(i)$ for each $i \in \{0,1,\dots,n-1\}$, proving \ref{lem:upfa}. Part \ref{lem:upfb} follows from the orthonormality of the rows of $G^{-1}$ and the columns of $G$: since the $i$th row of $\Pi G^{-1}$ is equal to the sum of the first $(i-1)$ rows of $G^{-1}$, each of which is orthogonal to the $i$th column of $G$, the $i$th row of $\Pi G^{-1}$ must itself be orthogonal to the $i$th column of $G$. Therefore there must be an even number of shared elements of $P(i)$ and $U(i)$. Part \ref{lem:upfc} follows from parts \ref{lem:upfa} and \ref{lem:upfb}, and the definition $R(i) = F(i) \, \triangle \, P(i)$.
		\end{proof}
	\end{lemma}

	\subsection{All classical encodings are equivalent to linear encodings}\label{sec:affine}

	In this section we prove Theorem \ref{thm:affine1}, which shows that the set of all classical encodings of the Fock basis that represent the Majorana operators with Pauli strings is equal to the set of all affine encodings. While this result appears in the literature, we include a comprehensive statement and proof here.  Corollary \ref{cor:affinelin}  incorporates our new definition of equivalence from Section \ref{sec:equiv} to show that all affine encodings are equivalent to linear encodings, with the only difference being the signs of some of the Pauli operators. These results play into our argument in Section \ref{sec:cbptree} relating ternary tree transformations to linear encodings of the Fock basis.

    \begin{theorem}\label{thm:affine1}\textit{(Cliffords that preserve the computational basis are equivalent to affine transformations of $\mathbb{F}_2^n$.)} 
    A Clifford operator $C \in \mathcal{C}_n$ that preserves the computational basis $C: \mathfrak{C}_n \rightarrow \mathfrak{C}_n$ must act as
        \begin{align}\label{eqn:affine3}
			C:\ket{\vec{f}} \longmapsto \ket{G( \vec{f} \oplus \vec{b})} \quad \text{for all} \quad \vec{f} \in \mathbb{F}_2^n\, ,
		\end{align}
		for some $G \in \text{GL}_n(\mathbb{F}_2)$ and some $\vec{b} \in \mathbb{F}_2^n$. 
     Likewise, for any $G \in \text{GL}_n(\mathbb{F}_2)$ and any $\vec{b} \in \mathbb{F}_2^n$, the unitary matrix that implements the affine map  of the form $\ket{\vec{f}} \mapsto \ket{G(\vec{f} \oplus \vec{b})}$ is necessarily a Clifford operator $C \in \mathcal{C}_n$.
    \begin{proof}
    Consider the subgroup of Cliffords $\{C \in \mathcal{C}_n \mid C:\mathfrak{C}_n \rightarrow \pm \mathfrak{C}_n \cup \pm i \mathfrak{C}_n\}$ that map computational basis states to computational basis states up to a possible phase. A set of generators for this subgroup are $\{\text{CNOT}_{ij}, \text{C}Z_{ij}, X_i, P_i \mid i \neq j \in [n]\}$ \cite{bravyi_hadamard-free_2021}, where
    \begin{align}
        \text{CNOT} = \begin{pmatrix} 1 & 0 & 0 & 0 \\ 0 & 1 & 0 & 0 \\ 0 & 0 & 0 & 1 \\ 0 & 0 & 1 & 0 \end{pmatrix} \, , \quad \text{C}Z = \begin{pmatrix} 1 & 0 & 0 & 0 \\ 0 & 1 & 0 & 0 \\ 0 & 0 & 1 & 0 \\ 0 & 0 & 0 & -1 \end{pmatrix}\, , \quad P = \begin{pmatrix} 1 & 0 \\ 0 & i \end{pmatrix} \, .
    \end{align}
    The only generators to introduce a phase change are C$Z_{ij}$ and $P_i$, and thus the CNOT$_{ij}$ and $X_i$ gates alone generate the Clifford subgroup  $\{C \in \mathcal{C}_n \mid C : \mathfrak{C}_n \rightarrow \mathfrak{C}_n\}$ that preserves the computational basis.
    Due to the identities \cite{harrison2023reducingqubitrequirementjordanwigner}
    \begin{align}
    \begin{quantikz}
    & \ctrl{1} & \gate{X} & \qw\\
    & \targ & \qw & \qw & \qw
    \end{quantikz} &= \begin{quantikz} & \gate{X} & \ctrl{1} & \qw\\
    & \gate{X} & \targ & \qw & \qw \end{quantikz} \\
    \begin{quantikz}
    & \ctrl{1} & \qw & \qw\\
    & \targ & \qw & \gate{X} & \qw
    \end{quantikz} &= \begin{quantikz} & \qw & \ctrl{1} & \qw\\
    & \gate{X} & \targ & \qw & \qw \end{quantikz}\, ,
    \end{align}
    it is possible to express a Clifford $C=C_GC_{\vec{b}}$ that preserves the computational basis as a sequence  of CNOT gates, which we label $C_G$, followed by a sequence of $X$ gates on qubits, which we label $C_\vec{b}$ where $b_i \in \mathbb{F}_2$ is 1 if and only if $C_\vec{b}$ acts with an $X$ gate on qubit $i$.

    The effect of $C_\vec{b}$ on the computational basis is simply to implement $C_\vec{b} : \ket{\vec{f}} \mapsto \ket{\vec{f} \oplus \vec{b}}$.
    The effect of $C_G$ is to implement $C_G : \ket{\vec{f}} \mapsto \ket{G \vec{f}}$ for some $G \in \text{GL}_n(\mathbb{F}_2)$. To understand the latter statement, consider the Clifford subgroup with generators $\{\text{CNOT}_{ij} \mid i,j \in [n]\}$, of which $C_G$ is a member. From \cite{bataille_quantum_2020}, this subgroup is isomorphic to $\mathrm{GL}_n(\mathbb{F}_2)$, the set of $n {\times} n$ invertible binary matrices. There is therefore a representation of the CNOT--only Clifford subgroup as invertible binary matrices that act directly on the binary string representations of the qubit states. For example, the $\mathrm{GL}_n(\mathbb{F}_2)$ isomorphism represents the equation $\mathrm{CNOT}\ket{10} = \ket{11}$ as
        \begin{equation}
        \begin{bmatrix}
        1 & 0 \\
        1 & 1
        \end{bmatrix}\begin{bmatrix} 1\\0 \end{bmatrix} = \begin{bmatrix} 1\\1 \end{bmatrix}\, .
        \end{equation}
	Therefore any Clifford operator $C$ that preserves the computational basis must implement an affine map of the form $\ket{\vec{f}} \mapsto \ket{\vec{G(\vec{f} \oplus \vec{b}})}$. 

For the reverse statement, one can implement $X$ gates on any subset of qubits to implement $C_\vec{b}$ and compose any number of CNOT gates to construct any invertible binary matrix transformation $C_G$ of the computational basis vector strings \cite{harrison_sierpinski_2024}. Thus, affine transformations of the form in Equation \ref{eqn:affine3} are always Clifford transformations.
    \end{proof}
    \end{theorem}

	\begin{corollary}\label{cor:affinelin}
		Every classical encoding of the Fock basis that represents the Majorana operators with Pauli strings is equivalent, in the sense of Definition \ref{defn:templates}, to a linear encoding.
		\begin{proof}
			Let $\mathfrak{m}=((\Gamma_{2i}, \Gamma_{2i+1}))_{i=0}^{n-1}$ be a classical encoding of the Fock basis with $\Gamma_i \in \mathcal{P}_n$. Therefore the mapping $C_\mathfrak{m} : \ket{\vec{f}}  \mapsto \ket{\vec{f}_\mathfrak{m}}$ is a Clifford operator, and, by Theorem \ref{thm:affine1}, with $C_\mathfrak{m} \ket{\vec{f}} = \ket{G(\vec{f} \oplus \vec{b})}$ for some $G \in \text{GL}_n(\mathbb{F}_2)$ and $\vec{b} \in \mathbb{F}_2^n$.
		
            Now consider the linear encoding of the Fock basis $\mathfrak{m}'=((\Gamma_{2i}', \Gamma_{2i+1}'))_{i=0}^{n-1}$ with Fock states $\ket{\vec{f}_{\mathfrak{m}'}} = \ket{G\vec{f}}$. Theorem \ref{thm:affine1} guarantees that the operator $C_{\mathfrak{m}'}$ implementing $C_{\mathfrak{m}'}:\ket{\vec{f}} \mapsto \ket{G\vec{f}}$ is a Clifford operator. Specifically, $C_{\mathfrak{m}'} = C_\mathfrak{m} C_\vec{b}$, where $C_\vec{b} = \prod_{b_i = 1} X_i$. The representations of the Majorana operators under $\mathfrak{m}'$ are the Pauli strings $\Gamma_i' = 
            C_{\mathfrak{m}'} \gamma_i C_{\mathfrak{m}'}^\dagger = C_\mathfrak{m} (C_\vec{b} \gamma_i C_\vec{b}) C_\mathfrak{m}^\dagger$. As $C_\vec{b}$ is a string of $X$ operators, then $C_\vec{b} \gamma_i C_\vec{b} =  \pm \gamma_i$ and so $\Gamma_i' = \pm C_\mathfrak{m} \Gamma_i C_\mathfrak{m}^\dagger = \pm \Gamma_i$.
            Therefore the two mappings $\mathfrak{m}$ and $\mathfrak{m}'$ are equivalent in the sense of Definition \ref{defn:templates}, i.e.\ $\mathfrak{m}' \in [\mathfrak{m}]$, because their Pauli operators differ by at most a sign change.
			\end{proof}
		\end{corollary}

In Appendix \ref{sec:appendix}, Theorem \ref{thm:affine} reproduces this result in the absence of prior knowledge about Clifford operators such as \cite{bataille_quantum_2020}. Corollary \ref{cor:affinemap} also gives the exact form of the Pauli representations of the Majorana operators for affine encodings of the Fock basis.

\section{Ternary tree transformations}\label{sec:treebased}

This section presents a refined and generalised definition of ternary tree transformations \cite{Vlasov_2022, miller2023bonsai, miller2024treespilation, jiang_optimal_2020}, which are fermion--qubit mappings that derive Pauli representations of the Majorana operators from ternary tree graphs. The diagrammatic notation of \cite{jiang_optimal_2020} efficiently depicts a fermion--qubit mapping with the minimal average Pauli weight. However, it does not specify the ordered pairing scheme, the signs of the Pauli operators, or the order of the fermionic labels, which are crucial to defining a ternary tree transformation to the same level of specificity as a linear encoding of the Fock basis.

Related work \cite{Vlasov_2022, miller2023bonsai, miller2024treespilation} not only expanded the definition of ternary tree transformations to incorporate any input ternary tree graph, but also explicitly defined a pairing scheme for the Pauli operators to ensure the vacuum state was $\ket{0}^{\otimes n}$. In Section \ref{sec:pptt} we make a more robust and general assertion: given the Pauli operators arising from a ternary tree, there is a unique scheme to pair and sign the operators that ensures the vacuum state can be any product stabiliser state.

\begin{definition} \textit{(Ternary trees).} \label{defn:tt}
	An \textit{$n$--vertex ternary tree} $T$ is an $n$--vertex tree graph, where each vertex has a unique label in $[n]$ and at most three children. Define the \textit{root} as the vertex with no parent. We adopt the convention of orienting tree graphs such that the root is the leftmost vertex.
\end{definition}

\begin{definition}\textit{(Unsigned Pauli operators.)} \label{defn:unsigned}
	The Pauli operator $P \in \mathcal{P}_n$ is an \textit{unsigned Pauli operator} if it belongs to the set $\{\mathds{1}, X, Y, Z \}^{\otimes n}$.
	The normal subgroup $K = \{\pm \mathds{1}, \pm i \mathds{1}\}$ of the $n$--qubit Pauli group $\mathcal{P}_n$ forms a quotient group that is isomorphic to the set of unsigned Pauli operators,
	\begin{align}
		\mathcal{P}_n/K = \{\{\pm P, \pm i P\} \mid P \in \mathcal{P}_n\} \cong \{\mathds{1}, X, Y, Z \}^{\otimes n}\, . \label{eqn:congruence}
	\end{align}
	In reference to Equation \ref{eqn:congruence}, the statement $P \in \mathcal{P}_n/K$ denotes that the Pauli operator $P$ is unsigned.
\end{definition}

\begin{definition}\textit{(Ternary--tree--based set of unsigned, anticommuting Pauli operators \cite{Vlasov_2022,jiang_optimal_2020}.)} \label{defn:treepaulis} Given an $n$--vertex ternary tree $T$, suppose the root of $T$ has label $r$. We define \textit{the $T$--based set of unsigned, anticommuting Pauli operators $\widetilde{\mathcal{G}}_T$} via the following procedure.
	\begin{enumerate}
		\item Draw additional edges so that each labelled vertex has three children, adding unlabelled vertices, which form the \textit{leaves} of the tree, if necessary.
		\item For each labelled vertex of $T$, label its rightward edges $X$, $Y$ and $Z$ from top-to-bottom. Associate each root-to-leaf path with the unsigned Pauli string that the path spells out, in the same manner as in fermion--qubit mapping diagrams of Definition \ref{defn:diagram}.
	\end{enumerate}
	Let $\widetilde{\mathcal{G}}_T \subset \mathcal{P}_n/K$ be the set of unsigned Pauli strings that arise from each of the $2n{+}1$ root-to-leaf paths. The elements of $\widetilde{\mathcal{G}}_T$ are anticommuting, because any two root-to-leaf paths diverge on a single vertex; the set $\widetilde{\mathcal{G}}_T$ is maximally anticommuting, because the maximum number of mutually anticommuting elements of $\mathcal{P}_n$ is $2n{+}1$ \cite{anticommuting2021sarkar}.
\end{definition}

\begin{figure}[btp]
	\centering
	\includegraphics[width=0.75\linewidth]{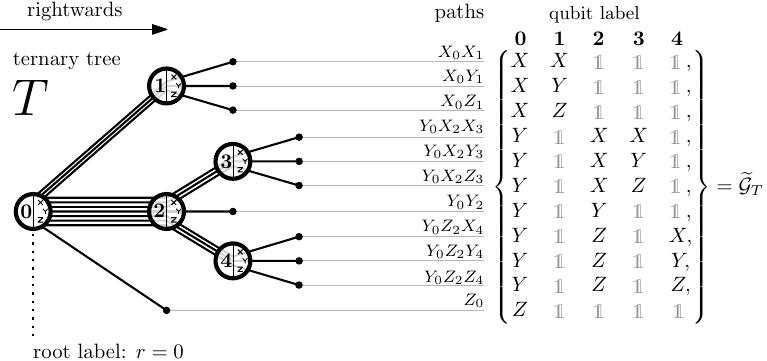}
	\caption{A 5--vertex ternary tree $T$ and the anticommuting set $\widetilde{\mathcal{G}}_T$  of unsigned Pauli operators.}
	\label{fig:tree-based-mappings-bonsai}
\end{figure}

\begin{exmp}\label{exmp:treec}
Figure \ref{fig:tree-based-mappings-bonsai} demonstrates a 5--vertex ternary tree $T$ and the $T$--based set  of unsigned, anticommuting of unsigned Pauli operators $\widetilde{\mathcal{G}}_T$.
\end{exmp}

\begin{definition} \textit{($T$--based mappings and ternary tree transformations.)} \label{defn:treebased}
	Let $\widetilde{\mathcal{G}}_T$ be the $T$--based set of unsigned, anticommuting Pauli operators. For any set $\{\Gamma_i\}_{i=0}^{2n-1}$ consisting of $2n$ distinct, signed, enumerated elements of $\widetilde{\mathcal{G}}_T$, i.e.\ $\pm\Gamma_i \in \widetilde{\mathcal{G}}_T$ for all $i=0,1,\dots, 2n-1$, we call the (ancilla--free) fermion--qubit mapping
	\begin{align}
		\mathfrak{m} = \big( ({\Gamma}_0, {\Gamma}_1), ({\Gamma}_2, {\Gamma}_3), \dots, ({\Gamma}_{2n-2}, {\Gamma}_{2n-1} ) \big) \label{eqn:treepaulis}
	\end{align}
	a \textit{ternary tree transformation}. Specifically, we call $\mathfrak{m}$ a $T$--\textit{based mapping}. Collectively, for all ternary trees $T$, we call the set of $T$--based mappings the \textit{ternary tree transformations}.
\end{definition}

\begin{exmp} \textit{(Original proposition \cite{jiang_optimal_2020, Vlasov_2022}.)}
For any $n \in \mathbb{N}$, the complete $n$--vertex ternary tree $T$ yields Pauli operators $\widetilde{\mathcal{G}}_T$ of equal weight ${\sim}\lceil \log_3(2n+1)\rceil$. The minimum value for the average weight of the Pauli operators of an $n$--mode ancilla--free fermion--qubit mapping $\mathfrak{m}= ((\Gamma_{2i}, \Gamma_{2i+1)}))_{i=0}^{n-1}$ is also $\lceil \log_3(2n+1)\rceil$, and so the complete ternary tee $T$ yields fermion--qubit mappings with the minimum average Pauli weight.
\end{exmp}

For an arbitrary choice of both the signs $\pm\Gamma_i \in \widetilde{\mathcal{G}}_T$ and the enumeration scheme $i=0,1,\dots,2n-1$ of the elements in $\{\Gamma_{i}\}_{i=0}^{2n-1}$, there is no guarantee that the Fock basis of a $T$--based mapping will consist of product states, let alone computational basis states.


\subsection{Uniqueness of product--preserving ternary tree transformations} \label{sec:pptt}

This section defines product--preserving and product--breaking fermion--qubit mappings. Previously work has made clear that there is always at least one $T$--based mapping with vacuum state equal to $\ket{0}^{\otimes n}$ for any $n$--qubit ternary tree $T$ \cite{Vlasov_2022, miller2023bonsai}, although the uniqueness of that mapping has not been clear. In Lemma \ref{lem:ttpairing}, we present a  general algorithm to produce a $T$--based for any $n$--vertex ternary tree $T$ with any $n$--qubit product stabiliser state as its vacuum -- not just a mapping with the vacuum state $\ket{0}^{\otimes n}$. Our algorithm lists all $T$--based mappings that have same vacuum state, and shows that they are equivalent up to the symmetries of Pauli braids and fermionic relabelling, effectively confirming the uniqueness of the existing product--preserving pairings of Pauli operators from ternary trees.

\begin{definition}\textit{(Product--preserving and product--breaking mappings.)} \label{defn:prod}
Let $\mathfrak{m}$ be an (ancilla--free) fermion--qubit mapping. By Equation \ref{eqn:occno}, the vacuum state $\ket{\vec{0}_\mathfrak{m}}$ is entangled if and only if all Fock states $\ket{\vec{f}_\mathfrak{m}}$ are entangled.
\begin{enumerate}[label=\alph*)]
	\item We say that $\mathfrak{m}$ is \textit{product--preserving} if $\ket{\vec{0}_\mathfrak{m}}$ is a product state.
	\item We say that $\mathfrak{m}$ is \textit{product--breaking} if $\ket{\vec{0}_{\mathfrak{m}}}$ is an entangled state.
\end{enumerate}
\end{definition}

\begin{figure}[btp]
	\centering
	\includegraphics[width=0.95\linewidth]{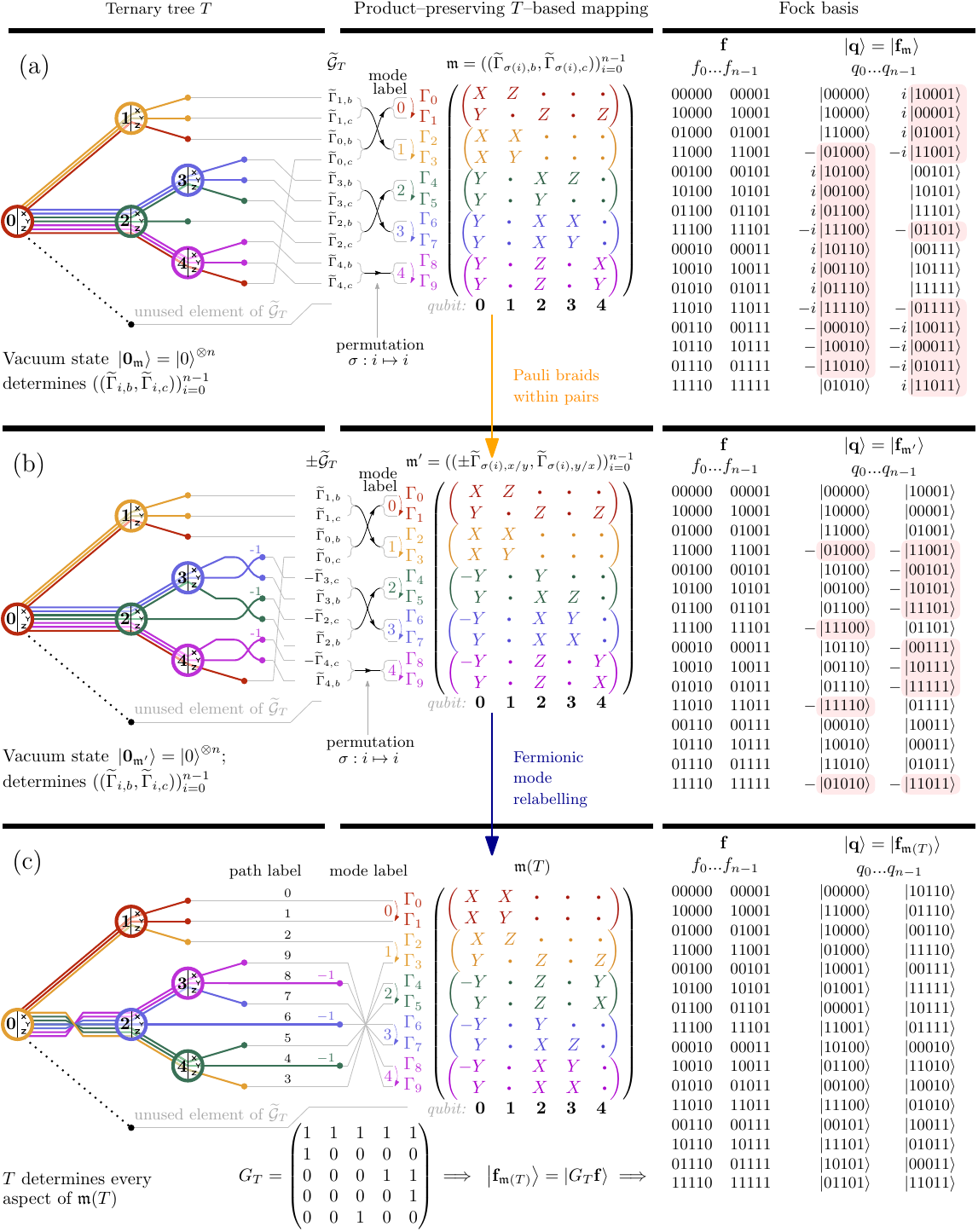}
	\caption{(a): The first prescription in \cite{Vlasov_2022, miller2023bonsai} for ternary--tree--based mappings produces a mapping $\mathfrak{m}$ with vacuum state $\ket{0}^{\otimes n}$ and a complex Fock basis. (b): The second prescription in \cite{miller2023bonsai} produces a mapping $\mathfrak{m}'$ with vacuum state $\ket{0}^{\otimes n}$ and a real Fock basis. (c): Our prescription produces the unique $T$--based mapping $\mathfrak{m}(T)$ that linearly encodes the Fock basis.}
	\label{fig:bonsai_pairing_all}
\end{figure}

\begin{exmp}\textit{(Examples of product--preserving and product--breaking mappings.)} \label{exmp:prod}
\begin{enumerate}[label=\alph*)]
	\item The mappings $\mathfrak{m}_\text{JW}$, $\mathfrak{m}_\text{BK}$ and $\mathfrak{m}_1$---$\mathfrak{m}_5$ in Figure \ref{fig:somemappings} are product--preserving, and the mapping $\mathfrak{m}_6$ is product--breaking.
	\item \label{exmp:oldpair}The ternary tree transformation $\mathfrak{m}$ in Figure \ref{fig:bonsai_pairing_all} (a) is a product--preserving mapping which results from applying the pairing algorithm from \cite{miller2023bonsai} to the ternary tree $T$ from Example \ref{exmp:treec}.
\end{enumerate}
\end{exmp}

Example \ref{exmp:prod} \ref{exmp:oldpair} made use of the Pauli operator pairing algorithm from \cite{miller2023bonsai}, which for any ternary tree $T$ yields a $T$--based mapping with vacuum state $\ket{0}^{\otimes n}$, demonstrates the existence of product--preserving ternary tree transformations. However, for the sake of showing equivalence between product--preserving ternary tree transformations and linear encodings of the Fock basis in Section \ref{sec:cbptree}, it is essential to find all $T$--based mappings with product stabiliser states as their vacuum states.

Lemma \ref{lem:ttpairing} generalises the previous pairing algorithm to produce a $T$--based mapping with any product vacuum state, for any ternary tree $T$. It also shows that all $T$--based mappings to satisfy this property are equivalent to the same mapping, with only the fermionic labelling, pair braiding, and Pauli sign change differentiating them. Section \ref{sec:cbptree} makes use of the uniqueness of the $T$--based mapping with a given product vacuum state up to equivalence.

\begin{lemma}\label{lem:ttpairing} \textit{(Uniqueness of product--preserving ternary tree transformations.)}
	Let $T$ be an $n$--vertex ternary tree. Then, for any product stabiliser state $\bigotimes_{i=0}^{n-1} \ket{a_i}_i$, where each $\ket{a_i} \in \{\ket{0}, \ket{1}, \ket{+}, \ket{-}, \ket{+i}, \ket{-i}\}$ is a $\pm 1$--eigenstate of a Pauli matrix $P_i \in \{Z,X,Y\}$, the following holds:
	\begin{enumerate}[label=\alph*)]
		\item There is an algorithmic pairing $\{(\widetilde{\Gamma}_{i,b}, \widetilde{\Gamma}_{i,c})\}_{i=0}^{n-1}$ of the elements in $\widetilde{\mathcal{G}}_T$ such that the mapping $\mathfrak{m} = ((\widetilde{\Gamma}_{i,b}, \widetilde{\Gamma}_{i,c}))_{i=0}^{n-1}$ has vacuum state $\ket{\vec{0}_\mathfrak{m}} = \bigotimes_{i=0}^{n-1}\ket{a_i}_i$. In fact, any $T$--based mapping $\mathfrak{m} = ((\Gamma_{2i},\Gamma_{2i+1}))_{i=0}^{n-1}$ with all $\Gamma_i \in \widetilde{\mathcal{G}}_T$ and with vacuum state $\ket{\vec{0}_\mathfrak{m}} = \bigotimes_{i=0}^{n-1}\ket{a_i}_i$ must be of the form $\mathfrak{m} = ((\widetilde{\Gamma}_{\sigma(i), b}, \widetilde{\Gamma}_{\sigma(i), c}))_{i=0}^{n-1}$ where the permutation $\sigma \in S_n$ is a fermionic labelling scheme.
		\item The only $T$--based mappings with vacuum state $\bigotimes_{i=0}^{n-1} \ket{a_i}_i$ are those of the form $\mathfrak{m} = ((\Gamma_{2i}, \Gamma_{2i+1}))_{i=0}^{n-1}$ where, for some permutation $\sigma \in S_n$ acting as a fermionic labelling scheme, the Pauli operator pairs are
		\begin{align} \label{eqn:avacform}
			(\Gamma_{2i}, \Gamma_{2i+1}) &= \begin{cases}
				(\pm\widetilde{\Gamma}_{\sigma(i), b}, \pm\widetilde{\Gamma}_{\sigma(i), c})\, , & \text{as in part a), up to a global sign, or} \\
				(\pm\widetilde{\Gamma}_{\sigma(i), c} , \mp \widetilde{\Gamma}_{\sigma(i), b})\, , & \text{a braid of the $i$th mode from part a).}
			\end{cases}
		\end{align}
	\end{enumerate}
	\begin{proof}
		
		\begin{figure}[btp]
			\centering
			\includegraphics[width=\linewidth]{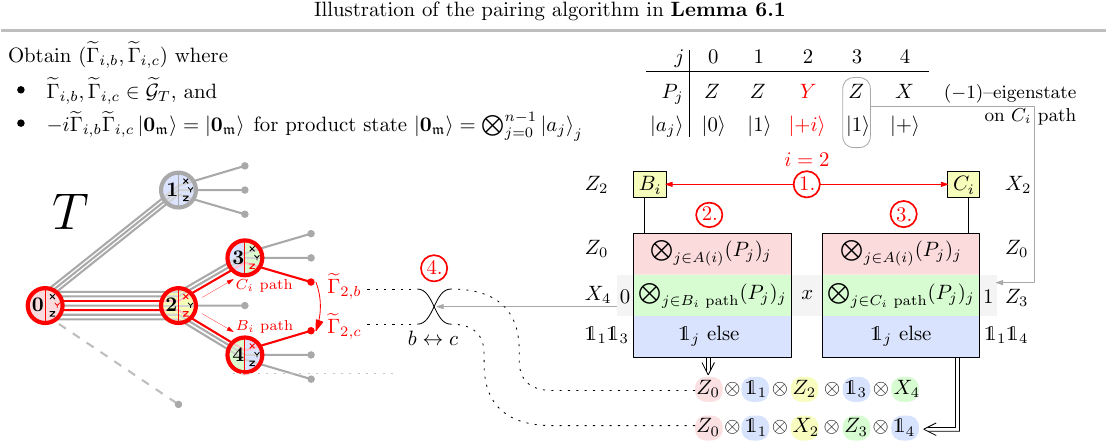}
			\caption{Demonstration of the pairing algorithm in Lemma \ref{lem:ttpairing} for a 5--vertex ternary tree $T$. The algorithm identifies the pairing of the elements of $\widetilde{\mathcal{G}}_T$ that \textit{must} comprise any $T$--based mapping with the vacuum state $\ket{0}\ket{1}\ket{+i}\ket{1}\ket{+}$.}
			\label{fig:alg_demonstration}
		\end{figure}

		For part a), the task is, for all $i \in [n]$, to find two operators $\widetilde{\Gamma}_{i,b}, \widetilde{\Gamma}_{i,c} \in \widetilde{\mathcal{G}}_T$ such that
		\begin{align}\label{eqn:prodreq}
			-i\widetilde{\Gamma}_{i,b} \widetilde{\Gamma}_{i,c} \left( \bigotimes_{j=0}^{n-1} \ket{a_j}_j\right) = \bigotimes_{j=0}^{n-1} \ket{a_j}_j\, .
		\end{align}
		Our four-step algorithm identifies such a pair of operators.
		
		Let $i \in [n]$, and recall that the operators in $\widetilde{\mathcal{G}}_T$ correspond to root-to-leaf paths in $T$. We will construct a pair of operators $(\widetilde{\Gamma}_{i,b},\widetilde{\Gamma}_{i,c})$ satisfying Equation \ref{eqn:prodreq} whose corresponding root-to-leaf paths diverge on the vertex of $T$ with label $i$.  Note that any two paths diverging on vertex $i$ will share vertex $i$ and all of its ancestors in $T$, which we denote by $A(i)$, and that the paths share no other vertices.
		
		The first step of our algorithm identifies how the operators $\widetilde{\Gamma}_{i,b}$ and $\widetilde{\Gamma}_{i,c}$ must act on the $i$th qubit, should they exist. A necessary condition for Equation  \ref{eqn:prodreq} to hold is for $-i\widetilde{\Gamma}_{i,b}\widetilde{\Gamma}_{i,c}$ to locally preserve the eigenstate $\ket{a_i}_i$. Since the paths diverge on the $i$th vertex, the operators must act with different single--qubit Paulis:
		\begin{enumerate}[label=\textcolor{red}{\arabic*.}]
			\item Let $B, C \in \{X,Y,Z\}$ be the single--qubit Pauli operators that satisfy $-iBC\ket{a_i} = \ket{a_i}$. Then, the operator $\widetilde{\Gamma}_{i,b}$ must act with $B_i$ on the $i$th qubit, and the operator $\widetilde{\Gamma}_{i,c}$ with $C_i$ on the $i$th qubit. \label{step:1}
		\end{enumerate}
		The second and third steps of our algorithm show the existence of operators $\widetilde{\Gamma}_{i,b}$ and $\widetilde{\Gamma}_{i,c}$ that satisfy step \ref{step:1}\ while also satisfying another necessary condition of Equation \ref{eqn:prodreq}: that the product state $\bigotimes_{j=0}^{n-1}\ket{a_j}_j$ is a $(\pm 1)$--eigenstate of $-i\widetilde{\Gamma}_{i,b}\widetilde{\Gamma}_{i,c}$.
		\begin{enumerate}[label=\textcolor{red}{\arabic*.}]
			\setcounter{enumi}{1}
			\item Consider the unique root-to-leaf path in $T$ that leaves the vertex with label $i$ via its rightward edge with label $B$, and then passes through each subsequent vertex $j$ via the rightward edge with label $P_j$ until terminating at an unlabelled leaf vertex. Define the \textit{$B_i$ path} to be the segment of this root-to-leaf path to the right of vertex $i$, and define $\widetilde{\Gamma}_{i,b}$ to be the unsigned Pauli string that arises from this path via the construction in Definition \ref{defn:treepaulis}.\label{step:2}
			\item Similarly, define $\widetilde{\Gamma}_{i,c}$ to be the Pauli string in $\widetilde{\mathcal{G}}_T$ that matches the unique root-to-leaf path in $T$ that leaves vertex $i$ via its rightward edge with label $C$ and passes through each subsequent vertex $j$ via the rightward edge with label $P_j$. Define the $C_i$ \textit{path} to be the segment of this root-to-leaf path to the right of vertex $i$.\label{step:3}
		\end{enumerate}
		The operator $-i\widetilde{\Gamma}_{i,b} \widetilde{\Gamma}_{i,c}$ acts trivially on all qubits with labels in $A(i)$, as $\widetilde{\Gamma}_{i,b}$ and $\widetilde{\Gamma}_{i,c}$ act identically upon them. Steps \textcolor{red}{\ref{step:2}}\ and \textcolor{red}{\ref{step:3}}\ ensure that $\bigotimes_{j=0}^{n-1}\ket{a_j}_j$ is a $(\pm 1)$--eigenstate of $-i\widetilde{\Gamma}_{i,b}\widetilde{\Gamma}_{i,c}$ via
		\begin{align}
			-i\widetilde{\Gamma}_{i,b} \widetilde{\Gamma}_{i,c} \left(\bigotimes_{j=0}^{n-1} \ket{a_j}_j \right) &= -i \left(\bigotimes_{j \in (B_i \text{ path})} (P_j)_j \right) B_i  \left(\bigotimes_{j \in (C_i \text{ path})} (P_j)_j \right) C_i \left(\bigotimes_{j=0}^{n-1} \ket{a_j}_j \right) \\
			&= \left(\bigotimes_{j \in (B_i \text{ path}) \cup (C_i \text{ path})} (P_j \ket{a_j})_j\right) (-iBC\ket{a_i})_i \left(\bigotimes_{j \text{ else}} \ket{a_j}\right) \\
			&= (-1)^{x} \bigotimes_{j=0}^{n-1} \ket{a_j}_j\, ,
		\end{align}
		where $x$ is the number of Pauli $(-1)$--eigenstates in $\{\ket{a_j}\mid j \in (B_i \text{ path} ) \cup (C_i \text{ path})\}$, i.e.\ the number of vertices in the $B_i$ and $C_i$ paths with labels $j$ that satisfy $P_j \ket{a_j} = -\ket{a_j}$. The algorithm thus requires a corrective step:
		\begin{enumerate}[label=\textcolor{red}{\arabic*.}]
			\setcounter{enumi}{3}
			\item If the number of Pauli $(-1)$--eigenstates in $\{\ket{a_j} \mid j \in (B_i \text{ path}) \cup (C_i \text{ path}) \}$ is odd, exchange the symbols $b$ and $c$. \label{step:4}
		\end{enumerate}

		\begin{figure}[btp]
			\centering
			\includegraphics[width=\linewidth]{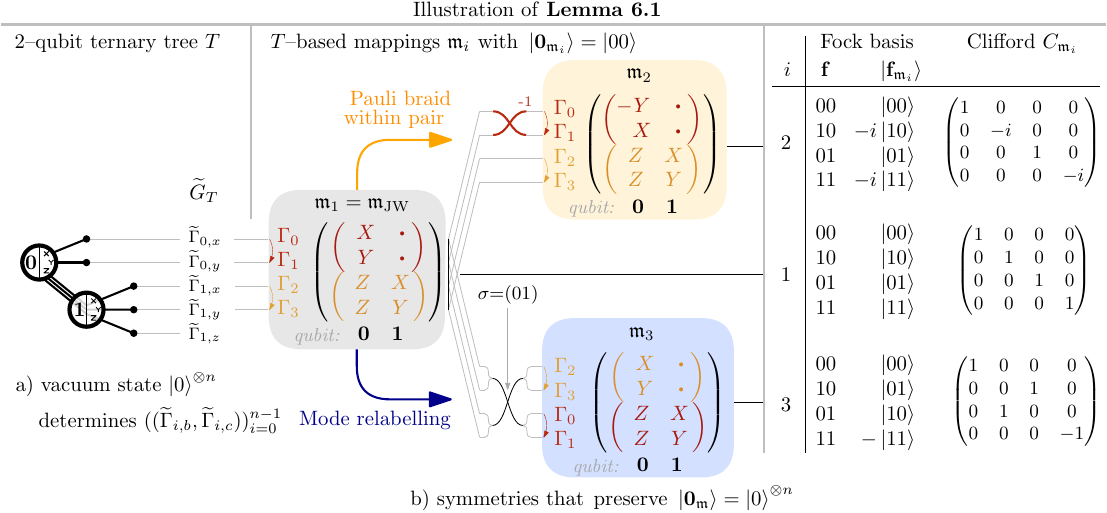}
			\caption{A two--vertex ternary tree $T$ and the Jordan--Wigner transformation $\mathfrak{m}_1=\mathfrak{m}_{\text{JW}}$,  which is a $T$--based mapping with vacuum state $\ket{00}$. The mappings $\mathfrak{m}_2$ and $\mathfrak{m}_3$ also have vacuum state $\ket{00}$, and are equivalent to $\mathfrak{m}_1$. Lemma \ref{lem:ttpairing} stipulates that any $T$--based mapping with vacuum state $\ket{00}$ may only differ from $\mathfrak{m}_1$ via  Pauli pair braids and fermionic relabelling.}
			\label{fig:braid_demonstration}
		\end{figure}

		Steps \ref{step:1}--\ref{step:4} identify a pair of operators $\widetilde{\Gamma}_{i,b}$ and $\widetilde{\Gamma}_{i,c}$  satisfying Equation \ref{eqn:prodreq}, which correspond to two  unique root-to-leaf paths of $T$. The uniqueness of these paths reflects the stronger statement that these are the only Pauli operators to anticommute on qubit $i$ and also act with either $P_j$ or $\mathds{1}_j$ on all qubits $j \notin A(i)$. This guarantees the existence and uniqueness of the pair of operators $(\widetilde{\Gamma}_{i,b}, \widetilde{\Gamma}_{i,c})$ for each $i \in [n]$. Thus
		the set $\{(\widetilde{\Gamma}_{j,b}, \widetilde{\Gamma}_{j,c})\}_{j=0}^{n-1}$ is the \textit{only} set of Pauli operator pairs satisfying Equation \ref{eqn:prodreq} in which the operators of the $i$th pair anticommute on the $i$th qubit.
		
		At first this seems to leave open the possibility of sets of operator pairs with a different structure, in which at least two pairs $(\widetilde{\Gamma}_{1}, \widetilde{\Gamma}_{2}), (\widetilde{\Gamma}_3, \widetilde{\Gamma}_4)$ are such that $\widetilde{\Gamma}_1$ anticommutes with $\widetilde{\Gamma}_2$ on qubit $i$, and $\widetilde{\Gamma}_{3}$ and $\widetilde{\Gamma}_4$ also anticommute on qubit $i$. But because there are only three mutually anticommuting single--qubit Pauli matrices,  there would be elements from distinct pairs -- say $\widetilde{\Gamma}_1$ and $\widetilde{\Gamma}_3$ -- which could not anticommute on qubit $i$ but would instead anticommute one of its children, say qubit $i'$. In this scenario, it is impossible for the product state $\bigotimes_{j=0}^{n-1} \ket{a_j}_j$ to be an eigenstate of both $-i\widetilde{\Gamma}_1\widetilde{\Gamma}_2$ and $-i\widetilde{\Gamma}_3\widetilde{\Gamma}_4$, because the two products act with different single--qubit Pauli operators on qubit $i'$.

		Therefore, there is a single degree of freedom remaining in identifying a set of pairs of operators in $\widetilde{\mathcal{G}}_T$ that satisfy Equation \ref{eqn:prodreq}: the bijection between fermionic modes and qubit labels. For any $\sigma \in S_n$, the $T$--based fermion--qubit mapping $\mathfrak{m} = ((\widetilde{\Gamma}_{\sigma(i),b}, \widetilde{\Gamma}_{\sigma(i),c}))_{i=0}^{n-1}$ has vacuum state $\ket{\vec{0}_\mathfrak{m}} = \bigotimes_{i=0}^{n-1} \ket{a_i}_i$, because any permutation of the index $i$ in $-i\widetilde{\Gamma}_{i,b}\widetilde{\Gamma}_{i,c}$ upholds Equation \ref{eqn:prodreq}. 
		
		Part b) relaxes the search for $T$--based mappings of strictly unsigned Pauli operators with vacuum state $\bigotimes_{i=0}^{n-1} \ket{a_i}_i$ to permit operators of the form $\pm \Gamma_i \in \widetilde{\mathcal{G}}_T$; the task thus amounts to investigating how to interchange and sign operators from part a) while preserving the vacuum state. Clearly the pair $(-\widetilde{\Gamma}_{\sigma(i), b}, -\widetilde{\Gamma}_{\sigma(i),c})$ satisfies Equation \ref{eqn:prodreq}, and due to the anticommutation $\{\widetilde{\Gamma}_{i,b}, \widetilde{\Gamma}_{i,c}\} = 0$, so do the Pauli braids of the operator pair:
		\begin{align}
			-i\left(-\widetilde{\Gamma}_{i,c}\right)\widetilde{\Gamma}_{i,b} \left(\bigotimes_{j=0}^{n-1} \ket{a_j}_j \right) = \left(\bigotimes_{j=0}^{n-1} \ket{a_j}_j \right)\, , \quad 
			-i\widetilde{\Gamma}_{i,c}\left(-\widetilde{\Gamma}_{i,b}\right) \left(\bigotimes_{j=0}^{n-1} \ket{a_j}_j \right) = \left(\bigotimes_{j=0}^{n-1} \ket{a_j}_j \right)\, .
		\end{align}
		Therefore, for any $\sigma \in S_n$, a mapping $\mathfrak{m} = ((\Gamma_{2i}, \Gamma_{2i+1}))_{i=0}^{n-1}$ with pairs of the form
		\begin{align} \label{eqn:braidsandall}
			(\Gamma_{2i}, \Gamma_{2i+1}) &= \begin{cases}
				(\pm \widetilde{\Gamma}_{\sigma(i), b}, \pm\widetilde{\Gamma}_{\sigma(i), c}) & \text{or} \\
				(\pm\widetilde{\Gamma}_{\sigma(i), c} , \mp \widetilde{\Gamma}_{\sigma(i), b} ) &
			\end{cases}
		\end{align}
		is a $T$--based mapping with vacuum state $\ket{\vec{0}_\mathfrak{m}} = \bigotimes_{i=0}^{n-1} \ket{a_i}_i$. These modifications of the pairs in mappings from part a) collectively describe every $T$--based mapping with vacuum state $\bigotimes_{i=0}^{n-1}\ket{a_i}_i$.
	\end{proof}
\end{lemma}

\begin{exmp}\textit{(Demonstration of Lemma \ref{lem:ttpairing}).}
	Let $T$ be the two--vertex ternary tree in Figure \ref{fig:braid_demonstration}. The $T$--based set of unsigned, anticommuting Pauli operators is  $\widetilde{\mathcal{G}}_T = \{X_0, Y_0, Z_0X_1, Z_0Y_1, Z_0Z_3\}$.  Lemma \ref{lem:ttpairing} a) identifies the Jordan--Wigner transformation $\mathfrak{m}_1 = \mathfrak{m}_\text{JW} ((X_0, Y_0), (Z_0 X_1, Z_0Y_1))$ as a $T$--based transformation with vacuum state $\ket{00}$ and unsigned Pauli strings; the only other $T$--based mappings with vacuum state $\ket{00}$ are equivalent to $\mathfrak{m}_1$. Two examples of equivalent $T$--based mappings with vacuum state $\ket{00}$ are $\mathfrak{m}_2$ and $\mathfrak{m}_3$, which Figure \ref{fig:braid_demonstration} also depicts.
\end{exmp}

\begin{exmp}\label{exmp:tree1}
	\textit{(Pauli operator pairing algorithm from \cite{miller2023bonsai, Vlasov_2022} in the language of Lemma \ref{lem:ttpairing}.)} \label{exmp:type1}
	Applying the algorithm for generating a product--preserving ternary--tree--based mapping from \cite{miller2023bonsai, Vlasov_2022} to an $n$--vertex ternary tree $T$ generates the unique $T$--based mapping of the form $\mathfrak{m} = ((\widetilde{\Gamma}_{\sigma(i),b}, \widetilde{\Gamma}_{\sigma(i),c}))_{i=0}^{n-1}$ with vacuum state $\ket{\vec{0}_\mathfrak{m}} = \ket{0}^{\otimes n}$, for some $\sigma \in S_n$. Revisiting Example \ref{exmp:prod}\ref{exmp:oldpair}, Figure \ref{fig:bonsai_pairing_all} (a) demonstrates the mapping of this kind for the 5--vertex ternary tree from Example \ref{exmp:treec}, with $\sigma$ equal to the trivial permutation. Note that the Fock states of this mapping are computational basis states with complex coefficients, i.e.\ $\ket{\vec{f}_{\mathfrak{m}}} \in \pm \mathfrak{C} \cup \pm i \mathfrak{C}$.
\end{exmp}

The authors  of \cite{miller2023bonsai} and the related work \cite{miller2024treespilation} suggest a refinement of the algorithm in Example \ref{exmp:type1}. Example \ref{exmp:tree2} details the modification, which braids Pauli operator pairs to produce mappings with Fock states that are strictly in the real computational basis.

\begin{exmp}\label{exmp:tree2} \textit{(Pauli operator pairing algorithm from \cite{miller2023bonsai, miller2024treespilation}.)}
	Let $T$ be an $n$-vertex ternary tree and, for some $\sigma \in S_n$, consider the $T$--based mapping $\mathfrak{m} = ((\widetilde{\Gamma}_{\sigma(i),b}, \widetilde{\Gamma}_{\sigma(i),c}))_{i=0}^{n-1}$ with vacuum state $\ket{\vec{0}_{\mathfrak{m}}} = \ket{0}^{\otimes n}$ arising from the prescription in Example \ref{exmp:type1}. By definition, for each $i \in \{0,1,\dots,n-1\}$, the pair of Pauli operators $(\Gamma_{2i}, \Gamma_{2i+1})$ $=$  $(\widetilde{\Gamma}_{\sigma(i), b} , \widetilde{\Gamma}_{\sigma(i), c})$ act with $X$ and $Y$ on the qubit with label $\sigma(i)$, and with $Z$ operators on all qubits in the $B_{\sigma(i)}$ and $C_{\sigma(i)}$ paths. Therefore, the pair consists of an operator that acts with a $Y$ matrix on an even number of qubits, and an operator that acts with a $Y$ matrix on an odd number of qubits. If $\widetilde{\Gamma}_{\sigma(i),b}$ acts with an odd number of $Y$ operators, define a new mapping $\mathfrak{m}' = ((\Gamma_{2i}',\Gamma_{2i+1}'))_{i=0}^{n-1}$ with operator pairs:
	\begin{align}\label{eqn:tacticalbraiding}
		\begin{cases}
			\Gamma_{2i}' = \phantom{-}\widetilde{\Gamma}_{\sigma(i),b}, \quad \Gamma_{2i+1}' = \widetilde{\Gamma}_{\sigma(i), c} & \text{if the number of $Y$s in $\widetilde{\Gamma}_{\sigma(i), b}$ is even}\, ,\\
			\Gamma_{2i}' = -\widetilde{\Gamma}_{\sigma(i),c}, \quad \Gamma_{2i+1}' = \widetilde{\Gamma}_{\sigma(i), b} & \text{if the number of $Y$s in $\widetilde{\Gamma}_{\sigma(i), b}$ is odd}\, .
		\end{cases}
	\end{align}
	Because $\mathfrak{m}'$ differs from $\mathfrak{m}$ only by Majorana braids, the vacuum state of $\mathfrak{m}'$ is $\ket{\vec{0}_{\mathfrak{m}'}} = \ket{\vec{0}_{\mathfrak{m}}} = \ket{0}^{\otimes n}$. Equation \ref{eqn:tacticalbraiding} braids the pair $(\Gamma_{2i}, \Gamma_{2i+1}) \mapsto (-\Gamma_{2i+1}, \Gamma_{2i}) = (\Gamma_{2i}', \Gamma_{2i+1}')$ from $\mathfrak{m}$ if and only if $\Gamma_{2i}$ acts with an odd number of $Y$ operators, which ensures that real computational bases $\pm\mathfrak{C}_n$ contain the Fock basis $\{\ket{\vec{f}_{\mathfrak{m}'}} \mid \vec{f} \in \mathbb{F}_2^n\}$ since $\ket{\vec{f}_\mathfrak{m}} = \Gamma_0^{f_0} \Gamma_2 ^{f_1} \dots \Gamma_{2n-2}^{f_{n-1}}\ket{0}^{\otimes n}$. Figure \ref{fig:bonsai_pairing_all} (b) demonstrates the mapping $\mathfrak{m}'$ arising from the modification of the mapping $\mathfrak{m}$ from Example \ref{exmp:tree1}.
\end{exmp}

\section{Product--preserving ternary tree transformations are equivalent to linear encodings of the Fock basis} \label{sec:cbptree}

Proof that ternary tree transformations yield the minimum--weight Pauli representations of the Majorana operators \cite{jiang_optimal_2020} showcased their potential, and
the demonstration that every ternary tree $T$ can yield a product--preserving $T$--based mapping \cite{Vlasov_2022, miller2023bonsai} removed the threat of entangled Fock basis states. In Section \ref{sec:treebased}, we established a broader class of product--preserving ternary tree transformations, which we will now demonstrate is equivalent to the class of linear encodings of the Fock basis.

Despite our generalisation of ternary tree transformations, the full relationship between ternary tree transformations and the linear encodings of Section \ref{app:ibm} remains unclear. Example \ref{exmp:tree1} illustrated that product--preserving ternary tree transformations with vacuum states of $\ket{0}^{\otimes n}$ are not necessarily linear encodings of the Fock basis, and can employ Fock states in both $\pm \mathfrak{C}_n$ and $\pm i \mathfrak{C}_n$. Example \ref{exmp:tree2} demonstrated how to apply Pauli braids to produce $T$--based mappings with Fock states in $\pm \mathfrak{C}_n$, which suggests that ternary tree transformations are close in definition to linear encodings.

In this section, we combine our universal notation for fermion--qubit mappings from Section \ref{sec:prelim} with our notion for equivalence from Section \ref{sec:equiv} to show that every product--preserving ternary tree transformation from Section \ref{sec:treebased} is indeed equivalent to a linear encoding of the Fock basis from Section \ref{app:ibm}. We detail a procedure to recover the invertible binary matrix of the unique linear encoding for each ternary tree input, illustrating that linear encodings contain product--preserving ternary tree transformations within the classical/affine/linear encoding hierarchy of the Fock basis. 

Section \ref{sec:ppttbm} contains crucial results that build towards Theorem \ref{thm:cbptbmappings}:
\begin{itemize}
	\item \textbf{Lemma \ref{lem:equiv}} shows that for any product stabiliser state $\bigotimes_{i=0}^{n-1} \ket{a'_i}_i$, and any $T$--based mapping $\mathfrak{m}$ with vacuum state $\ket{\vec{0}_\mathfrak{m}} = \bigotimes_{i=0}^{n-1}\ket{a_i}_i$, there is a $T'$--based mapping $\mathfrak{m}'$ with vacuum state $\ket*{\vec{0}_{\mathfrak{m}'}}$ that differs from $\mathfrak{m}$ only by locally relabelling Pauli operators, where $T'$ only differs from $T$ by edge relabelling.
	
	\item \textbf{Lemma \ref{lem:compbasistt}} establishes that there is a unique Clifford operator $C_T \in\mathcal{C}_n$ determining a $T$--based mapping $\mathfrak{m}(T)= ((C_T \gamma_{2i} C_T^\dagger, C_T \gamma_{2i+1}C_T^\dagger ))_{i=0}^{n-1}$ with the properties that 
	
	1) $\mathfrak{m}(T)$ is a $T$--based mapping (i.e.\ $C_T \gamma_i C_T^\dagger \in \pm \widetilde{\mathcal{G}}_T$ for all $i \in [2n]$), and
	
	2) $\mathfrak{m}(T)$ is a classical encoding of the Fock basis (i.e.\  $\ket*{\vec{f}_{\mathfrak{m}(T)}} \in \mathfrak{C}_n$ for all $\vec{f}\in \mathbb{F}_2^n$) with $\ket*{\vec{0}_{\mathfrak{m}(T)}} = \ket{{0}}^{\otimes n}$.

	Theorem \ref{thm:affine1} thus implies that $\mathfrak{m}(T)$ is a linear encoding of the Fock basis.
	
	\item \textbf{Lemma \ref{lem:findmatrix}} finds the invertible binary matrix $G_T \in \text{GL}_n(\mathds{F}_2)$ such that $\ket*{\vec{f}_{\mathfrak{m}(T)}} = \ket{G_T \vec{f}}$.
\end{itemize}

Section \ref{sec:mainthm} presents the main result:

\begin{itemize}	
	\item \textbf{Theorem \ref{thm:cbptbmappings}:}
	\begin{enumerate}[label=\alph*)]
		\item 	For each ternary tree $T$, there is a unique $T$--based mapping  $\mathfrak{m}(T)$ that is also a linear encoding of the Fock basis.
		\item  Every $n$--mode product--preserving ternary tee transformation is equivalent to a mapping in the set $\{\mathfrak{m}(T) \mid T \, \text{is an $n$--vertex ternary tree}\}$.
		\end{enumerate}
\end{itemize}
In Section \ref{sec:sierpinski}, we apply Theorem 2 to the complete ternary tee graph. In doing so, we recover the pruned Sierpinski tree transformation \cite{harrison2024mapping}, demonstrating its equivalence to complete ternary tree transformation.

\subsection{Supporting results for Theorem \ref{thm:cbptbmappings}}\label{sec:ppttbm}

In Section \ref{sec:treebased}, Lemma \ref{lem:ttpairing} showed that that all $T$--based mappings with the same product vacuum state are equivalent to the same mapping, up to fermionic labelling and the signs and braiding of the Pauli operator pairs.  Lemma \ref{lem:equiv} uses this result to show that all product--preserving $T$--based mappings are equivalent to the same mapping, up to local Pauli relabelling.

\begin{figure}[btp]
	\centering
	\includegraphics[width=\linewidth]{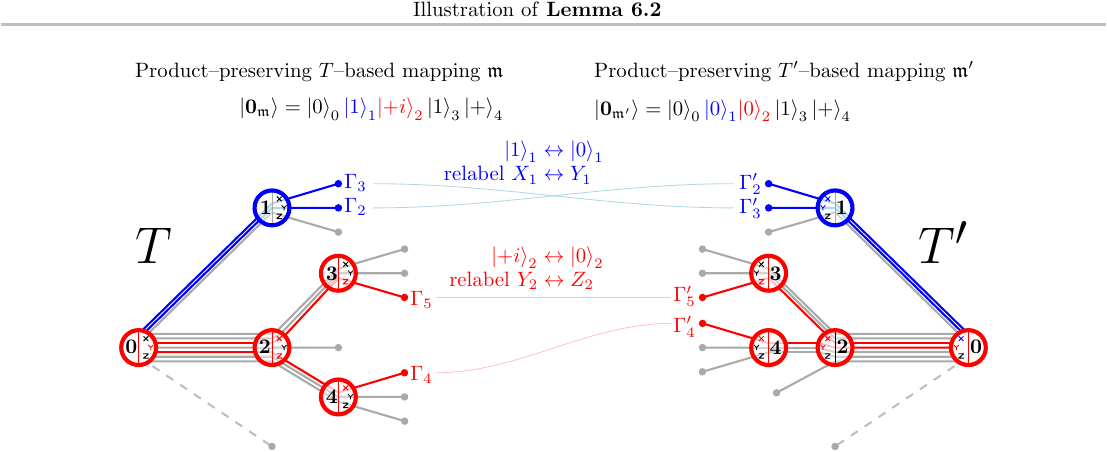}
	\caption{Adjusting the labels of the local Pauli matrices adjusts the vacuum state of a ternary tree transformation. This figure demonstrates two such adjustments of a $T$--based mapping for a particular 5--vertex ternary tree: exchanging $X$ and $Y$ labels on the 1st qubit modifies the vacuum state locally from $\ket{1}_1$ to $\ket{0}_0$, and exchanging $Y$ with $Z$ on the 2nd qubit modifies $\ket{+i}_2$ to $\ket{0}_2$. }
	\label{fig:product_preserving}
\end{figure}

\begin{lemma}\label{lem:equiv}
	\textit{(For each product stabiliser state, there is a mapping based upon any ternary tree with that product state as its vacuum state, up to local relabelling of edges.)} Let $T$ be an $n$--vertex ternary tree, and let $\mathfrak{m}$ be a $T$--based mapping with vacuum state $\ket{\vec{0}_{\mathfrak{m}}} = \bigotimes_{i=0}^{n-1} \ket{a_i}_i$. Then, for any product stabiliser state $\bigotimes_{i=0}^{n-1} \ket{a_i'}_i$, there exists an $n$--vertex ternary tree $T'$ and a  $T'$--based mapping $\mathfrak{m}'$ with vacuum state $\bigotimes_{i=0}^{n-1} \ket{a_i'}_i$, where the operators of $\mathfrak{m'}$ differ from those of $\mathfrak{m}$ only by local permutations of the Pauli labels $\{X,Y,Z\}$.
	\begin{proof}
		From Lemma \ref{lem:ttpairing}, the mapping $\mathfrak{m} = ((\Gamma_{2i}, \Gamma_{2i+1}))_{i=0}^{n-1}$ must be of the form in Equation \ref{eqn:avacform} for some $\sigma \in S_n$. 
		The following process implements a modification of $T$:
		
		Let $i \in [n]$. If the operator pair $(\Gamma_{2i},\Gamma_{2i+1})$ is of the form $(\pm\widetilde{\Gamma}_{\sigma(i),b}, \pm\widetilde{\Gamma}_{\sigma(i),c})$, let $B,C \in \{X,Y,Z\}$ be the Pauli  matrices satisfying $-iBC\ket{a_i} = \ket{a_i}$ and let $B',C' \in \{X, Y, Z\}$ be such that $-iB'C' \ket{a_i'} = \ket{a_i'}$. Otherwise, if $(\Gamma_{2i},\Gamma_{2i+1})$ is of the form $(\mp \widetilde{\Gamma}_{\sigma(i),c}, \pm \widetilde{\Gamma}_{\sigma(i),b})$, let $B,C \in \{X,Y,Z\}$ be the Pauli matrices satisfying $i B C \ket{a_i} = \ket{a_i}$, and let $B',C'$ be such that $iBC \ket{a_i'} =\ket{a_i'}$.
		
		Consider the permutation $\rho \in S_{\{X,Y,Z\}}$ of the Pauli labels with $\rho(D) = D'$ for $D' \in \{B,C\}$. Define $T'$ to be the the ternary tree that arises from applying $\rho$ to each of the labels of the immediately rightward edges of vertex $i$ in the ternary tree $T$. For each $j \in [n]$, define $\Gamma_j'$ to be the Pauli operators $\widetilde{\mathcal{G}}_{T'}$ that results from applying $\rho$ to the labels of the local Pauli operators acting on the $i$th qubit in $\Gamma_j$.  The vacuum state of the resulting $T'$--based mapping $\mathfrak{m}' = ((\Gamma_{2j}',\Gamma_{2j+1}'))_{j=0}^{n-1}$ is
		\begin{align}
			\ket{\vec{0}_{\mathfrak{m}'}} = \bigotimes_{j \neq i} \ket{a_j}_j \otimes \ket{a_i'}_i\, .
		\end{align}
		Repeatedly defining $\mathfrak{m} \leftarrow \mathfrak{m}'$, $T \leftarrow T'$ and applying this modification for all $i \in [n]$ produces a ternary tree $T'$ and a $T'$--based mapping $\mathfrak{m}'$ with vacuum state $\ket{\vec{0}_{\mathfrak{m}'}} = \bigotimes_{i=0}^{n-1} \ket{a_i'}_i$, as required. Figure \ref{fig:product_preserving} depicts an example of two modifications to $T$ and $\mathfrak{m}$ that could occur during this process.
	\end{proof}
\end{lemma}

A consequence of Lemma \ref{lem:equiv} is that, for any ternary tree $T$, each product--preserving $T$--based mapping has an equivalent mapping with any other product stabiliser vacuum state. In other words, we can choose the mapping with vacuum state $\ket{0}^{\otimes n}$ to be the sole representative of the template of product--preserving $T$--based mappings; the only difference between the product--preserving $T$--based mappings are the labels of the Pauli matrices on each qubit. One powerful conclusion that we can draw from this is that the existing suite of mappings in the literature \cite{Vlasov_2022, miller2023bonsai, miller2024treespilation} is sufficient to describe all product--preserving ternary tree transformations. What we cannot yet conclude is whether for an arbitrary ternary tree $T$ there is always a $T$--based mapping that is a linear encoding of the Fock basis.

Eschewing the intrinsic pairing structure of Lemma \ref{lem:ttpairing}, Lemma \ref{lem:compbasistt} proves the existence and uniqueness of a Clifford operator $C_T \in \mathcal{C}_n$ so that the mapping
\begin{align}
	\mathfrak{m}(T) = \left( \left(C_T \gamma_{2i} C_T^\dagger , C_T \gamma_{2i+1} C_T^\dagger \right) \right)_{i=0}^{n-1}
\end{align}
is a $T$--based mapping that linearly encodes the Fock basis. Equivalently, the Clifford operator $C_T$ effects the linear transformation
\begin{align}
	C_T : \ket{\vec{f}} \longmapsto \ket{G_T \vec{f}} \quad \text{for all} \quad \vec{f}\in \mathbb{F}_2^n
\end{align}
for some invertible binary matrix $G_T \in \text{GL}_n(\mathbb{F}_2)$.  Lemma \ref{lem:findmatrix} details how to determine $G_T$ and $C_T$ given the ternary tree $T$.

\begin{lemma}\label{lem:compbasistt} \emph{(Unique $T$--based mapping $\mathfrak{m}(T)$ to linearly encode the Fock basis.)} 
	Let $T$ be an $n$--vertex ternary tree. Then there is a unique  Clifford operator $C_T \in \mathcal{C}_n$ such that the mapping
    \begin{align}
    \mathfrak{m}(T) = ((C_T \gamma_{2i} C_T^\dagger, C_T \gamma_{2i+1} C_T^\dagger))_{i=0}^{n-1}
    \end{align}
    has the following properties:
	\begin{enumerate}[label=\arabic*)]
		\item  $\mathfrak{m}(T)$ is a $T$--based mapping, i.e.\ $\pm C_T \gamma_i C_T^\dagger \in  \widetilde{\mathcal{G}}_T$ for all $i \in [2n]$, and
		\item  $\mathfrak{m}(T)$ is a classical encoding of the Fock basis, i.e.\ $\ket*{\vec{f}_{\mathfrak{m}(T)}} \in \mathfrak{C}_n$ for all $\vec{f} \in \mathbb{F}_2^n$, with vacuum state $\ket*{\vec{0}_{\mathfrak{m}(T)}} = \ket{0}^{\otimes n}$.
	\end{enumerate}
	Theorem \ref{thm:affine} proved that every classical encoding is affine, and so the second property of $\mathfrak{m}(T)$ guarantees that it is a linear encoding of the Fock basis.

	\begin{figure}[bpt]
		\centering
		\includegraphics[width=0.85\linewidth]{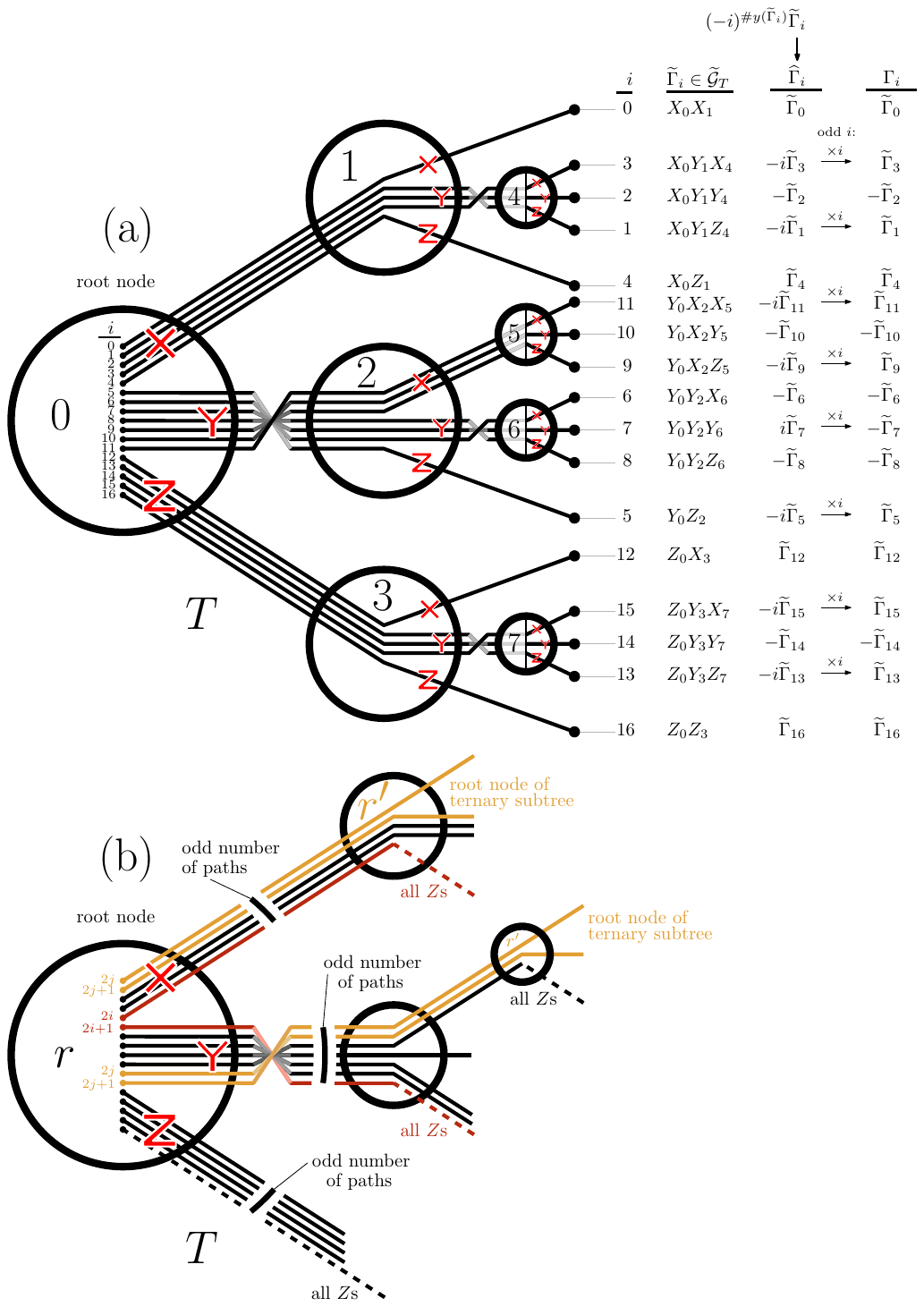}
		\caption{(a) Deriving the Pauli operators $\{\Gamma_i\}_{i=0}^{2n-1}$ from a ternary tree $T$ such that $\mathfrak{m}(T) = ((\Gamma_{2i}, \Gamma_{2i+1}))_{i=0}^{n-1}$ is a classical encoding of the Fock basis. (b) Visual guide to the proof that $\widehat{\Gamma}_{2i} \widehat{\Gamma}_{2i+1} \ket{0}^{\otimes n} = \ket{0}^{\otimes n}$.}
		\label{fig:pauli_enumeration}
	\end{figure}
	
	\begin{proof}
		Inspired by \cite{harrison_sierpinski_2024}, the following sequence of instructions details how to obtain the mapping $\mathfrak{m}(T)$ from the ternary tree $T$, with Figure \ref{fig:pauli_enumeration} (a) serving as a reference. Arrange $2n+1$ points vertically within the root vertex, labelled in ascending order from top to bottom. Draw lines representing the root-to-leaf paths of ${T}$ via the following scheme, which associates an index $i \in \{0,1,\dots, 2n\}$ to each path:
		\begin{itemize}
			\item Starting at the point labelled $i$ in the root, trace the line along the root-to-leaf path that ends in the $i$th highest unlabelled vertex on the right-hand-side of the tree.
			\item For each vertex, invert the subsequent top-to-bottom ordering of the lines that pass through a $Y$ branch of that vertex for the remainder of the tree.
		\end{itemize}
		Define the unsigned Pauli operator $\widetilde{\Gamma}_i$ to be the element of $\widetilde{\mathcal{G}}_T$ that corresponds to the $i$th root-to-leaf path according to this numbering system. Let $\#y:\mathcal{P}_n/K \rightarrow [n]$ denote the number of single--qubit $Y$ matrices appearing in an unsigned Pauli operator. Finally, define the Pauli operators
		\begin{align} \label{eqn:hats}
			\widehat{\Gamma}_i = (-i)^{\#y(\widetilde{\Gamma}_i)} \widetilde{\Gamma}_i
			\in \{\mathds{1}, X, -iY, Z \}^{\otimes n}\, .
		\end{align}
		The operator $\widehat{\Gamma}_i$ is anti--Hermitian if $\#y(\widetilde{\Gamma}_i)$ is odd. These operators cannot be representations of the Majorana operators in a $T$--based mapping; however, they do satisfy a useful relation:
		\begin{align}\label{eqn:k=1}
			\widehat{\Gamma}_i\ket{0}^{\otimes n} \in (-i)^{\#y(\widetilde{\Gamma}_i)} \left(\prod_{j \, : \, \widetilde{\Gamma}_i \vert _{\{j\}} = Y} i \right) \mathfrak{C}_n = \mathfrak{C}_n.
		\end{align}
		
		\begin{claim}
			The Pauli operators $\{\widehat{\Gamma}_i\}_{i=0}^{2n}$ satisfy 
			\begin{align}
				\widehat{\Gamma}_{2i} \widehat{\Gamma}_{2i+1} \ket{0}^{\otimes n} &= \ket{0}^{\otimes n}  \text{for all} \quad i=0,1,\dots,n-1\, , \label{eqn:0vac}
			\end{align}
			\begin{claimproof}
				Note that the bottom-most root-to-leaf path of a ternary tree takes only $Z$ branches.
				Suppose that the paths with labels $2i$ and $2i{+}1$ diverge on the root vertex. Because there are an odd number of paths departing the root via each of the $X$, $Y$ and $Z$ branches, the only way for the consecutive paths $2i$ and $2i{+}1$ to diverge at the root is for the $(2i)$th path to be the bottom-most path taking the $X$ branch, and for the $(2i{+}1)$th path to be the topmost to take the $Y$ branch, as Figure \ref{fig:pauli_enumeration} (b) illustrates. These paths become the bottom-most root-to-leaf paths of their respective ternary subtrees of $T$, and therefore subsequently to the root, the paths take only $Z$ branches. This means that:
				\begin{align}
					\widehat{\Gamma}_{2i} \widehat{\Gamma}_{2i+1} \ket{0}^{\otimes n} = (-iXY)_{r} \ket{0}_{r} \otimes \left( \bigotimes_{s \neq r} \ket{0}_s \right) = \ket{0}^{\otimes n}\, .
				\end{align}
				
				Consider another pair of paths, with labels $2j$ and $2j{+}1$, that do not diverge at the root of $T$. Figure \ref{fig:pauli_enumeration} (b) highlights two possible candidates for $j$. Apply the argument of the previous paragraph to the ternary subtree of $T$ whose root is the vertex $r'$ upon which paths $2j$ and $2j{+}1$ \textit{do} diverge. The vertical ordering of the two paths depends on the number of $Y$ branches in their shared path up to $r'$; let this number be $y$. Then, since the topmost of the two paths departs vertex $r'$ via an $X$ branch, and the bottommost via a $Y$ branch,
				\begin{align}
					\widehat{\Gamma}_{2j} \widehat{\Gamma}_{2j+1} \ket{0}^{\otimes n} = \left((-1)^y\right)^2 (-iXY)_{r'} \ket{0}_{r'} \otimes \left(\bigotimes_{s \neq r'} \ket{0}\right) = \ket{0}^{\otimes n}\, ,
				\end{align}
				where one factor of $(-1)^y$ is due to anticommuting $\widehat{\Gamma}_{2j}$ and $\widehat{\Gamma}_{2j+1}$ so that the action on qubit $r'$ is $-iXY$,  and the other factor of $(-1)^y$ is due to the shared $(-iY)$ operators in $\widehat{\Gamma}_{2j}$ and $\widehat{\Gamma}_{2j+1}$. This proves the claim.		
			\end{claimproof}
		\end{claim}
		
		\begin{claim} The Pauli operators $\{\widehat{\Gamma}_i\}_{i=0}^{2n}$ satisfy
			\begin{align}
				\widehat{\Gamma}_{i_1} \widehat{\Gamma}_{i_2} \dots \widehat{\Gamma}_{i_k} \ket{0}^{\otimes n}  & \in  \mathfrak{C}_n \qquad \text{for all} \quad 0 \leq i_1 < i_2 < \dots < i_k \leq 2n \quad \text{and} \quad  k = 1, 2, \dots, 2n\, . \label{eqn:pauliorder}
			\end{align}
			\begin{claimproof} 
				Observe that the repeated action of $(-iY)$ operators on a single-qubit state $\ket{0}$ is
				\begin{align}
					(-iY)^{l} \ket{0} = (-1)^{t_{l-1}} \ket{l \text{ mod } 2}\, , \quad (-iY)^{l}\ket{1} = (-1)^{t_{l}}\ket{ (l+1)\text{ mod } 2}\, ,
				\end{align}
				and so for general $l_1, l_2 \in \mathbb{N}$,
				\begin{align}
					(-iY)^{l_1} \ket{l_2 \text{ mod } 2} &= (-1)^{t_{(l_1+(l_2 \text{ mod } 2)-1)}} \ket{(l_1+l_2) \text{ mod } 2} \in (-1)^{t_{(l_1+(l_2 \text{ mod } 2)-1)}} \mathfrak{C}_1\, . \label{eqn:ygate}
				\end{align}
				
				As part of a proof by induction, assume the following is true for some $k$: \textit{For any $n$--vertex ternary tree $T$, the Pauli operators $\{\widehat{\Gamma}_{i}\}_{i=0}^{2n}$ in Equation \ref{eqn:hats}
					satisfy the following statement: let $k \in \{1,2, \dots, 2n\}$ and let the integer sequence $i_1,i_2,\dots,i_k$ be ascending, i.e.\ $0 \leq i_1 < i_2 < \dots < i_k \leq 2n$. Then, the action of the operators in order of descending labels $\widehat{\Gamma}_{i_k}, \widehat{\Gamma}_{i_{k-1}}, \dots, \widehat{\Gamma}_{i_2}, \widehat{\Gamma}_{i_1}$ upon the state $\ket{0}^{\otimes n}$ satisfies}
				\begin{align}
					\widehat{\Gamma}_{i_1} \widehat{\Gamma}_{i_2} \dots \widehat{\Gamma}_{i_k} \ket{0}^{\otimes n} \in \mathfrak{C}_n \label{eqn:inductivestep}\, .
				\end{align}
				By Equation \ref{eqn:k=1}, the statement is true for $k=1$.	Note that, as a consequence of Equation \ref{eqn:inductivestep}, reversing the order of the operators introduces a sign of $(-1)^{t_{k-1}}$, where $t_k$ is the $k$th triangular number and $t_1=1$:
				\begin{align}
					\widehat{\Gamma}_{i_k} \widehat{\Gamma}_{i_{k-1}} \dots \widehat{\Gamma}_{i_2} \widehat{\Gamma}_{i_1} \ket{0}^{\otimes n} = 
					(-1)^{t_{k-1}}
					\widehat{\Gamma}_{i_1} \widehat{\Gamma}_{i_2} \dots \widehat{\Gamma}_{i_k} \ket{0}^{\otimes n} \in (-1)^{t_{k-1}} \mathfrak{C}_n\, , \label{eqn:inductiveflip}
				\end{align}
				
				Now, suppose the inductive statement is true for some $k \in \{1,2,\dots,2n\}$ and for some sequence of integers $i_1, i_2, \dots, i_k$ with $0<i_1<i_2 < \dots <i_k \leq 2n$. Then, there exists an integer $0 \leq i_0<i_1$; now consider the quantity
				\begin{align}
					\widehat{\Gamma}_{i_0} \widehat{\Gamma}_{i_1} \widehat{\Gamma}_{i_2} \dots \widehat{\Gamma}_{i_k} \ket{0}^{\otimes n}  \, . \label{eqn:mysterystate}
				\end{align}
				
				In order to determine whether the state in Equation \ref{eqn:mysterystate} is in the computational basis, introduce the following notation. Let $\mathcal{S} \subseteq \{0,1,\dots,n-1\}$ be a subset of qubit labels. Recall that each $\widehat{\Gamma}_i = ( -i) ^ {\#y(\widetilde{\Gamma}_i)} \widetilde{\Gamma}_i$ for some $\widetilde{\Gamma}_i \in \widetilde{\mathcal{G}}_T$. Define $\mathcal{S}_X(\widehat{\Gamma})$, $\mathcal{S}_Y(\widehat{\Gamma})$ and $\mathcal{S}_Z(\widehat{\Gamma})$ to be the subsets of $\mathcal{S}$ containing the labels of qubits upon which $\widetilde{\Gamma}$ acts with $X$, $Y$ and $Z$ matrices, respectively. This allows a well--defined notion of the restriction $\widehat{\Gamma} |_\mathcal{S}$ of  $\widehat{\Gamma}$ to the subset of qubits with labels in $\mathcal{S}$; specifically,
				\begin{align}
					\widehat{\Gamma} \bigg\vert_{\mathcal{S}}  \coloneqq \left(\bigotimes_{s \in \mathcal{S}_X(\widehat{\Gamma})} X_s \right)
					\left(\bigotimes_{s \in \mathcal{S}_Y(\widehat{\Gamma})} (-iY)_s \right)
					\left(\bigotimes_{s \in \mathcal{S}_Z(\widehat{\Gamma})} Z_s \right)\, . \label{eqn:restriction}
				\end{align}
				Using the notation of Equation \ref{eqn:restriction}, define $\ket{0}_{\mathcal{S}} = \bigotimes_{s \in \mathcal{S}} \ket{0}_s$ and define the product states $\ket{\mathcal{S}}$ via
				\begin{align}
					\ket{\mathcal{S}} \coloneqq \left(\widehat{\Gamma}_{i_0}\widehat{\Gamma}_{i_1}\widehat{\Gamma}_{i_2} \dots \widehat{\Gamma}_{i_k} \right)\bigg\vert_{\mathcal{S}} \ket{0}_{\mathcal{S}}\, .
				\end{align}
				Then, for any partition $\Pi$ of the qubit labels $\{0,1,\dots,n -1\}$,
				the state in Equation \ref{eqn:mysterystate} is equal to
				\begin{align}
					\widehat{\Gamma}_{i_0} \widehat{\Gamma}_{i_1} \widehat{\Gamma}_{i_2} \dots \widehat{\Gamma}_{i_k} \ket{0}^{\otimes n}  = \bigotimes_{\mathcal{S} \in \Pi} \ket{\mathcal{S}}\, .
				\end{align}

				\begin{figure}[btp]
					\centering
					\includegraphics[width=\linewidth]{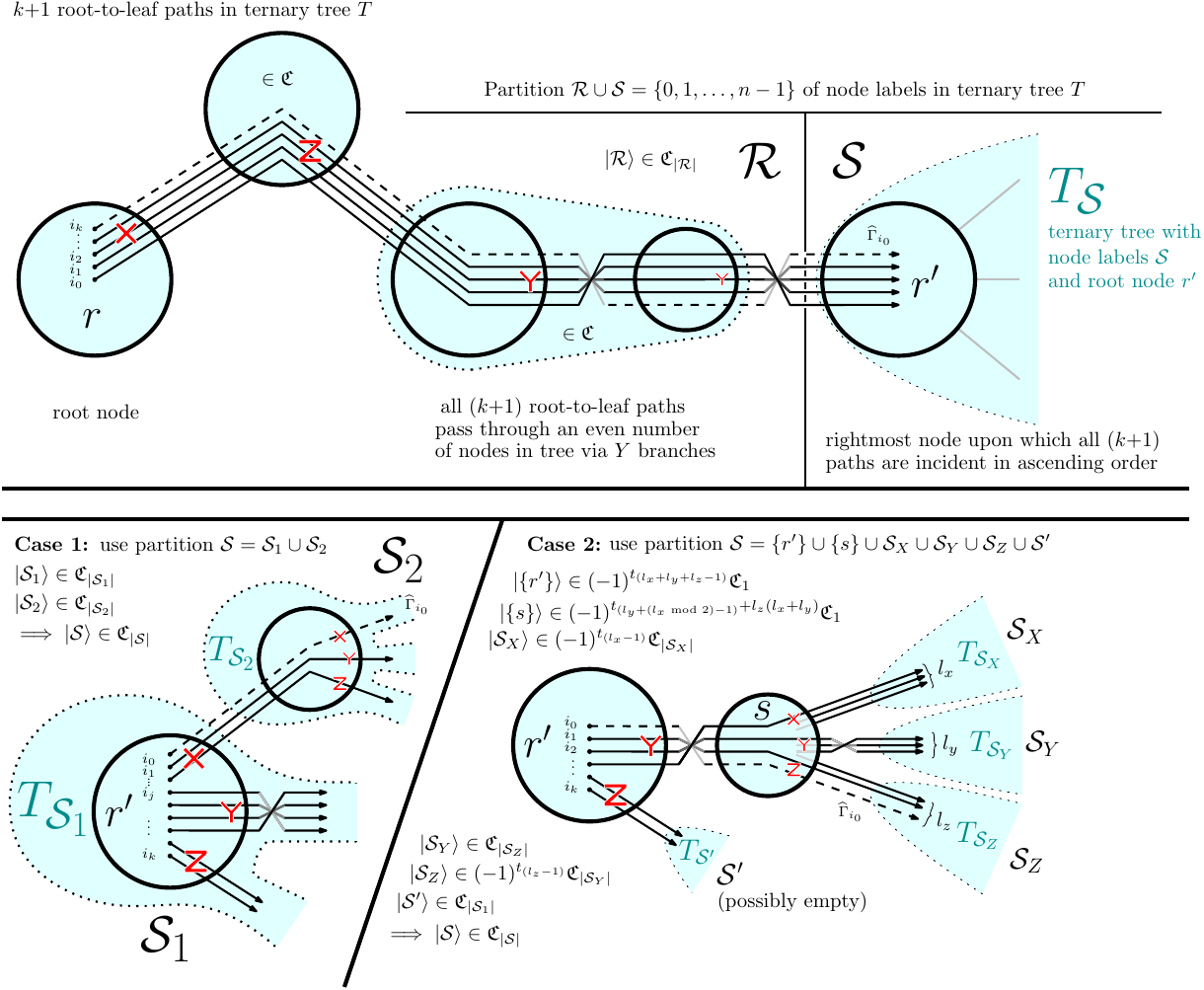}
					\caption{Part of the proof of Lemma \ref{lem:compbasistt}. The inductive statement is that for any $n$--vertex ternary tree $T$, the state $\widehat{\Gamma}_{i_1} \widehat{\Gamma}_{i_2} \dots \widehat{\Gamma}_{i_k} \ket{0}^{\otimes n}$ resides in the computational basis, where the Pauli operators $\widehat{\Gamma}_i \in \{ \mathds{1}, X, -iY, Z\}^{\otimes n}$ act with labels in descending order $0 \leq i_1 < i_2 < \dots < i_k \leq 2n$. This statement thus applies to any sub-tree of $T$ on the vertex subset $\mathcal{S} \subseteq \{0,1,\dots,n-1\}$ and any product of at most $k$ operators, with labels in descending order, acting on the state $\ket{0}_\mathcal{S}$.}
					\label{fig:pauli-cases}
				\end{figure}

				Let the root of ${T}$ have label $r$, and let $r'$ be the label of the rightmost vertex through which all $k{+}1$ of the root-to-leaf paths pass such that their labels are in ascending order from top-to-bottom. This vertex is the last upon which all paths are `in order'. Either $r=r'$, as is the case in Figure \ref{fig:pauli_enumeration}, or all $k{+}1$ paths pass identically through the vertices of a segment of the tree, as in Figure \ref{fig:pauli-cases}. Let $\mathcal{R}$ be the possibly empty subset of vertex labels in this tree segment, up to and excluding $r'$; let $\mathcal{S}$ be the remaining subset of vertices including and to the right of $r'$, so that $\mathcal{R} \cup \mathcal{S} = \{0,1,\dots, n-1\}$ partitions the vertex labels. The state is thus
				\begin{align}\label{eqn:rs}
					\widehat{\Gamma}_{i_0} \widehat{\Gamma}_{i_1} \dots \widehat{\Gamma}_{i_k} \ket{0}^{\otimes n} = \ket{\mathcal{R}} \otimes \ket{\mathcal{S}}\, .
				\end{align}
				
				Consider the product state $\ket{\mathcal{R}}$. Because $X^{k+1} \ket{0} = \ket{(k+1) \text{ mod } 2} \in \mathfrak{C}_1$ and $Z^{k+1} \ket{0} = \ket{0} \in \mathfrak{C}_1$, any vertex $r_a \in \mathcal{R}$ upon which the paths take an $X$ or $Z$ branch correspond to a qubit that remains in the computational basis, i.e.\ $(\widehat{\Gamma}_{i_0} \widehat{\Gamma}_{i_1} \dots \widehat{\Gamma}_{i_k}) \vert_{\{r_a\}} \ket{0}_{\{r_a\}} \in \mathfrak{C}_1$.
				The $k{+}1$ paths appear in their original ascending order from top-to-bottom on vertex $r'$, and so there must be an even number of vertices in $\mathcal{R}$ upon which the paths take a $Y$ branch, since each such occurrence inverts the order of the paths. For any pair of vertices $\{r_a, r_b\} \in \mathcal{R}$ upon which all $k{+}1$ paths take a $Y$ branch, 
				\begin{align}
					(-iY)^{k+1}_{r_a} \otimes (-iY)^{k+1}_{r_b} \ket{0}_{\{r_a,r_b\}} =\left((-1)^{t_{k}}\right)^2 \ket{(k+1) \text{ mod 2}}_{\{r_a,r_b\}} = \ket{(k+1) \text{ mod } 2}_{\{r_a, r_b\}} \in \mathfrak{C}_2\, , 
				\end{align}
				and so $\widehat{\Gamma}_{i_0} \widehat{\Gamma}_{i_1} \dots \widehat{\Gamma}_{i_k} \vert_{\{r_a,r_b\}} \ket{0}_{\{r_a,r_b\}} \in \mathfrak{C}_2$. Therefore $\ket{\mathcal{R}} \in \mathfrak{C}_{|\mathcal{R}|}$.
				
				Now consider the product state $\ket{\mathcal{S}}$. There are two scenarios describing how the paths might continue on from vertex $r'$, as Figure \ref{fig:pauli-cases} highlights. In both cases, we demonstrate that $\ket{\mathcal{S}} \in \mathfrak{C}_{|\mathcal{S}|}$. 
				\begin{itemize}
					\item \textbf{Case 1:} $\big(\widehat{\Gamma}_{i_0}\vert_{\{r'\}}=X\big)$: there are root-to-leaf paths extending along the $X$ branch from  vertex $r'$.
					\item \textbf{Case 2:} $\big(\widehat{\Gamma}_{i_0}\vert_{\{r'\}}=-iY\big)$: there are no root-to-leaf paths extending along the $X$ branch from vertex $r'$.
				\end{itemize}
				
				In \textbf{Case 1}, partition $\mathcal{S}$ into the set $\mathcal{S}_1$ of all vertices in the root-to-leaf paths leaving vertex $r'$ via the $Y$ or $Z$ branches, and the set $\mathcal{S}_2$ consisting of the vertices in the root-to-leaf paths leaving vertex $r'$ via the $X$ branch. This partitions the ternary tree with root $r'$ into two ternary subtrees of $T$: the ternary tree $T_{\mathcal{S}_1}$ with root $r'$, which contains vertices with labels in $\mathcal{S}_1$, and the ternary tree $T_{\mathcal{S}_2}$ containing vertices with labels in $\mathcal{S}_2$.
				
				We wish to investigate $\ket{\mathcal{S}} = \ket{\mathcal{S}_1} \otimes \ket{\mathcal{S}_2}$. The state $\ket{\mathcal{S}_2}$ is in the computational basis by the inductive statement Equation \ref{eqn:inductivestep} applied to the $|\mathcal{S}_2|$--vertex ternary tree $T_{\mathcal{S}_2}$, because $\ket{\mathcal{S}_2}$ is the result of at most $k$ strings with descending labels in $\{\mathds{1}, X, -iY, Z \}^{\otimes |\mathcal{S}_2|}$ acting on the state $\ket{0}_{\mathcal{S}_2}$. To show that the state $\ket{\mathcal{S}_1}$ is also in the computational basis, suppose that the first $j\in \{1, 2, \dots, k-1\}$ operators $\widehat{\Gamma}_{i_0}, \widehat{\Gamma}_{i_1}, \widehat{\Gamma}_{i_2}, \dots, \widehat{\Gamma}_{i_{j-1}}$ act with $X$ upon qubit $r'$, with the remaining operators acting on qubit $r'$ with either $(-iY)$ or $Z$. Then,
				\begin{align}
					\ket{\mathcal{S}_1} &=
					\left(
					\widehat{\Gamma}_{i_0} \widehat{\Gamma}_{i_1} \widehat{\Gamma}_{i_2} \dots \widehat{\Gamma}_{i_k} \right) \bigg\vert_{\mathcal{S}_1} \ket{0}_{\mathcal{S}_1}
					= 
					\left( X_r\right)^j 
					\underbrace{\left( \widehat{\Gamma}_{i_j} \widehat{\Gamma}_{i_{j+1}} \dots \widehat{\Gamma}_{i_k}\right) \bigg\vert_{\mathcal{S}_1}
						\ket{0}_{\mathcal{S}_1}}_{\mathclap{\in \mathfrak{C}_{|\mathcal{S}_1|} \text{ by inductive statement on $T_{\mathcal{S}_1}$}}}
					\in
					\mathfrak{C}_{|\mathcal{S}_1|}\, ,
				\end{align}
				applying the inductive statement to the $|\mathcal{S}_1|$--vertex ternary tree $T_{\mathcal{S}_1}$ as there are at most $k$ operators in descending order acting on the state $\ket{0}_{\mathcal{S}_1}$, and using the fact that the $(X_r)^j$ operators fix the computational basis. Therefore $\ket{\mathcal{S}} = \ket{\mathcal{S}_1} \otimes \ket{\mathcal{S}_2} \in \mathfrak{C}_{|\mathcal{S}|}$.

				In \textbf{Case 2}, partition $\mathcal{S}$ into six sets: the sets $\{r'\}$ and $\{s\}$, where $s$ is the label of the vertex immediately to the right of $r'$ along the $Y$ branch, the possibly empty sets $\mathcal{S}_X$, $\mathcal{S}_Y$ and $\mathcal{S}_Z$ containing the labels of the vertices in the root-to-leaf paths leaving $s$ via its $X$, $Y$ and $Z$ branches, and the possibly empty set $\mathcal{S}'$ containing the labels of the vertices in the root-to-leaf paths leaving $r'$ via its $Z$ branch. Note that this partition induces ternary subtrees $T_{\mathcal{S}_X}$, $T_{\mathcal{S}_Y}$, $T_{\mathcal{S}_Z}$ and $T_{\mathcal{S}'}$ of the ternary tree, with vertex sets $\mathcal{S}_X$, $\mathcal{S}_Y$, $\mathcal{S}_Z$ and $\mathcal{S}'$, respectively.
				
				Consider
				\begin{align} \label{eqn:kets}
					\ket{\mathcal{S}} = \ket{\{r'\}} \otimes \ket{\mathcal{S}'} \otimes \ket{\{s\}} \otimes \ket{\mathcal{S}_X} \otimes \ket{\mathcal{S}_Y} \otimes \ket{\mathcal{S}_Z}\, .
				\end{align}
				Applying the inductive statement to $T_{\mathcal{S}'}$, Equation \ref{eqn:inductivestep} implies that $\ket{\mathcal{S}'} \in \mathfrak{C}_{|\mathcal{S}'|}$. Suppose that $l_x$, $l_y$ and $l_z$ of the root-to-leaf paths pass through the $X$, $Y$ and $Z$ edges of the vertex with label $s$, respectively. Applying Equation \ref{eqn:inductiveflip} in the inductive statement to ternary trees $T_{\mathcal{S}_X}$, $T_{\mathcal{S}_Y}$ and $T_{\mathcal{S}_Z}$ then implies that $\ket{\mathcal{S}_X} \in (-1)^{t_{(l_x-1)}}\mathfrak{C}$, $\ket{\mathcal{S}_Y} \in \mathfrak{C}$ and $\ket{\mathcal{S}_Z} \in (-1)^{t_{(l_z-1)}}\mathfrak{C}$. 
				
				Equation \ref{eqn:ygate} establishes that $\ket{\{r'\}} \in (-1)^{t_{(l_x+l_y+l_z-1)}}\mathfrak{C}_1$ and that the state of the qubit with label $s$ is
				\begin{align}
					\ket{\{s\}} &=
					\underbrace{\widetilde{\Gamma}_{i_0} \widehat{\Gamma}_{i_1} \dots \widehat{\Gamma}_{i_{(l_z-1)}}}_{l_z \text{ operators}}
					\underbrace{\widehat{\Gamma}_{i_{l_z}} \dots \widehat{\Gamma}_{i_{(l_z+l_y-1)}}}_{l_y \text{ operators}}
					\underbrace{\widehat{\Gamma}_{i_{(l_z+l_y)}} \dots \widehat{\Gamma}_{i_{(l_x+l_y+l_z-1)}}}_{l_x \text{ operators}}
					\bigg\vert_{\{s\}}
					\ket{0}_s \\
					&= Z^{l_z} (-iY)^{l_y} X^{l_x} \ket{0}_s \\
					&= Z^{l_z} (-iY)^{l_y} \ket{ l_x \text{ mod }2} \\
					&= Z^{l_z} (-1)^{t_{(l_x + (l_x \text{ mod } 2) - 1)}} \ket{(l_x + l_y) \text{ mod } 2} \\
					&= (-1)^{t_{(l_y + (l_x \text{ mod } 2 )-1)} + l_z(l_x+l_y)} \ket{(l_x + l_y) \text{ mod } 2}\\
					&\in (-1)^{t_{(l_y + (l_x \text{ mod } 2 )-1)} + l_z(l_x+l_y)} \mathfrak{C}_1\, .
				\end{align}
				Substituting the factors of $\ket{\mathcal{S}}$ that are not necessarily in the computational basis into Equation \ref{eqn:kets},
				\begin{align}
					\ket{\mathcal{S}} 
					\in \underbrace{(-1)^{t_{(l_x + l_y + l_z - 1)}}}_{\text{from }\ket{\{r\}}} \underbrace{(-1)^{t_{(l_y + (l_x \text{ mod } 2 )-1)}}+ l_z(l_x+l_y)}_{\text{from }\ket{\{s\}}} \underbrace{(-1)^{t_{(l_x-1)}}}_{\text{from }\ket{\mathcal{S}_X}} \underbrace{(-1)^{t_{(l_z-1)}}}_{\text{from }\ket{\mathcal{S}_Z}} \mathfrak{C}_{|\mathcal{S}|} = (-1)^{A}\mathfrak{C}_{|\mathcal{S}|}\, ,
				\end{align}
				where, using the property $t_{l_1+l_2} = t_{l_1} + t_{l_2} + l_1 l_2$ of the triangular numbers and working modulo 2,
				\begin{align}
					A &=  t_{(l_x+l_y+l_z-1)} + t_{(l_y + (l_x \text{ mod } 2) - 1)} + l_z(l_x + l_y) + t_{(l_x-1)} + t_{(l_z-1)} &&\mod 2 \\
					&= \underbrace{t_{(l_x+l_y+l_z)} + (l_x+l_y+l_z)}_{t_{l-1}= t_l - l} +
					\underbrace{t_{l_y + (l_x \text{ mod }2)} + (l_y + l_x)}_{t_{l-1}= t_l - l} + 
					l_z(l_x+l_y) +
					\underbrace{t_{l_x} + l_x}_{\mathclap{t_{l-1} = t_l - l}} + \underbrace{t_{l_z} + l_z}_{\mathclap{t_{l-1} = t_l - l}} 
					&&\mod 2 \\
					&= \underbrace{t_{l_y} +t_{(l_x \text{ mod } 2)} + l_y(l _x\text{ mod 2})}_{t_{(l_1+l_2)} = t_{l_1} + t_{l_2} + l_1l_2} + t_{(l_x + l_y + l_z)} + t_{l_x} + t_{l_z} + l_z(l_x+l_y) + l_x &&\mod 2\\
					&= t_{(l_x+l_y+l_z)} + \underbrace{t_{l_x} + t_{l_y} + t_{l_z} + (l_xl_y + l_y l_z + l_z l_x)}_{=t_{(l_x + l_y + l_z)}} + \underbrace{t_{(l_x \text{ mod } 2)}}_{=l_x \text{ mod } 2} + l_x &&\mod 2 \\
					&= 0\, ,
				\end{align}
				and so $\ket{\mathcal{S}} \in \mathfrak{C}_{|\mathcal{S}|}$. Using Equation \ref{eqn:rs} the state is equal to $\ket{\mathcal{R}} \otimes \ket{\mathcal{S}}$ in the computational basis.
				
				Since Equation \ref{eqn:0vac} shows that the inductive statement is true for $k=1$, this proves the inductive step: for any sequence of $(k+1)$ integers $0 \leq i_0 \leq  i_1 <  \dots < i_k \leq 2n$, 
				\begin{align}
					\widehat{\Gamma}_{i_0} \widehat{\Gamma}_{i_1} \dots \widehat{\Gamma}_{i_k} \ket{0}^{\otimes n} \in \mathfrak{C}_n\, .
				\end{align}
			\end{claimproof}
		\end{claim}
		
		Define the operators $\Gamma_{2i} = \widehat{\Gamma}_{2i}$ and $\Gamma_{2i+1} = -i \widehat{\Gamma}_{2i+1}$. 	For all $i \in [n]$, the operator $\widehat{\Gamma}_{2i}$ acts on an even number of qubits with $(-iY)$, and the operator $\widehat{\Gamma}_{2i+1}$ acts on an odd number of qubits with $(-iY)$; thus each of the $\Gamma_i$ is Hermitian. Define the fermion--qubit mapping $\mathfrak{m}({T}) = ((\Gamma_{2i},\Gamma_{2i+1}))_{i=0}^{n-1}$. Via Equation \ref{eqn:0vac}, the vacuum state of the mapping is $\ket*{\vec{0}_{\mathfrak{m}({T})}} = \ket{0}^{\otimes n}$, and the occupation number basis states are
		\begin{align}
			\ket*{\vec{f}_{\mathfrak{m}(T)}} &= \Gamma_0 \Gamma_2 \dots \Gamma_{2n-2} \ket{0}^{\otimes n} = \widehat{\Gamma}_0 \widehat{\Gamma}_2 \dots \widehat{\Gamma}_{2n-2} \ket{0}^{\otimes n} \in \mathfrak{C}_n \quad \text{for all} \quad \vec{f} \in \mathbb{F}_2^n\, .
		\end{align}
		Therefore $\mathfrak{m}(T)$ is a $T$--based mapping that classically encodes the Fock basis. Because the representations of the Majorana operators are Pauli strings $\Gamma_i \in \mathcal{P}_n$, the mapping $\mathfrak{m}(T)$ must be an affine encoding of the Fock basis, via Theorem \ref{thm:affine1}. Since the vacuum state $\ket*{\vec{0}_{\mathfrak{m}(T)}} = \ket{0}^{\otimes n}$, the mapping must in fact be a linear encoding of the Fock basis, and therefore the Clifford operator $C_T \coloneqq C_{\mathfrak{m}(T)}$ with $C_T : \ket{\vec{f}} \mapsto \ket*{\vec{f}_{\mathfrak{m}(T)}}$ satisfies $C_T \gamma_i C_T^\dagger = \Gamma_i$ for all $i \in [2n]$, as required.
	\end{proof}
\end{lemma}

Since $\mathfrak{m}(T)$ is a linear encoding of the Fock basis, there is an invertible binary matrix $G_T \in \text{GL}_n(\mathbb{F}_2)$ with $\ket*{\vec{f}_{\mathfrak{m}(T)}} = \ket*{G_T \vec{f}}$ for all $\vec{f}\in \mathbb{F}_2^n$; Lemma \ref{lem:findmatrix} demonstrates how to find this matrix.

\begin{lemma} \textit{(Determining $G_T$ and $C_T$ from $\mathfrak{m}(T)$.)} \label{lem:findmatrix}
	Given an $n$--vertex ternary tree $T$, let $G_T \in \text{GL}_n(\mathbb{F}_2)$ be the invertible binary matrix satisfying  $\ket*{\vec{f}_{\mathfrak{m}(T)}} = \ket*{G_T \vec{f}}$ for all $\vec{f} \in \mathbb{F}_2^n$. Then,
	\begin{align}
		(G_T)_{ij} = \begin{cases}
			1 & \text{if $\Gamma_{2j}$ acts on qubit $i$ with $X_i$ or $Y_i$}\\
			0 & \text{if $\Gamma_{2j}$ acts on qubit $i$ with $\mathds{1}_i$ or $Z_i$}\, .
		\end{cases} \label{eqn:gdef}
	\end{align}
	The Clifford operator $C_T \in \mathcal{C}_n$ has the following stabiliser tableau\footnote{see Definition \ref{defn:symplectic}}, which identifies $C_T$ up to a global phase:
	\begin{align}
		[C_G] = \left[\begin{array}{ccc|ccc|c}
			&   &   &  0  & \dots     & 0  & 0\\ 
			&G_T & & \vdots & \ddots & \vdots& \vdots \\ 
			&    &    &  0 & \dots     & 0 & 0 \\
			\hline
			0&\dots &0 & & & &0\\ 
			\vdots& \ddots & \vdots  &\multicolumn{3}{c|}{(G_T^{-1})^\top}  & \vdots\\ 
			0& \dots  &0 & &  & & 0
		\end{array}\right]\, . \label{eqn:ctstab}
	\end{align}
    Taking the phaseless representative of this tableau, which implements $\ket{\vec{f}} \mapsto \ket{G_T \vec{f}}$, identifies $C_T$ exactly.
	\begin{proof}
		As a linear transformation on $\mathbb{F}_2^n$, the $j$th column of $G_T$ is the image of $\vec{1}_j$ under $G_T$. Note that
		\begin{align}
			\ket{G_T\vec{1}_j} = \ket{(\vec{1}_j)_{\mathfrak{m}(T)}} = \Gamma_{2j} \ket{0}^{\otimes n} = \left(\bigotimes_{\widehat{\Gamma}_{2j} \vert_{\{i\}} = X_i \text{ or } Y_i} \ket{1}_i \right) \otimes \left( \bigotimes_{\widehat{\Gamma}_{2j} \vert_{\{i\}} = \mathds{1}_i \text{ or } Z_i} \ket{0}_i \right)\, ,
		\end{align}
		which corroborates Equation \ref{eqn:gdef}. The result of Equation \ref{eqn:ctstab} appears in the literature \cite{dehaene_clifford_2003}. In Appendix \ref{app:affine} we present Theorem \ref{thm:affine}, which is a self-contained proof of the statement in Equation \ref{eqn:ctstab}. 
	\end{proof}
\end{lemma}

\begin{figure}[tp]
	\centering
	\includegraphics[width=\linewidth]{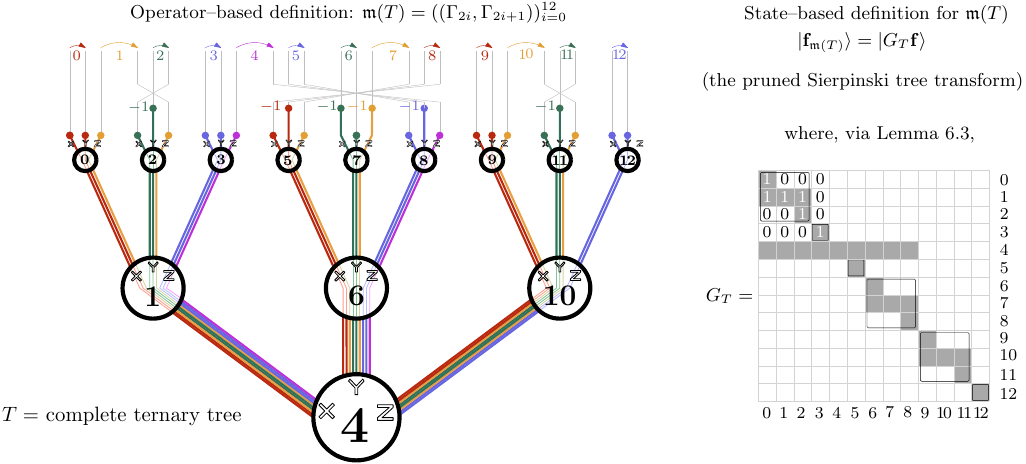}
	\caption{The mapping $\mathfrak{m}(T)$, where $T$ is a complete 13--vertex ternary tree, is equal to the pruned Sierpinski tree transform on 13 qubits. The diagram for the mapping reveals its operator--based definition in terms of the Pauli operators $\{\Gamma_i\}_{i=0}^{12}$. The state--based definition reveals that $\mathfrak{m}(T)$ is a linear encoding of the Fock basis of the form $\ket*{\vec{f}_{\mathfrak{m}(T)}} = \ket{G_T \vec{f}}$ for the invertible binary matrix $G_T \in \text{GL}_{13}(\mathbb{F}_2)$ in the bottom-right of the figure, which is the pruned Sierpinski tree transform.}
	\label{fig:fulltt}
\end{figure}

\subsection{Main theorem: equivalence of ternary tree transformations to linear encodings}\label{sec:mainthm}

Theorem \ref{thm:cbptbmappings} summarises all the results of Section \ref{sec:ppttbm} into one statement.

\begin{theorem}\label{thm:cbptbmappings}\textit{(Every product--preserving ternary tree transformation is equivalent to a linear encoding of the Fock basis.)}
	\begin{enumerate}[label=\alph*)]
	\item \textit{Existence and uniqueness of $\mathfrak{m}(T)$:}
	For every $n$--vertex ternary tree $T$, there is a unique ancilla--free fermion--qubit mapping that both is a $T$--based mapping and linearly encodes the Fock basis.
	
	\item \textit{Completeness of $\mathfrak{m}(T)$:}
	Every $n$--mode product--preserving ternary tree transformation is equivalent to a mapping in the set $\{\mathfrak{m}(T) \mid  T\text{ is an $n$--vertex ternary tree}\}$.
	
\end{enumerate}
	\begin{proof}
		Let $T$ be an $n$--vertex ternary tree.
		\begin{enumerate}[label=\alph*)]
			\item Lemma \ref{lem:compbasistt} guarantees the existence of $\mathfrak{m}(T)$. Because $\mathfrak{m}(T)$ has vacuum state $\ket*{\vec{0}_{\mathfrak{m}(T)}} = \ket{0^{\otimes n}}$, all other product--preserving $T$--based mappings with vacuum state $\ket{0}^{\otimes n}$, and hence all $T$--based mappings that are linear encodings of the Fock basis, are equivalent to $\mathfrak{m}(T)$ up to Pauli braids, sign change of pairs, and fermionic labelling, via Lemma \ref{lem:ttpairing}. However, performing any of these operations would disturb the Fock basis:
			\begin{alignat}{4}
				\text{(Pauli braid $\Gamma_{2j} \mapsto \pm \Gamma_{2j+1}$)} : &\ket{(\vec{1}_j)_{\mathfrak{m}(T)}} &&= \Gamma_{2j} \ket{0}^{\otimes n} \longmapsto & \ket*{(\vec{1}_j)_{\mathfrak{m}'}} =  &&\pm\Gamma_{2j+1} \ket{0}^{\otimes n} &\in (\pm i) \mathfrak{C}_n \\
				\text{(sign change $\Gamma_{2j} \mapsto -\Gamma_{2j}$)}: &\ket{(\vec{1}_j)_{\mathfrak{m}(T)}} &&= \Gamma_{2j} \ket{0}^{\otimes n} \longmapsto & \ket*{(\vec{1}_j)_{\mathfrak{m}'}} = && {}_{\phantom{+1}}-\Gamma_{2j} \ket{0}^{\otimes n} &\in -\mathfrak{C}_n\, .
			\end{alignat}
			For any permutation $\sigma \in S_n$ of the fermionic mode labels, there exist $i,j \in [n]$ with $i<j$ such that $\sigma(i) > \sigma(j)$. In this situation:
			\begin{align}
				\text{(fermionic labelling)}:  \ket*{(\vec{1}_i+\vec{1}_j)_{\mathfrak{m}(T)}} =  \Gamma_{2i} \Gamma_{2j} \ket{0}^{\otimes n} \mapsto \ket*{\vec{f}_{\mathfrak{m}'}} =  \Gamma_{2\sigma(i)} \Gamma_{2\sigma(i)} \ket{0}^{\otimes n} \in -\mathfrak{C}_n\, .
			\end{align}
			Therefore $\mathfrak{m}(T)$ is the unique $T$--based mapping to linearly encode the Fock basis.
			\item The only product--preserving $T$--based mappings that are not equivalent to $\mathfrak{m}(T)$ via Pauli braids, sign changes of pairs, and fermionic mode relabelling are those that feature the operator $\widetilde{\Gamma}_{2n} \coloneqq \prod_{i=0}^{2n-1} \widetilde{\Gamma}_i$ as one of the Majorana representations. As a tensor product of only the  single--qubit matrices $\{\mathds{1}, Z\}$, this operator arises from the bottom-most root-to-leaf path in $T$. By Lemma \ref{lem:ttpairing}, a $T$--based mapping with $\widetilde{\Gamma}_{2n}$ as one of the Majorana representations will have a vacuum state that is not equal to $\ket{0}^{\otimes n}$. By Lemma \ref{lem:equiv}, any such mapping is equivalent to a $T'$--based mapping with vacuum state $\ket{0}^{\otimes n}$ for some $n$--vertex ternary tree $T'$ which differs from $T$ only by local Pauli relabelling of the edges of $T$. Therefore, any such mapping is equivalent to $\mathfrak{m}(T')$; hence the set of mappings $\{\mathfrak{m}(T) \mid T \text{ is an $n$--vertex ternary tree}\}$ describes all product--preserving $n$--qubit ternary--tree--based mappings. \qedhere
	\end{enumerate}\end{proof}
\end{theorem}

\begin{figure}[btp]
	\centering
	\includegraphics[width=0.7\linewidth]{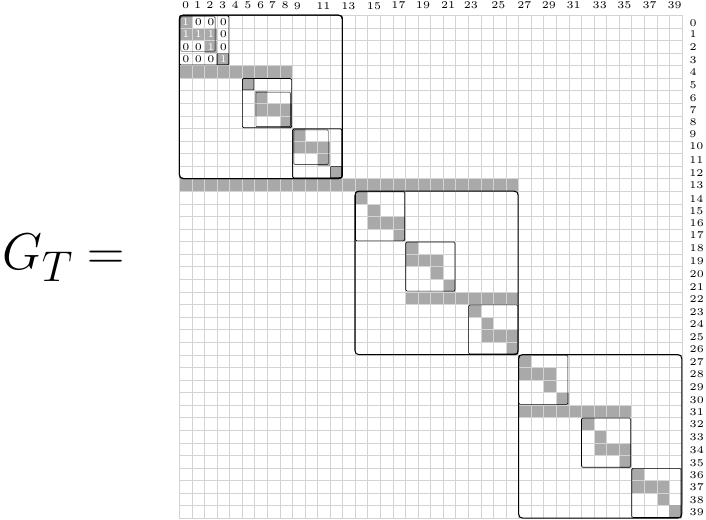}
	\caption{The invertible binary matrix $G_T$ for the 40--vertex complete ternary tree $T$, which describes the linear encoding $\mathfrak{m}(T)$ of the Fock basis $\ket*{\vec{f}_{\mathfrak{m}(T)}} = \ket{G_T \vec{f}}$ for all $\vec{f} \in \mathbb{F}_2^{40}$. Complete ternary trees have $1,4,13,40,..., 3k+1$ vertices for $k \in \mathbb{N}$. The outlined squares are a visual guide to the recursive definition of $G_T$ as the number of vertices in the complete ternary tree $T$ grows.}
	\label{fig:ttmat}
\end{figure}

\begin{exmp}\textit{(A ternary tree transformation that linearly encodes the Fock basis.)}\label{exmp:type3}
	Figure \ref{fig:bonsai_pairing_all} (c) demonstrates the mapping $\mathfrak{m}(T)$ that arises from the 5--vertex ternary tree from Example \ref{exmp:treec}. Note that all Fock basis vectors are strictly in the computational basis, and that the matrix $G_T$ is the one that arises from Lemma \ref{lem:findmatrix}.
\end{exmp}

\subsection{The pruned Sierpinski Tree as a ternary tree transformation}\label{sec:sierpinski}

An initial value proposition of ternary tree transformations was that the complete ternary tree graph yielded Pauli operators with the minimum possible average Pauli weight of ${\sim}\lceil \log_3(2n+1)\rceil$ qubits. Examples \ref{exmp:tree1} and \ref{exmp:tree2} demonstrated a caveat: that ternary tree transformations are not necessarily linear encodings of the Fock basis. However, Theorem \ref{thm:cbptbmappings} has now allayed this concern by showing that every ternary tree transformation is \textit{equivalent} to a linear encoding, in the spirit of Definition \ref{defn:templates}. In particular, this means that there is some invertible binary matrix that implements mappings where every Pauli operator has weight ${\sim}\lceil \log_3(2n+1)\rceil$.

In \cite{harrison2024mapping}, the authors find a linear encoding of the Fock basis via the pruned Sierpinski tree data structure \cite{harrison2024sierpinskitriangledatastructure} which produces Pauli operators with the minimum average Pauli weight ${\sim}\lceil \log_3(2n+1)\rceil$, matching the promise of ternary tree transformations. Using Theorem \ref{thm:cbptbmappings} shows that these two classes of mappings are equivalent: for the complete ternary tree $T$, we find the invertible binary matrix $G_{T}$ that implements the $T$--based linear encoding $\mathfrak{m}(T)$, and note that it exactly matches the definition of the pruned Sierpinski tree transform. Figure \ref{fig:fulltt} displays the equivalence for $n=13$, and Figure \ref{fig:ttmat} shows the recursive definition of the matrix $G_T$ for increasing $n=1,4,13,40,...$\ .

\section{Conclusion}

This paper presents a comprehensive study of fermion--qubit mappings with ancilla--free application, introducing what we believe are several practical advancements in the field. The first half of this paper provides several definitional and notational updates, which allow for a more streamlined understanding of fermion--qubit mappings. Section \ref{sec:prelim} introduces a unified definition which bridges the gap between the operator--based and state--based mappings that appear in the literature. In Section \ref{sec:equiv}, we established a practical equivalence relation for fermion--qubit mappings with Pauli representations of the Majorana operators, which make up virtually all of the mappings of interest in the field, addressing the redundancies in their notation and highlighting the symmetries that should not affect fermionic simulation algorithms. The notion of equivalence partitions these mappings into templates, distinguishing mappings that differ meaningfully in the Pauli structure of the Majorana representations. We introduced classical encodings in Section \ref{app:ibm} and showed that classical encodings with Pauli strings for Majorana representations are affine encodings, which we showed are equivalent to linear encodings.

The second half of this paper focused our definitions, notation, and concept of equivalence upon ternary tree transformations. While ternary tree transformations are already notable for offering the minimum average Pauli operator weight and optimisation algorithms for restricted qubit architectures, we showed that product--preserving ternary tree transformations also benefit from being equivalent to linear encodings in the Fock basis. Our proof not only improves the existing product--preserving Pauli operator pairing algorithm in Section \ref{sec:treebased}, but also provides a formula to determine the unique linear encoding corresponding to each ternary tree graph in Section \ref{sec:cbptree}.

\subsection{Future work}

This work opens new directions for discovery and optimisation of fermion--qubit mappings. In the theoretical direction,  separating mappings that represent Majorana operators with Pauli strings into templates under our equivalence relation could lead to a deeper understanding of fermion--qubit mappings in general. We suspect that the group of symmetries is undocumented in the literature. Creating a database of the templates of $n$--mode fermion--qubit mappings would produce a practical classification scheme, where rather than searching over all configurations of Pauli strings or encodings of the Fock basis, one need only search through the catalogue of templates for the best mapping on a case-by-case scenario. Additional insights lie in exploring the less well--behaved mappings: product--preserving mappings that are not classical encodings or even product--breaking mappings themselves could have unique properties with unforeseen advantages in quantum simulation.

On the practical side, our findings provide important context for algorithmic search for optimal fermion--qubit mappings. We have demonstrated that any search for linear encodings inherently includes ternary tree transformations as a subset, indicating that current algorithms might already be identifying these mappings without realising it. Our results provide the necessary tools to identify whether any linear encoding is a ternary tree transformation. Computational searches are often tailored to real--world problems, such as studying Hamiltonians in quantum chemistry applications, and it would be interesting to see if ternary tree transformations tended to outperform other types of linear encodings in practical applications.

\section{Acknowledgements}

Invaluable conversations with James Whitfield and Andrew Projanksy about the Sierpinski tree data structure alerted us to the possibility that more ternary tree transformations might be linear encodings. Majorana--themed discussions with Campbell McLauchlan and Wilfred Salmon played a large part in the definition of equivalence classes for fermion--qubit mappings. MC received the support of a Cambridge Australia Allen \& DAMTP Scholarship during the preparation of this document. BH was supported by the US NSF grants PHYS-1820747 and NSF (EPSCoR-1921199) as well as the Office of Science, Office of Advanced Scientific Computing Research under program Fundamental Algorithmic Research for Quantum Computing. BH was also supported by the ``Quantum Chemistry for Quantum Computers'' project sponsored by the DOE, Award DE- SC0019374. SS acknowledges support from the Royal Society University Research Fellowship and ``Quantum simulation algorithms for quantum chromodynamics" grant (ST/W006251/1) and EPSRC Reliable and Robust Quantum Computing grant
(EP/W032635/1).

\newpage
\bibliographystyle{unsrt}
\bibliography{citations}

\newpage
\appendix \vspace{-2cm}
\section{Glossary}\label{sec:appendix}
\begin{table*}[h]
	\centering
	\label{table:fermionic}
	\begin{tabular}{p{2.5 cm} p{3 cm} p{9.5cm}}

		\textit{symbol} & \textit{object type} & \textit{description} \\

        \toprule \\ 
        \multicolumn{3}{c}{\textbf{Section \ref{sec:prelim}}}   \\ & & \\
		
		$\mathcal{H}_\text{fermion}$ & $\sim \bigoplus_{i=0}^{n-1} A(\mathcal{H}_2^{\otimes i})$  & The $2^n$--dimensional state space of an $n$--mode fermionic system. The notation $A(\mathcal{H}_2^{\otimes i})$ denotes the space of antisymmetrised $i$--fermion quantum states.  \\ & & \\
		
		
		$\hat{a}_i^{(\dagger)}$ & $\mathcal{H}_\text{fermion}$ operator & Annihilation (creation) operator for a fermion in the $i$th mode, for $i\in [n]$; they collectively satisfy the canonical anticommutation relations $\{\hat{a}_i, \hat{a}_j^\dagger \} = \delta_{ij} \mathds{1}$, $\{\hat{a}_i, \hat{a}_j\}= 0$. \\ & & \\

		

		$\ketf{\Omega_\text{vac}}$ & $\in \mathcal{H}_\text{fermion}$ & \textit{(Definition \ref{defn:fock}.)} The fermionic vacuum state, a simultaneous 0--eigenstate of the number operators $\{\hat{a}_i^\dagger \hat{a}_i \}_{i=0}^{n-1}$ and $(+1)$--eigenstate of the operators $\{-i\hat{\gamma}_{2i} \hat{\gamma}_{2i+1}\}_{i=0}^{n-1}$). Note that any simultaneous 0--eigenstate of the operators $\{a_i^\dagger a_i\}_{i=0}^{n-1}$ must be of the form $e^{i \phi} \ketf{\Omega_\text{vac}}$ for some $\phi \in [0,2\pi)$. \\ & & \\
		
		$\ketf{\vec{f}}$ & $\in \mathcal{H}_\text{fermion}$ &
		\textit{(Definition \ref{defn:fock}.)} Fock basis state corresponding to the occupation number vector $\vec{f} = (f_0,f_1,\dots, f_{n-1}) \in \mathbb{F}_2^n$. The Fock state $\ketf{\vec{f}}$ relates to the Majorana operators and vacuum state via $\ketf{\vec{f}} = (\hat{\gamma}_0)^{f_0} (\hat{\gamma}_2)^{f_1} \dots (\hat{\gamma}_{2n-2})^{f_{n-1}} \ketf{\Omega_\text{vac}}$ \textit{(Lemma \ref{lem:occnobasis})}; it is an $f_i$--eigenstate of $\hat{a}_i^\dagger \hat{a}_i$ and a $(-1)^{f_i}$--eigenstate of $-i\hat{\gamma}_{2i} \hat{\gamma}_{2i+1}$. \\ & & \\

		$\hat{\gamma}_{2i}, \hat{\gamma}_{2i+1}$ & $\mathcal{H}_\text{fermion}$ operator & \textit{(Definition \ref{defn:maj}.)} Majorana operators defined via $\hat{\gamma}_{2i} = \hat{a}_i+ \hat{a}_i^\dagger$, $\hat{\gamma}_{2i+1} = -i(\hat{a}_i - \hat{a}_i^\dagger )$ or equivialently $\hat{a}_i = \frac{1}{2} (\hat{\gamma}_{2i} + i \hat{\gamma}_{2i+1})$ for $i\in [n]$. They satisfy the canonical anticommutation relations $\hat{\gamma}_i^\dagger = \hat{\gamma}_i$, $\{\hat{\gamma}_i, \hat{\gamma}_j \} = 2\delta_{ij} \mathds{1}$ for $i,j \in [2n]$.\\ & & \\

		$\mathcal{H}_2$ & $\sim \mathbb{C}_2$  & Two--dimensional complex Hilbert space which describes the state of a single qubit or fermionic mode.   \\ & & \\		

  		$\mathcal{H}_2^{\otimes n}$ & $\sim \mathbb{C}_2^{\otimes n}$ & The $2^n$--dimensional state space of an $n$--qubit system. \\ && \\

		$\mathcal{U}_n$ & subgroup of $2^n{\times}2^n$ matrices  & The $n$--qubit unitary matrices, which satisfy $U^\dagger U = \mathds{1}^{\otimes n}$ for all $U \in \mathcal{U}_n$. \\ & & \\

        $\mathds{1}, X, Y, Z$ & $\in \mathcal{U}_1$ & The identity matrix and single--qubit Pauli matrices
		$\mathds{1} = \begin{pmatrix} 1 & 0 \\ 0 & 1 \end{pmatrix}$,
		$X = \begin{pmatrix} 0 & 1 \\ 1 & 0 \end{pmatrix}$,
		$Y = \begin{pmatrix} 0 & -i \\ i & 0 \end{pmatrix}$,
		$Z = \begin{pmatrix} 1 & 0 \\ 0 & -1 \end{pmatrix}$.
		\\& & \\ 

	\end{tabular}
\end{table*}

        \vspace{-2cm}

\begin{table*}[hbtp]
	\centering
	\label{table:qubit}
	\begin{tabular}{p{2 cm} p{3.1 cm} p{9.9cm}}
		\textit{symbol} & \textit{object type} & \textit{description}  \\ & & \\
		
        $\ket{\psi}$ & $\in \mathcal{H}_2^{\otimes n}$ & $2^n$--dimensional $n$--qubit state vector. \\ & & \\
  
		$\mathcal{P}_n$ & subgroup of $\mathcal{U}_n$ & The $n$--qubit Pauli group $\langle X, Y, Z \rangle^{\otimes n}$. We call the elements of $\mathcal{P}_n$ Pauli strings. \\ & & \\

        $\{\gamma_i\}_{i=0}^{2n-1}$ & $\subset \mathcal{P}_n$ & \textit{(Definition \ref{defn:jw}.)} The Pauli strings $\gamma_{2i} = (\bigotimes_{k=0}^{i-1} Z_k ) X_i$, $\gamma_{2i+1} = (\bigotimes_{k=0}^{i-1} Z_k) Y_i$ which are the representations of the Majorana operators $\hat{\gamma_i}$ under the Jordan--Wigner transformation. \\ & & \\

		$\mathfrak{m}$ & fermion--qubit \hspace{2em} mapping & \textit{(Definition \ref{defn:mapping}.)} The operator--based definition of a fermion--qubit mapping $\mathfrak{m} = ((\Gamma_{2i}, \Gamma_{2i+1}))_{i=0}^{n-1}$ as an ordered list of ordered pairs of anticommuting Hermitian qubit operators $\{\Gamma_i\}_{i=0}^{2n-1}$. The operator $\Gamma_i$ represents the Majorana operator $\hat{\gamma}_i$.\\& & \\

		$\{\Gamma_i\}_{i=0}^{2n-1}$ & $\subset \mathcal{P}_n$ & The representations of the Majorana operators under the fermion--qubit mapping $\mathfrak{m}$. A set of anticommuting, Hermitian qubit operators, they satisfy $\{\Gamma_i,\Gamma_j\} = 2\delta_{ij}\mathds{1}$, $\Gamma_i^\dagger = \Gamma_i$. \\ & & \\

  		$\mathcal{C}_n$ & subgroup of $\mathcal{U}_n$ & The $n$--qubit Clifford group, all $2^n{\times}2^n$ unitary matrices that normalise the Pauli group: $\mathcal{C}_n = \{C \in \mathcal{U}_n \mid C \mathcal{P}_n C^\dagger = \mathcal{P}_n\}$. \\ & & \\

		$U_\mathfrak{m}$  $(C_\mathfrak{m})$ & $\in \mathcal{U}_n$ $(\mathcal{C}_n)$ & \textit{(Definition \ref{defn:mapping}, Lemma \ref{lem:Cfix}.)} The unique unitary matrix with the property $U_\mathfrak{m} : \ket{\vec{f}} \mapsto \ket{\vec{f}_{\mathfrak{m}}}$ for all $\vec{f} \in \mathbb{F}_2^n$.  The matrix satisfies $U_\mathfrak{m} \gamma_i U_{\mathfrak{m}}^\dagger = \Gamma_i$ for all $i \in [2n]$. The Majorana representations are Pauli strings $\Gamma_i \in \mathcal{P}_n$ if and only if $U_\mathfrak{m} \in \mathcal{C}_n$ is a Clifford operator, in which case we denote the matrix by $C_\mathfrak{m}$.\\ & & \\

  		$\{A_i^{(\dagger)}\}_{i=0}^{n-1}$ & $\subset \text{M}_{2^n}(\mathbb{F}_2)$ & The representations $A_i^{(\dagger)} = \frac{1}{2} (\Gamma_{2i} \pm i \Gamma_{2i+1})$ of the annihilation (creation) operators under the fermion--qubit mapping $\mathfrak{m}$.\\ & & \\

  		$\mathfrak{m}_{\text{JW}}$ & fermion--qubit \hspace{2em} mapping & The Jordan--Wigner transformation as a list of anticommuting Hermitian Pauli strings $\mathfrak{m}_\text{JW} = (({\gamma}_{2i}, {\gamma}_{2i+1}))_{i=0}^{2n-1}$. \\ & & \\

		$\ket{\vec{0}_\mathfrak{m}}$ & $\in \mathcal{H}_2^{\otimes n}$ & \textit{(Definition \ref{defn:occno}.)} The representation of $\ketf{\Omega_\text{vac}}$ under $\mathfrak{m}$, which is a simultaneous 0--eigenstate of the number operators $\{A_i^\dagger A_i\}_{i=0}^{n-1}$ and $(+1)$--eigenstate of the vacuum stabilisers $\{-i\Gamma_{2i}\Gamma_{2i+1}\}_{i=0}^{n-1}$. Note that any simultaneous 0--eigenstate of $A_i^\dagger A_i$ must be of the form $e^{i \phi} \ket{\vec{0}_{\mathfrak{m}}}$ for some $\phi \in [0,2\pi)$. \\ & & \\
		
		$\ket{\vec{f}_\mathfrak{m}}$ & $ \in \mathcal{H}_2^{\otimes n}$ & \textit{(Definitions \ref{defn:occno}, \ref{defn:statesfirst}.)} The state--based definition of $\mathfrak{m}$ in terms of its representation of the Fock basis $\{\ketf{\vec{f}} \mid \vec{f} \in\mathbb{F}_2^n\}$. The Fock state $\ket{\vec{f}_\mathfrak{m}}$ of $\mathfrak{m}$ relates to the Pauli operators of $\mathfrak{m}$ via $\ket{\vec{f}_\mathfrak{m}} = (\Gamma_0)^{f_0} (\Gamma_2)^{f_1} \dots (\Gamma_{2n-2})^{f_{n-1}} \ket{\vec{0}_\mathfrak{m}}$; it is an $(f_i)$--eigenstate of $A_i^\dagger A_i$ and a $\left((-1)^{f_i}\right)$--eigenstate of $-i\Gamma_{2i}\Gamma_{2i+1}$. \\ & & \\ \bottomrule  \\
  
	\end{tabular}
\end{table*}

\begin{table*}[hbtp]
	\centering
	\label{table:mappings}
	\begin{tabular}{p{2 cm} p{3.2 cm} p{9.4cm}}

		\textit{symbol} & \textit{object type} & \textit{description}  
        \\
        \toprule \\

          \multicolumn{3}{c}{\textbf{Section \ref{sec:equiv}}}  \\ & & \\
  
		 $[\mathfrak{m}]$ & equivalence class & \textit{(Definition \ref{defn:templates}).} The template for a fermion--qubit mapping $\mathfrak{m}$ that represents the Majorana operators with the Pauli strings $\Gamma_i \in \mathcal{P}_n$ for all $i \in [2n]$. The template $[\mathfrak{m}]$  consists of all mappings with identical  Pauli string Majorana representations to $\mathfrak{m}$ up to qubit, local basis and fermionic labels, and Pauli signs and ordering within pairs.
		 \\  \\
		 \toprule & & \\
  
          \multicolumn{3}{c}{\textbf{Section \ref{app:ibm}}} \\ \\
  
		$\mathfrak{C}_n$ & $\subset \mathcal{H}_2^{\otimes n}$ & \textit{(Definition \ref{defn:compbasis}.)} The computational basis for $\mathcal{H}_2^{\otimes n}$: $\mathfrak{C}_n = \{\ket{00\dots0}, \ket{10\dots 0}, \dots ,$ $ \ket{11\dots 1}\}$. \\ & & \\

        $e^{i \theta}\mathfrak{C}_n$, $\pm \mathfrak{C}_n$ & $\subset \mathcal{H}_2^{\otimes n}$ & Computational bases with phase offsets for $\theta \in [0, 2\pi)$: $e^{i \theta}\mathfrak{C}_n = \{e^{i \theta}\ket{00\dots0}, \dots ,$ $ e^{i \theta}\ket{11\dots 1}\}$ and $\pm \mathfrak{C}_n = \mathfrak{C}_n \cup (-1) \mathfrak{C}_n$. \\ & & \\

        \multicolumn{2}{l}{\textit{``classical encoding of the Fock basis $\mathfrak{m}$"}} & \textit{(Definition \ref{defn:classlin}.)} A fermion--qubit mapping $\mathfrak{m}$ for which $\ket{\vec{f}_\mathfrak{m}} \in \mathfrak{C}_n$ for all $\vec{f} \in \mathbb{F}_2^n$. \\ & & \\
        
        \multicolumn{2}{l}{\textit{``affine encoding of the Fock basis $\mathfrak{m}$"}} &  \textit{(Definition \ref{defn:classlin}.)} A fermion--qubit mapping $\mathfrak{m}$ for which $\ket{\vec{f}_\mathfrak{m}} = \ket{G(\vec{f}\oplus \vec{b})}$ for all $\vec{f} \in \mathbb{F}_2^n$, where $G \in \text{GL}_n(\mathbb{F}_2)$ and $\vec{b} \in \mathbb{F}_2^n$. Affine encodings are linear encodings that represent the Majorana operators with Pauli strings, and vice versa \textit{(Theorem \ref{thm:affine1})}.\\ & & \\

		\multicolumn{2}{l}{\textit{``linear encoding of the Fock basis $\mathfrak{m}$"}} & \textit{(Definition \ref{defn:classlin})} A fermion--qubit mapping $\mathfrak{m}$ for which $\ket{\vec{f}_\mathfrak{m}} = \ket{G \vec{f}}$ for all $\vec{f} \in \mathbb{F}_2^n$, where $G \in \text{GL}_n(\mathbb{F}_2)$. Every affine encoding of the Fock basis is equivalent to a linear encoding \textit{(Corollary \ref{cor:affinelin})}. \\ & & \\

        $C_G$ & $\in \mathcal{C}_n$ & \textit{(Proof of Theorem \ref{thm:affine1}.)} The unique Clifford operator $C_G \in \mathcal{C}_n$ to implement the basis transformation $\ket{\vec{f}} \mapsto \ket{G \vec{f}}$, where $G \in \text{GL}_n(\mathbb{F}_2)$. Its stabiliser tableau is $[C_G] = \left[\begin{array}{cc|c} G & 0 & 0 \\ 0 & (G^{-1})^\top & 0 \end{array}\right]$.\\ & & \\

        \bottomrule \\ 

        \multicolumn{3}{c}{\textbf{Section \ref{sec:treebased}}} \\ \\

		$T$ & ternary tree graph & \textit{(Definition \ref{defn:tt}).} A ternary tree, which is a labelled tree graph with $n$ vertices, each of which has at most three children.\\ & & \\
   
		$\mathcal{P}_n/K$ & group isomorphic to $\{\mathds{1}, X, Y, Z \}^{\otimes n}$ & \textit{(Definition \ref{defn:unsigned})}. Notation for the unsigned Pauli operators $\{\mathds{1}, X, Y, Z\}^{\otimes n}$. A normal subgroup of $\mathcal{P}_n$ is $K = \{\pm \mathds{1}, \pm i \mathds{1}\}^{\otimes n}$, and the quotient group $\mathcal{P}_n/K$ consists of cosets $[P] = \{\pm P, \pm i P\}$ for $P \in \{\mathds{1}, X, Y, Z\}^{\otimes n}$; we make the identification $[P] \leftrightarrow P$. \\ & & \\


	\end{tabular}
\end{table*}

\begin{table*}[hbtp]
	\centering
	\label{table:trees}
	\begin{tabular}{p{2 cm} p{3.4 cm} p{9.6cm}}

		\textit{symbol} & \textit{object type} & \textit{description}  \\ \\

		$\widetilde{\mathcal{G}}_T$ & $\subset \mathcal{P}_n/K$ & \textit{(Definition \ref{defn:treepaulis}).} The maximally anticommuting set $\widetilde{\mathcal{G}}_T$ of unsigned Pauli operators that arises from converting the root-to-leaf paths of ${T}$ into Pauli strings. \\ & & \\
  		
		\multicolumn{2}{l}{``\textit{$T$--based mapping $\mathfrak{m}$}"} & \textit{(Definition \ref{defn:treebased}).} A fermion--qubit mapping $\mathfrak{m}$ for which $\pm\Gamma_i \in \widetilde{\mathcal{G}}_T$ for the ternary tree $T$. \\ & & \\

		\multicolumn{2}{l}{``\textit{ternary tree transformation $\mathfrak{m}$}"} & \textit{(Definition \ref{defn:treebased}).} A fermion--qubit mapping $\mathfrak{m}$ for which there exists some ternary tree $T$ such that $\mathfrak{m}$ is a $T$--based mapping. \\ & & \\

        \multicolumn{2}{l}{\textit{``product--preserving mapping $\mathfrak{m}$"}}
		& \textit{(Definition \ref{defn:prod}).} A fermion--qubit mapping $\mathfrak{m}$ for with Fock states $\ket{\vec{f}_\mathfrak{m}}$ that are tensor products of the single--qubit $X$, $Y$ and $Z$ eigenstates for all $\vec{f} \in \mathbb{F}_2^n$. A sufficient condition is for $\ket{\vec{0}_{\mathfrak{m}}}$ to be a product state. \\ & & \\

        \toprule \\ 

        \multicolumn{3}{c}{\textbf{Section \ref{sec:cbptree}}} \\ \\

		$C_T$ & $\in \mathcal{C}_n$ & \textit{(Lemma \ref{lem:compbasistt}.)} The unique Clifford operator $C_T \in \mathcal{C}_n$ such that $C_T \gamma_i C_T^\dagger = \Gamma_i$, where $\mathfrak{m}(T) = ((\Gamma_{2i}, \Gamma_{2i+1}))_{i=0}^{n-1}$. \\ & & \\

		$\mathfrak{m}({T})$ & fermion--qubit \hspace{2em} mapping & \textit{(Lemma \ref{lem:compbasistt}.)} The unique $T$--based mapping  $\mathfrak{m}(T) = ((\Gamma_{2i}, \Gamma_{2i+1}))_{i=0}^{n-1}$ that linearly encodes the Fock basis.\\ & & \\

		$\{\widetilde{\Gamma}_i\}_{i=0}^{2n}$ & $\subset \mathcal{P}_n/K$ & \textit{(Proof of Lemma \ref{lem:compbasistt}.)} A subset of $\mathcal{\widetilde{G}}_T$ with a specific enumeration scheme.\\ & & \\
  
		$\#y$ & map
        $\mathcal{P}_n/K \rightarrow [n]$ & \textit{(Proof of Lemma \ref{lem:compbasistt}.} A function that counts the number of $Y$ operators in unsigned Pauli strings. \\ & & \\

	   $\{\widehat{\Gamma}_i\}_{i=0}^{2n-1}$ & $\subset \{\mathds{1}, X, -iY, Z\}^{\otimes n}$ & \textit{(Proof of Lemma \ref{lem:compbasistt}.)} A set of anticommuting Pauli strings $\{\widehat{\Gamma}_i, \widehat{\Gamma}_j \} = 2\delta_{ij} \mathds{1}$, where $\widehat{\Gamma}_i = (-1)^{ \#y(\widetilde{\Gamma}_i)} \widetilde{\Gamma}_i$.\\ & & \\

        $T_{\mathcal{S}}$ & tree graph & \textit{(Proof of Lemma \ref{lem:compbasistt}.)} The ternary subtree of $T$ on the vertices with labels $\mathcal{S} \subseteq [n]$.\\ & & \\

  		$G_T$ & $\in \text{GL}_n(\mathbb{F}_2)$ & \textit{(Lemma \ref{lem:findmatrix}.)} The unique invertible binary matrix $G \in \text{GL}_n(\mathbb{F}_2)$ such that $\ket{G_T \vec{f}} = \ket*{\vec{f}_{\mathfrak{m}(T)}}$ for all $\vec{f} \in \mathbb{F}_2^n$. \\ & & \\

        \bottomrule \\

        \multicolumn{3}{c}{\textbf{Appendix \ref{app:affine}}} \\ \\

  		$[C]$ & $2n \times (2n+1)$ tableau & \textit{(Definition \ref{defn:symplectic}.)} The stabiliser tableau of the Clifford operator $C \in \mathcal{C}_n$. The first $n$ columns are the $2n$--bit symplectic representations of (unsigned) $CX_i C^\dagger$, and the second $n$ columns of (unsigned) $C Z_i C^\dagger$, for $i \in [n]$. The final column stores the signs of the operators $C X_i C^\dagger$ in its first $n$ entries, and of $C Z_i C^\dagger$ in its final $n$ entries. If $[C_1] = [C_2]$, then $C_1 = e^{i \theta} C_2$ for some $\theta \in [0,2\pi)$. \\ & & \\

		

		
		\bottomrule
	\end{tabular}
\end{table*}

\newpage

\section{Stabiliser tableaux of affine encodings}\label{app:affine}

	\begin{definition}\textit{(Symplectic notation for Pauli  and Clifford operators.)} \label{defn:symplectic}
		
		\begin{enumerate}[label=\alph*)]
		\item \textit{Symplectic representation of Pauli operators.} Let $P \in \mathcal{P}_n/K$ be an unsigned Pauli operator, and let $P\vert_{\{i\}}$ denote the Pauli matrix acting on the $i$th qubit within the overall action of $P$ (e.g.\ $(X_0Y_1)\vert_{0} = X$). The \textit{symplectic representation of $P$} is the 
			$2n$--bit vector $\vec{v}(P) \in \mathds{F}_2^{2 n}$ with coordinates
			\begin{align}
			(v_i(P), v_{i+n}(P)) = \begin{cases}
				(0,0) & \quad \text{if } P|_{\{i\}}  = \mathds{1}\, ,\\
				(1,0) & \quad \text{if } P|_{\{i\}}  = X \, ,\\
				(0,1) & \quad \text{if } P|_{\{i\}}  = Z \, ,\\
				(1,1) & \quad \text{if } P|_{\{i\}}  = Y\, 
			\end{cases} \quad \text{for all } i \in [n]\, .
			\end{align}

	\item \textit{Symplectic inner product.} \label{defn:symplecticinnerproduct}
			The \textit{symplectic inner product} of two unsigned Pauli operators $P, Q \in \mathcal{P}_n/K$ is the quantity
			\begin{align}
				\vec{v}(P)^\top \begin{pmatrix} 0 & \mathds{1}_n\\ \mathds{1}_n & 0 \end{pmatrix} \vec{v}(Q)\, .
			\end{align}
			For $P, Q \in \mathcal{P}_n/K$, the symplectic inner product is equal to
			\begin{align}
				\vec{v}(P)^\top \begin{pmatrix} 0 & \mathds{1}_n \\ \mathds{1}_n & 0 \end{pmatrix} \vec{v}(Q) &= ( \underbrace{\phantom{|}\dots\dots \phantom{|}}_{\mathclap{P|_{\{i\}} = Y \text{ or } Z} } | \overbrace{\phantom{|}\dots\dots \phantom{|}}^{\mathclap{P|_{\{i\}} = X \text{ or } Z}}    )
				\quad 
				\begin{pmatrix} \phantom{\bigg\{}\vdots \phantom{\bigg\}} \\ \hline \phantom{\bigg\{}\vdots \phantom{\bigg\}} \\ \end{pmatrix}
				\begin{matrix} \big\} \, \scriptstyle{Q|_{\{i\}} = X \text{ or } Y}\\  \\ \big\} \, \scriptstyle{Q|_{\{i\}} = Y \text{ or } Z} \end{matrix} \\
				&= \abs{ \left\{ \, i \, \big\vert \, P|_{\{i\}} =( X \text{ or } Y) \text{ and }  Q|_{\{i\}} = (Y \text{ or } Z) \right\}}  \\
				&\phantom{=}\, + \abs{ \left\{ \, i \, \big\vert \, P|_{\{i\}} =( Y \text{ or } Z) \text{ and }  Q|_{\{i\}} = (X \text{ or } Y) \right\}}  \nonumber \\
				&= \abs{ \left\{ \, i \, \big\vert \, P|_{\{i\}} = X \text{ and }  Q|_{\{i\}} = Y \right\}} +\abs{ \left\{ \, i \, \big\vert \, P|_{\{i\}} = X \text{ and }  Q|_{\{i\}} = Z \right\}}  \\
				&\phantom{=} \, + \abs{ \left\{ \, i \, \big\vert \, P|_{\{i\}} = Y \text{ and }  Q|_{\{i\}} = X \right\}} +\abs{ \left\{ \, i \, \big\vert \, P|_{\{i\}} = Y \text{ and }  Q|_{\{i\}} = Z \right\}} \nonumber \\
				&= \begin{cases} 1 & \phantom{[}\{P,Q\}\phantom{]} = \phantom{]} 0 \\ 0 & \phantom{\{}[P,Q]\phantom{\}} =  \phantom{]}0 \, , \end{cases}
			\end{align}
			i.e.\ the symplectic inner product of $P$ and $Q$ is 1 if and only if $P$ and $Q$ anticommute. 
	
	\item \textit{Stabiliser tableau of Clifford operators.}
	Let $C \in \mathcal{C}_n$ be a Clifford operator. Then, there exists a vector $\vec{b} \in \mathbb{F}_2^{2n}$ and a set of unsigned Pauli operators $\{P_i\}_{i=0}^{2n-1} \subset \mathcal{P}_n/K$ such that $C X_i C^\dagger = (-1)^{b_i} P_i$ and $C Z_i C^\dagger = (-1)^{b_{i+n}} P_{i+n}$. The \textit{stabiliser tableau} of $C$ is the $(2n){\times}(2n+1)$ binary array
	\begin{align}
							[C] = \left[\begin{array}{ccc|ccc|c}&&&&&&\\ \vec{v}(P_0) & \dots & \vec{v}(P_{n-1}) & \vec{v}(P_n) & \dots & \vec{v}(P_{2n-1}) & \vec{b} \\ &&&&&& \end{array}\right]\, , \label{eqn:stabdef}
	\end{align}
	where the notation in Equation \ref{eqn:stabdef} implies that the first column of $[C]$ is $\vec{v}(P_0)$, and so on.
	\end{enumerate}
	\end{definition}
	
	\begin{lemma}
		Two Clifford operators with the same stabiliser tableau are identical up to a global phase. \label{lem:equivcliff}
	\begin{proof}
		This is simply a limitation of the symplectic representation for the  Clifford group \cite{dehaene_clifford_2003}.
	\end{proof}
	\end{lemma}

Theorem \ref{thm:affine} provides a self-contained derivation of the stabiliser tableaux of classical encodings of the Fock basis. The proof integrates Lemmas \ref{lem:a2}--\ref{lem:a5}.

\begin{theorem}\textit{(The stabiliser tableaux of affine encodings.)} \label{thm:affine}
For any $G \in \text{GL}_n(\mathbb{F}_2)$ and any $\vec{b} \in \mathbb{F}_2^n$, the unitary transformation $\ket{\vec{f}} \mapsto \ket{G(\vec{f} \oplus\vec{b})}$ is a Clifford transformation $C \in \mathcal{C}_n$, where $G \in \text{GL}_n(\mathbb{F}_2)$ and $\vec{b} \in \mathbb{F}_2^n$. The Clifford operator $C$ has the following stabiliser tableau:
		\begin{align} \label{eqn:thm2main}
			[C] = \left[\begin{array}{ccc|ccc|c}
				&       &       & 0      &\dots  & 0                & 0 \\
				\multicolumn{3}{c|}{G}& \vdots &\ddots & \vdots           & \vdots  \\
				&       &       & 0      &\dots  & 0                & 0 \\ \hline
				0       &\dots  & 0     &        &       &                  &   \\
				\vdots  &\ddots &\vdots & \multicolumn{3}{c|}{(G^{-1})^\top}               & \vec{b} \\
				0       &\dots  & 0     &        &       &                  &  
			\end{array}\right]\, .
		\end{align}
		
		\begin{proof}
			Theorem \ref{thm:affine} follows from Lemmas \ref{lem:a2}--\ref{lem:a5}. The supplementary materials of \cite{Picozzi_2023} have inspired this proof.
			\begin{lemma} \label{lem:a2}
				A Clifford operator $C \in \mathcal{C}_n$ that preserves the computational basis $\mathfrak{C}_n$ maps single--qubit $X$ and $Z$ operators to Pauli strings consisting of only $X$ or $Z$ matrices, respectively. That is, the elements in the off-diagonal quadrants of the stabiliser tablau of $C$ must be zero:
				\begin{align}
					[C] = \left[\begin{array}{ccc|ccc|c}
						&       &       & 0      &\dots  & 0        &   \\
						&       &       & \vdots &\ddots & \vdots   &   \\
						&       &       & 0      &\dots  & 0        &   \\ \hline
						0       &\dots  & 0     &        &       &          &   \\
						\vdots  &\ddots &\vdots &        &       &          & \\
						0       &\dots  & 0     &        &       &                  &  
					\end{array}\right]\, .
				\end{align}
				\begin{claimproof}
					Suppose that for some $j \in [n]$, the image of $X_j$ under $C$ is $C(X_j) = \pm P$ where $P \in \mathcal{P}_n/K$ is such that $P\vert_{\{k\}} = Y$ or $Z$. For $\vec{f} \in \mathbb{F}_2^n$,
					\begin{align}
						X_j \ket{\vec{f}} = \ket{\vec{f} \oplus \vec{1}_j} \in \mathfrak{C}_n  \implies C(X_j \ket{\vec{f}}) \in \mathfrak{C}_n\, ,
					\end{align}
					as $C$ preserves the computational basis. At the same time,
					\begin{align}
						C(X_j \ket{\vec{f}}) = C(X_j) C \ket{\vec{f}} = \pm P \ket{\sigma (\vec{f})}\, ,
					\end{align}
					and so $\pm P\ket{\sigma(\vec{f})} \in \mathfrak{C}_n$. However, this necessarily cannot hold for all bit-strings in $\mathbb{F}_2^n$, as
					\begin{align}
						P \ket{\sigma(\vec{f})} \in \left( \prod_{\substack{(\sigma(\vec{f}))_i=0 \\ P|_{\{i\}}=Y}} i \right) \left( \prod_{\substack{(\sigma(\vec{f}))_i=1 \\ P|_{\{i\}}=Y}} (-i) \right)\left( \prod_{\substack{(\sigma(\vec{f}))_i=1\\ P|_{\{i\}}=Z}} (-1) \right) \mathfrak{C}_n\, .
					\end{align}
					Indeed, because $C$ implements the permutation $\sigma$ on the computational basis, there must exist a unique $\vec{f}' \in \mathbb{F}_2^n$ with $(\sigma(\vec{f}'))_i = (\sigma(\vec{f}))_i$ for all $i \neq k$, and with $(\sigma(\vec{f}'))_k \neq (\sigma(\vec{f}))_k$. 	Regardless of whether $P$ acts with a $Y$ or $Z$ on qubit $k$, the image of the state $X_j \ket{\vec{f}'}$ under $C$ is not in the computational basis, because
					\begin{align}
						C(X_j \ket{\vec{f}'}) = C(X_j) C\ket{\vec{f}'} =  \pm P \ket{\sigma(\vec{f}')} \in -\mathfrak{C}_n\, .
					\end{align}
					This contradicts the assumption that $C$ preserves the computational basis. Therefore the images of single-qubit $X$ operators under $C$ must be Pauli strings consisting of $X$ matrices or the identity, which is equivalent to stating that every entry in the bottom-left quadrant of $[C]$ must be zero.
					
					Suppose that for some $j \in [n]$, the image of $Z_j$ under $C$ is $C(Z_j) = \pm P$ where $P \in \mathcal{P}_n/K$ is such that $P\vert_{\{k\}} =X$ or $Y$. For $\vec{f} \in \mathbb{F}_2^n$,
					\begin{align}
						Z_j \ket{\vec{f}} = (-1)^{f_j} \ket{\vec{f}} \implies 
						C(Z_j \ket{\vec{f}}) = (-1)^{f_j} = (-1)^{f_j} C\ket{\vec{f}} = \ket{\sigma(\vec{f})}\, . \label{eqn:contra1}
					\end{align}
					At the same time, because $P$ contains at least one $X$ or $Y$ matrix,
					\begin{align}
						C(Z_j \ket{\vec{f}}) = C(Z_j) C\ket{\vec{f}} = \pm P \ket{\sigma(\vec{f})} \neq \ket{\sigma(\vec{f})}\, ,
					\end{align}
					which contradicts Equation \ref{eqn:contra1}. Therefore the images of single--qubit $Z$ operators under $C$ must be Pauli strings consisting of $Z$ matrices or the identity, which is equivalent to stating that every entry in the top-right quadrant of $[C]$ must be zero.
				\end{claimproof}
			\end{lemma}
			
			\begin{lemma}\label{lem:a3}
				The on-diagonal quadrants of the stabiliser tableau of a Clifford operator $C \in \mathcal{C}_n$ that preserves the computational basis $\mathfrak{C}_n$ are matrix inverse transposes of each other. That is, for some $G \in \text{GL}_n(\mathbb{F}_2)$,
				\begin{align}
					[C] = \left[\begin{array}{ccc|ccc|c}
						&       &       & 0      &\dots  & 0                &  \\
						\multicolumn{3}{c|}{G}& \vdots &\ddots & \vdots           &   \\
						&       &       & 0      &\dots  & 0                &  \\ \hline
						0       &\dots  & 0     &        &       &                  &   \\
						\vdots  &\ddots &\vdots & \multicolumn{3}{c|}{(G^{-1})^\top}               & \\
						0       &\dots  & 0     &        &       &                  &  
					\end{array}\right]\, .
				\end{align}
				\begin{claimproof}
					Note that Clifford operations preserve commutation relations, and so, for all $P, Q \in \mathcal{P}$,
					\begin{align} \label{eqn:anticoms}
						\{P, Q\} = \{C(P)), C(Q)\}\, .
					\end{align}
					Define the Pauli operators
					\begin{align}
						P_i  = \begin{cases}
							X_i & 0 \leq i < n \\ Z_i & n \leq i < 2n\, .
						\end{cases}
					\end{align}
					Let $G_1, G_2 \in \text{GL}_n(\mathbb{F}_2)$ be the on-diagonal entries of the stabiliser tableau of $C$, i.e.
					\begin{align}
					[C] = 	\left[\begin{array}{ccc|ccc|c}
							&       &       & 0      &\dots  & 0       &          \\
							\multicolumn{3}{c|}{G_1}& \vdots &\ddots & \vdots     &        \\
							&       &       & 0      &\dots  & 0          &       \\ \hline
							0       &\dots  & 0     &        &       &        &            \\
							\vdots  &\ddots &\vdots & \multicolumn{3}{c|}{G_2}         &       \\
							0       &\dots  & 0     &        &       &                  &
						\end{array}\right]\, ,
					\end{align}
					so that therefore $\vec{v}(C(X_i))$ is equal to the $i$th column of $G_1$ followed by $n$ zeroes, and $\vec{v}(C(Z_i))$ is equal to $n$ zeroes followed by the $i$th column of $G_2$. Note that $\{P_i, P_j\} = 0$ if and only if $i = j \oplus n$; via Equation \ref{eqn:anticoms}, conjugation by $C$ preserves the anticommutation relations of the $P_i$ operators, and so
					\begin{align}
						\{C(P_i), C(P_j)\} = 0 &\iff i = j \oplus n\, . \label{eqn:deltaijn}
					\end{align}
					Note that Equation \ref{eqn:deltaijn} implies  that $\delta_{i,j\oplus n}$ is 1 if and only if $C(P_i)$ and $C(P_j)$ anticommute; thus, by Definition \ref{defn:symplectic}\ref{defn:symplecticinnerproduct}, it is equal to the symplectic inner product of $C(P_i)$ and $C(P_j)$:
					\begin{align}
						\vec{v}(C(P_i))^\top \begin{pmatrix} 0 & \mathds{1}_n \\ \mathds{1}_n & 0 \end{pmatrix} \vec{v}(C(P_j)) &= \delta_{i,j\oplus n} \\
						\begin{pmatrix} (G_1)^\top & 0 \\ 0 & (G_2)^\top \end{pmatrix} 
						\begin{pmatrix}0 & \mathds{1}_n \\ \mathds{1}_n & 0 \end{pmatrix}
						\begin{pmatrix}G_1 & 0 \\ 0 & G_2 \end{pmatrix} &= \begin{pmatrix}0 & \mathds{1}_n \\ \mathds{1}_n & 0 \end{pmatrix} \\
						\begin{pmatrix} 0 &  (G_1)^\top G_2 \\ (G_2)^\top G_1 & 0 \end{pmatrix} &= \begin{pmatrix}0 & \mathds{1}_n \\ \mathds{1}_n & 0 \end{pmatrix} \\
						G_2 &= (G_1^{-1})^\top\, ,
					\end{align}
					as required.
				\end{claimproof}
			\end{lemma}
			
			\begin{lemma}\label{lem:a4}
				A Clifford $C \in \mathcal{C}_n$ that preserves the computational basis also preserves the signs of single--qubit $X$ operators. That is,
				\begin{align}\label{eqn:cbptableau}
					[C] = \left[\begin{array}{ccc|ccc|c}
						&       &       & 0      &\dots  & 0                & 0 \\
						\multicolumn{3}{c|}{G}& \vdots &\ddots & \vdots           & \vdots \\
						&       &       & 0      &\dots  & 0                & 0 \\ \hline
						0       &\dots  & 0     &        &       &                  &   \\
						\vdots  &\ddots &\vdots & \multicolumn{3}{c|}{(G^{-1})^\top}               &\\
						0       &\dots  & 0     &        &       &                  &  
					\end{array}\right]\, .
				\end{align}
				\begin{claimproof}
					Suppose $C(X_i) = -P$ for some $P \in \mathcal{P}_n/K$. Now,
					\begin{align}
						X_i \ket{\vec{0}} = \ket{\vec{1}_i} \in \mathfrak{C}_n \implies C(X_i \ket{\vec{0}}) = C(X_i) C \ket{\vec{0}} = -P \ket{\sigma(\vec{0})} \in -\mathfrak{C}_n\, ,
					\end{align}
					which contradicts the assumption that $C$ preserves the computational basis.
				\end{claimproof}
			\end{lemma}
			
			\begin{lemma} \label{lem:a5}
				The family of unitary matrices of the form $\{e^{i \theta} U \mid \theta \in [0,2\pi)\}$ where $U: \ket{\vec{f}} \mapsto \ket{G(\vec{f} \oplus\vec{b})}$ is equal to the family of Cliffords with stabiliser tableaux
				\begin{align}
					[e^{i \theta}C] = \left[\begin{array}{ccc|ccc|c}
						&       &       & 0      &\dots  & 0                & 0 \\
						\multicolumn{3}{c|}{G}& \vdots &\ddots & \vdots           & \vdots \\
						&       &       & 0      &\dots  & 0                & 0 \\ \hline
						0       &\dots  & 0     &        &       &                  &   \\
						\vdots  &\ddots &\vdots & \multicolumn{3}{c|}{(G^{-1})^\top}               & \vec{b}\\
						0       &\dots  & 0     &        &       &                  &  
					\end{array}\right]\, . \label{eqn:affinetab}
				\end{align}
				\begin{claimproof}
						Let $C_k$ be a Clifford operator that implements $C_k(X_i) = X_i$, $C_k(Z_i) = (-1)^{\delta_{ik}} Z_i$ for all $i \in [n]$; all such Clifford operators are identical up to a global phase.  Note that $C_k(Z_i \ket{\vec{0}})= C_k\ket{\vec{0}}$, while also
						\begin{align}
							C_k(Z_i \ket{\vec{0}}) = C_k(Z_i) C_k\ket{\vec{0}} = (-1)^{\delta_{ik}} Z_i C_k \ket{\vec{0}}\, .
						\end{align}
						Therefore $C_k\ket{\vec{0}}$ is a simultaneous $(+1)$--eigenvector of the $n$ Pauli operators $\{(-1)^{\delta_{ik}} Z_i\}_{i=0}^{n-1}$, and must be of the form $C_k \ket{\vec{0}} = e^{i\phi}\ket{\vec{1}_k}$ for some $\phi \in [0,2\pi)$; fix the global phase by taking $C_k\ket{\vec{0}} = \ket{\vec{1}_k}$. Notice that
						\begin{align}
							C_k \ket{\vec{f}} &= C_k \left( \prod_{f_i = 1} X_i \ket{\vec{0}} \right) = \prod_{f_i=1} \left(C_k (X_i) \right) C_k \ket{\vec{0}} = \prod_{f_i = 1} X_i \ket{\vec{1}_k} = \ket{\vec{f} \oplus \vec{1}_k}\, ,
						\end{align}
						and so $C_k$ is also the unitary that implements the map $C_k : \ket{\vec{f}} \mapsto \ket{\vec{f} \oplus \vec{1}_k}$. Thus,
						\begin{align}
							\prod_{b_k=1}C_k : \ket{\vec{f}} \longmapsto \ket{\vec{f} \oplus\vec{b}}\, . \label{eqn:plusb}
						\end{align}
						The stabiliser tableaux of the $C_k$ reveal the tableau of the operator in Equation \ref{eqn:plusb} via
						\begin{align} \label{eqn:cbtab}
							[C_k] = \left[\begin{array}{ccc|ccc|c}
								&       &       & 0      &\dots  & 0                & 0 \\
								\multicolumn{3}{c|}{\mathds{1}_n}& \vdots &\ddots & \vdots           & \vdots \\
								&       &       & 0      &\dots  & 0                & 0 \\ \hline
								0       &\dots  & 0     &        &       &                  &   \\
								\vdots  &\ddots &\vdots & \multicolumn{3}{c|}{\mathds{1}_n}               & \vec{1}_k\\
								0       &\dots  & 0     &        &       &                  &  
							\end{array}\right] \implies 
							\left[\prod_{b_k=1} C_k\right] = \left[\begin{array}{ccc|ccc|c}
								&       &       & 0      &\dots  & 0                & 0 \\
								\multicolumn{3}{c|}{\mathds{1}_n}& \vdots &\ddots & \vdots           & \vdots \\
								&       &       & 0      &\dots  & 0                & 0 \\ \hline
								0       &\dots  & 0     &        &       &                  &   \\
								\vdots  &\ddots &\vdots & \multicolumn{3}{c|}{\mathds{1}_n}               & \vec{b}\\
								0       &\dots  & 0     &        &       &                  &  
							\end{array}\right]\, .
						\end{align}
						
						Consider the Clifford $C_G$ that implements $C_G(X_i) = X_{U(i)}$, $C_G(Z_i) = Z_{F(i)}$ where $U(i)$  and $F(i)$ are the update and flip sets of $G$ as in Definition \ref{defn:upf}. The operator $C_G$ has stabiliser tableau
						\begin{align} \label{eqn:cgtab}
							[C_G] = \left[\begin{array}{ccc|ccc|c}
								&       &       & 0      &\dots  & 0                & 0 \\
								\multicolumn{3}{c|}{G}& \vdots &\ddots & \vdots           & \vdots \\
								&       &       & 0      &\dots  & 0                & 0 \\ \hline
								0       &\dots  & 0     &        &       &                  &   0 \\
								\vdots  &\ddots &\vdots & \multicolumn{3}{c|}{(G^{-1})^\top}               & \vdots \\
								0       &\dots  & 0     &        &       &                  &  0
							\end{array}\right]\, ,
						\end{align}
						and so $C_G$ preserves the computational basis, by Lemma \ref{lem:a3}. Note that $C_G(Z_i \ket{\vec{0}}) = C_G \ket{\vec{0}}$, while also
						\begin{align}
							C_G(Z_i \ket{\vec{0}}) = C_G(Z_i) C_G \ket{\vec{0}} = Z_{F(i)} C\ket{\vec{0}}\, .
						\end{align}
						Therefore $C_G\ket{\vec{0}}$ is a $(+1)$--eigenstate of $Z_{F(i)}$ for all $i \in [n]$, which means that $C_G \ket{\vec{0}} = \ket{\vec{f}'}$ where $\vec{f}' \in \mathbb{F}_2^n$ is such that $\{i \mid f'_i = 1\} \cap F(i)$ is even. However, since $G$ is invertible, the only vector $\vec{f}'$ satisfying this requirement for all $i \in [n]$ is $\vec{0}$, and hence $C_G \ket{\vec{0}} = \ket{\vec{0}}$. Using this, the image under $C_G$ of an arbitrary vector $\ket{\vec{f}}$ is
						\begin{align}
							C_G \ket{\vec{f}} &= C_G\left(\prod_{f_i = 1} X_i \ket{\vec{0}} \right) = \prod_{f_i= 1} C_G(X_{U(i)}) C_G \ket{\vec{0}} = \prod_{f_i=1} X_{U(i)} \ket{\vec{0}} = \prod_{\substack{i : f_i = 1 \\ j:G_{ij}=1}} X_j \ket{\vec{0}} = \prod_{(G \vec{f})_i = 1} X_i \ket{\vec{0}} = \ket{G \vec{f}}\, ,
						\end{align}
						and therefore $C_G$ implements the transformation 
						\begin{align}
							C_G : \ket{\vec{f}} \longmapsto \ket{G\vec{f}}\, . \label{eqn:gtimes}
						\end{align}
						Combining Equations \ref{eqn:plusb} and \ref{eqn:gtimes}, we obtain
						\begin{align}
							C_G \left(\prod_{b_k=1} C_{k} \right) : \ket{\vec{f}} \longmapsto \ket{G(\vec{f} \oplus \vec{b})} \implies C_G \left( \prod_{b_k=1} C_k \right) = C\, .
						\end{align}
						Comparing the stabiliser tableaux in Equations \ref{eqn:cbtab} and \ref{eqn:cgtab} reveals that
						\begin{align}
							[C]=\left[C_G \left(\prod_{b_k=1} C_k \right) \right] =  \left[\begin{array}{ccc|ccc|c}
								&       &       & 0      &\dots  & 0                & 0 \\
								\multicolumn{3}{c|}{G}& \vdots &\ddots & \vdots           & \vdots \\
								&       &       & 0      &\dots  & 0                & 0 \\ \hline
								0       &\dots  & 0     &        &       &                  &   \\
								\vdots  &\ddots &\vdots & \multicolumn{3}{c|}{(G^{-1})^\top}               & \vec{b}\\
								0       &\dots  & 0     &        &       &                  &  
							\end{array}\right]\, ,
						\end{align}
						as required.
					\end{claimproof}
				\end{lemma}
				From Lemma \ref{lem:a4}, each Clifford that preserves the computational basis has a stabiliser tableau of the form in Equation \ref{eqn:cbptableau}, for some $G \in \text{GL}_n(\mathbb{F}_2)$. By Lemma \ref{lem:a5}, this is the tableau of the Clifford implementing the map $\ket{\vec{f}} \mapsto \ket{G(\vec{f} \oplus\vec{b})}$, for some $\vec{b} \in \mathbf{F}_2^n$. This completes the proof.
				
			\end{proof}
		\end{theorem}

		\begin{corollary}\label{cor:affinemap} [Generalisation of Lemma 7.1 in \cite{harrison_sierpinski_2024}]
			Let $G \in \text{GL}_n(\mathbb{F}_2)$ and $\vec{b} \in \mathbb{F}_2^n$. The Pauli representations $\Gamma_i \in \mathcal{P}_n$ of the Majorana operators of the ancilla--free fermion--qubit mapping $\mathfrak{m} = ((\Gamma_{2i}, \Gamma_{2i+1}))_{i=0}^{n-1}$ with the Fock basis encoding $\ket{\vec{f}_{\mathfrak{m}}} = \ket{G(\vec{f} \oplus \vec{b})}$ for all $\vec{f} \in \mathbb{F}_2^n$ are
			\begin{align}
				\Gamma_{2i} &= (-1)^{\sum_{k=0}^{i-1} b_k} X_{U(i)}  Z_{R(i)}\, ,  \label{eqn:affine1} \\
				\Gamma_{2i+1} &= i(-1)^{\sum_{k=0}^{i} b_k} X_{U(i)} Z_{R(i)}\, , \label{eqn:affine2}
			\end{align}
			where $U(i)$, $P(i)$ and $R(i)$ are the update, parity and remainder sets of $G$ as in Definition \ref{defn:upf}.
			\begin{proof}
				Let $C_\mathfrak{m} \in \mathcal{C}_n$ be the Clifford operator implementing $C_\mathfrak{m} : \ket{\vec{f}} \mapsto \ket{G(\vec{f} \oplus\vec{b})}$ for all $\vec{f}\in \mathbb{F}_2^n$. By Theorem \ref{thm:affine}, Equation \ref{eqn:thm2main} displays the stabiliser tableau of $C_\mathfrak{m}$, and therefore
				\begin{alignat}{2}
					C_\mathfrak{m}(Z_i) &= C_\mathfrak{m} Z_i C_\mathfrak{m}^\dagger &&= (-1)^{b_i}Z_{F(i)} \\
					C_\mathfrak{m}(X_i) &= C_\mathfrak{m} X_i C_\mathfrak{m}^\dagger &&= X_{U(i)}  \\
					C_\mathfrak{m}(Y_i) &= -i C_\mathfrak{m} (Z_i X_i) C_\mathfrak{m}^\dagger && = (-i)(-1)^{b_i} Z_{F(i)} X_{U(i)}\, .
				\end{alignat}
				For Equation \ref{eqn:affine1}, observe
				\begin{align}
					\Gamma_{2i} = C_\mathfrak{m} \gamma_{2i}C_\mathfrak{m}^\dagger = C_\mathfrak{m} (Z_0 Z_1 \dots Z_{i-1} X_i) &= C_\mathfrak{m}(Z_0) C_\mathfrak{m}(Z_1) \dots C_{\mathfrak{m}}(Z_{i-1}) C_\mathfrak{m}(X_i) \\
					&= (-1)^{\sum_{k=0}^{i-1}b_k} Z_{F(0)} Z_{F(1)} \dots Z_{F(i-1)} X_{U(i)} \\
					&= (-1)^{\sum_{k=0}^{i-1}b_k} Z_{P(i)} X_{U(i)}\\
					& = (-1)^{\sum_{k=0}^{i-1}b_k} X_{U(i)} Z_{P(i)}\, ,
				\end{align}
				using Lemma \ref{lem:upf2} \ref{lem:upfb} to rearrange the Paulis in the last line. Similarly, for Equation \ref{eqn:affine2},
				\begin{align}
					\Gamma_{2i+1} = C_\mathfrak{m} \gamma_{2i+1}C_\mathfrak{m}^\dagger = (-i)C_\mathfrak{m} (Z_0 Z_1 \dots Z_{i-1} Z_i X_i) &= (-i) C_\mathfrak{m}(Z_0) C_\mathfrak{m}(Z_1) \dots C_{\mathfrak{m}}(Z_{i}) C_\mathfrak{m}(X_i) \\
					&= (-i)(-1)^{\sum_{k=0}^{i-1}b_k} Z_{F(0)} Z_{F(1)} \dots Z_{F(i-1)} Z_{F(i)} X_{U(i)} \\
					&= (-i)(-1)^{\sum_{k=0}^{i-1}b_k} Z_{R(i)} X_{U(i)}\\
					& = i(-1)^{\sum_{k=0}^{i-1}b_k} X_{U(i)} Z_{R(i)}\, ,
				\end{align}
				using Lemma \ref{lem:upf2} \ref{lem:upfc} and $R(i) = P(i) \, \triangle \, F(i)$. \end{proof}
		\end{corollary}

\end{document}